\documentclass[11pt,a4paper]{article}
\pdfoutput=1
\usepackage{epsfig,amsfonts,amssymb}
\usepackage{graphicx}
\usepackage{mathrsfs}
\usepackage{comment}
\usepackage{cite}
\usepackage{graphicx}
\usepackage{float}
\usepackage[normalem]{ulem}
\usepackage[utf8]{inputenc}
\usepackage[T1]{fontenc}
\usepackage{lmodern, textcomp}
\usepackage[english]{babel}
\usepackage{csquotes,cancel}

\usepackage[left=1in,right=1in,top=.8in,bottom=.8in]{geometry}

\usepackage{eurosym, xspace}
\usepackage{graphicx, subfig, float}
\usepackage{verbatim}

\usepackage{amsmath, amsfonts, amssymb,upgreek}
\usepackage{cancel}
\usepackage{mathtools}
\usepackage{slashed}

\usepackage[usenames, dvipsnames]{xcolor}
\usepackage[amsmath, hyperref, framed]{}
\usepackage{hhline}
\usepackage{bm}
\usepackage{array, caption}
\usepackage{array,multirow}

\newcolumntype{C}[1]{>{\centering\arraybackslash}m{#1}}

\usepackage[pdftex, breaklinks=true, linktocpage,
	colorlinks=true, urlcolor=blue, linkcolor=blue, citecolor=red
]{hyperref}

\usepackage{cite}

\usepackage{tikz-feynman}
\usepackage{tikz, pgf}
\usetikzlibrary{shapes.misc}
\usetikzlibrary{snakes}
\usetikzlibrary{decorations.pathmorphing}
\usetikzlibrary{arrows.meta}
\usetikzlibrary{shapes}
\usetikzlibrary{fit}

\makeatletter
\newsavebox{\@brx}
\newcommand{\llangle}[1][]{\savebox{\@brx}{\(\m@th{#1\langle}\)}%
  \mathopen{\copy\@brx\kern-0.5\wd\@brx\usebox{\@brx}}}
\newcommand{\rrangle}[1][]{\savebox{\@brx}{\(\m@th{#1\rangle}\)}%
  \mathclose{\copy\@brx\kern-0.5\wd\@brx\usebox{\@brx}}}
\makeatother

\def\ZZZ{{\hbox{ Z\kern-1.6mm Z}}}
\def\RRR{{\hbox{ R\kern-2.4mm R}}}
\def\CCC{{\hbox{ C\kern-2.0mm C}}}
\def\zzz{{\hbox{z\kern-1mm z}}}

\newcommand{\qeq}{{\hbox{=\kern-2.3mm ? \kern.5mm }}}
\renewcommand{\qeq}{=}

\newcommand{\be}{\begin{eqnarray}}
\newcommand{\ee}{\end{eqnarray}}

\newcommand{\vp}{\varphi}

\newcommand{\ben}{\begin{eqnarray}\displaystyle}
\newcommand{\een}{\end{eqnarray}}

\newcommand{\p}{\partial}

\def\one{{\hbox{ 1\kern-.8mm l}}}
\def\zero{{\hbox{ 0\kern-1.5mm 0}}}

\newcommand{\bea}[1]{\begin{eqnarray}\label{#1} }
\newcommand{\eea}{\end{eqnarray}}

\usepackage{bm}

\newcommand\non{\nonumber}

\newcommand\f{\frac}

\def\figone{

\def\JPicScale{0.8}
\ifx\JPicScale\undefined\def\JPicScale{1}\fi
\unitlength \JPicScale mm
\begin{picture}(135,80)(0,0)
\linethickness{0.3mm}
\multiput(40,80)(0.12,-0.18){167}{\line(0,-1){0.18}}
\linethickness{0.3mm}
\multiput(30,70)(0.18,-0.12){167}{\line(1,0){0.18}}
\linethickness{0.3mm}
\put(30,50){\line(1,0){30}}
\linethickness{0.3mm}
\multiput(30,30)(0.18,0.12){167}{\line(1,0){0.18}}
\linethickness{0.3mm}
\multiput(40,20)(0.12,0.18){167}{\line(0,1){0.18}}
\linethickness{0.3mm}
\put(60,50){\line(1,0){40}}
\linethickness{0.3mm}
\multiput(100,50)(0.12,0.18){167}{\line(0,1){0.18}}
\linethickness{0.3mm}
\multiput(100,50)(0.18,0.12){167}{\line(1,0){0.18}}
\linethickness{0.3mm}
\put(100,50){\line(1,0){30}}
\linethickness{0.3mm}
\multiput(100,50)(0.18,-0.12){167}{\line(1,0){0.18}}
\linethickness{0.3mm}
\multiput(100,50)(0.12,-0.18){167}{\line(0,-1){0.18}}

\put(30,80){\makebox(0,0)[cc]{$\zeta Q_B A_1^c$}}

\put(25,70){\makebox(0,0)[cc]{$A_2^c$}}

\put(25,50){\makebox(0,0)[cc]{$A_n^c$}}

\put(20,30){\makebox(0,0)[cc]{$A_1^o$}}

\put(35,20){\makebox(0,0)[cc]{$A_p^o$}}

\put(35,60){\makebox(0,0)[cc]{$\vdots$}}

\put(40,30){\makebox(0,0)[cc]{$\vdots$}}

\put(125,80){\makebox(0,0)[cc]{$B_1^c$}}

\put(135,70){\makebox(0,0)[cc]{$B_2^c$}}

\put(135,50){\makebox(0,0)[cc]{$B_m^c$}}

\put(135,30){\makebox(0,0)[cc]{$B_1^o$}}

\put(120,15){\makebox(0,0)[cc]{$B_q^o$}}

\put(115,35){\makebox(0,0)[cc]{$\vdots$}}

\put(120,57){\makebox(0,0)[cc]{$\vdots$}}

\put(66,47){\makebox(0,0)[cc]{$\psi_r^c\vp_r$}}

\put(93,47){\makebox(0,0)[cc]{$\psi_s^c\vp_s$}}

\end{picture}

}

\def\figtwo{

\def\JPicScale{0.8}
\ifx\JPicScale\undefined\def\JPicScale{1}\fi
\unitlength \JPicScale mm
\begin{picture}(135,80)(0,0)
\linethickness{0.3mm}
\multiput(40,80)(0.12,-0.18){167}{\line(0,-1){0.18}}
\linethickness{0.3mm}
\multiput(30,70)(0.18,-0.12){167}{\line(1,0){0.18}}
\linethickness{0.3mm}
\put(30,50){\line(1,0){30}}
\linethickness{0.3mm}
\multiput(30,30)(0.18,0.12){167}{\line(1,0){0.18}}
\linethickness{0.3mm}
\multiput(40,20)(0.12,0.18){167}{\line(0,1){0.18}}
\linethickness{0.3mm}
\put(60,50){\line(1,0){40}}

\put(30,80){\makebox(0,0)[cc]{$\zeta Q_B A_1^c$}}

\put(25,70){\makebox(0,0)[cc]{$A_2^c$}}

\put(25,50){\makebox(0,0)[cc]{$B_m^c$}}

\put(20,30){\makebox(0,0)[cc]{$A_1^o$}}

\put(35,20){\makebox(0,0)[cc]{$B_q^o$}}

\put(35,60){\makebox(0,0)[cc]{$\vdots$}}

\put(40,30){\makebox(0,0)[cc]{$\vdots$}}

\put(66,47){\makebox(0,0)[cc]{$\psi_r^c\vp_r$}}

\put(93,45){\makebox(0,0)[cc]{$\tilde\psi_s^c\tilde\vp^s$}}

\put(100,50){\makebox(0,0)[cc]{$\times$}}

\end{picture}

}

\definecolor{armygreen}{rgb}{0.29, 0.33, 0.13}

\begin{document}

\baselineskip 24pt

\begin{center}
{\Large \bf  Flat space spinning massive amplitudes from momentum space CFT }

\end{center}

\vskip .6cm
\medskip

\vspace*{4.0ex}

\baselineskip=18pt

\begin{center}

{\large 
\rm  Raffaele Marotta$^a$, Kostas Skenderis$^b$ and Mritunjay Verma$^{b,c}$ }

\end{center}

\vspace*{4.0ex}

\centerline{\it \small $^a$Istituto Nazionale di Fisica Nucleare (INFN), Sezione di Napoli, }
\centerline{ \it \small Complesso Universitario di Monte S. Angelo ed. 6, via Cintia, 80126, Napoli, Italy}
\centerline{ \it \small $^b$ Mathematical Sciences and STAG Research Centre, University of Southampton,}
\centerline{\it \small  Highfield, Southampton SO17 1BJ, UK}

\centerline{\it \small $^c$ Indian Institute of Technology Indore, Khandwa Road, Simrol, Indore 453552, India}
\vspace*{1.0ex}
\centerline{\small E-mail:  raffaele.marotta@na.infn.it, k.skenderis@soton.ac.uk, 
mritunjay@iiti.ac.in }

\vspace*{5.0ex}

\centerline{\bf Abstract} \bigskip

We discuss the flat space limit of AdS using the momentum space representation of CFT correlators. 
The flat space limit involves sending the AdS radius and the dimensions of operators dual to massive fields to infinity while also scaling appropriately the sources of the dual operators.
In this limit, $d$-dimensional CFT correlators  become $(d+1)$-dimensional scattering amplitudes.
We exemplify our discussion with the computation of the flat-space limit of the CFT 3-point function of a conserved current, a non-conserved charged vector operator and its conjugate. The flat-space limit should yield the scattering amplitude of an Abelian gauge field with two massive vector fields.
This scattering amplitude computes the electromagnetic form factors of the electromagnetic current in a spin-1 state, and these form factors encode the electromagnetic properties of the massive vector field (charge, magnetic moment and quadruple moment). In terms of the CFT, the flat-space limit amounts to zooming in the infrared region of the {\it triple-K} integrals that determine the 3-point function, while also scaling to infinity the order of (some of) the Bessel functions that feature in the {\it triple-K} integrals. In this limit the {\it triple-K} integral becomes proportional to the energy-preserving delta function, and the flat space limit correctly yields the corresponding flat space scattering amplitude in complete detail.

\vfill

\vfill \eject

\baselineskip18pt

\tableofcontents

\section{Introduction } 
\label{s1}
The AdS/CFT gives a realization in string theory of the holographic principle, providing,  at least conceptually,  a non-perturbative formulation of string theory on AdS background in terms of a boundary conformal field theory \cite{9310026,9409089,9711200}. In its most general formulation, the correspondence is conjectured to be a duality between a quantum gravity theory formulated on a $(d+1)$-dimensional asymptotically locally AdS background (AlAdS) times a compact manifold and a $d$-dimensional quantum field theory located on the boundary of AlAdS \cite{9802109,9802150}. The strong/weak nature of this duality can be exploited to explore the strong-coupling regime of the dual conformal field theories which are dual to a weakly coupled classical bulk theory. 
A weakly coupled bulk theory corresponds to the large radius limit of AdS. As the AdS radius approaches infinity, the AdS geometry reduces to the flat space geometry\footnote{In the most well understood example of duality, namely when the bulk type IIB string theory is dual to $\mathcal{N}=4$ SYM, the relation between the AdS radius and the boundary parameters is 
\be
L\;\; \sim\;\; \bigl(g_{YM}^2N\bigl)^{\f{1}{4}}
\ee
In the 't Hooft limit, one simultaneously sends $N$ to infinity and $g_{YM}^2$ to zero keeping L large but fixed. For the flat space limit, one needs to consider the more subtle limit in which we again send $N$ to infinity but we now keep $g_{YM}^2$ fixed so that $L \to \infty$ \cite{Susskind:1998vk, 9901076}. \label{ft:limit}  } and, for consistency, the physics in AdS in this limit should match that of flat space (at least locally). In particular, we could obtain some insight about  quantum gravity in flat space by using the 
flat-space limit of the AdS/CFT correspondence.

Motivated by this there has been a body of work since the early days of the AdS/CFT correspondence discussing the flat limit of AdS results, starting from \cite{Susskind:1998vk, 9901076, Giddings:1999qu, 9907129}. Due to the AdS/CFT correspondence, the limit should also make sense on the CFT side at the level of CFT correlators, at least for holographic CFTs, and $(d+1)$ dimensional flat space-time should emerge from $d$-dimensional CFT correlator in a suitable limit. However, it was also clear from the very beginning that the limit is subtle, and it has been a challenge to make the plausible physical picture into a precise and mathematically well-defined limit. The limit has been analyzed in a variety of different formulations and setups: position space \cite{0903.4437, 1002.2641,Maldacena:2015iua, 2007.13745}, Mellin space \cite{Penedones:2010ue, Fitzpatrick:2011hu, Paulos:2016fap},  partial wave expansion \cite{Maldacena:2015iua, Paulos:2016fap}, momentum space \cite{1201.6449, 1201.6452, Hijano:2019qmi, Hijano:2020szl, Farrow:2018yni, 2204.06462}, see also \cite{2106.04606} for a comparison of the different formulations, and 
\cite{0904.3544, 0907.0151, Fitzpatrick:2010zm, Fitzpatrick:2011ia, Fitzpatrick:2011jn, Maldacena:2011nz,Goncalves:2014ffa, 1912.10046, Caron-Huot:2021kjy, Chandorkar:2021viw, vanRees:2022itk, Duary:2022pyv
} for further work. One outcome of these works is that the flat space limit is a singular limit. For example, in the momentum space approach of \cite{1201.6449,1201.6452}, $(d+1)$-dimensional flat space amplitudes involving gluons and gravitons were obtained from the coefficients of singular terms of the flat limit of $d$-dimensional CFT correlators involving the conserved currents and stress-energy tensor, respectively. 

The flat-space limit provides a link to flat space holography. There have been different approaches to flat space holography, including celestial holography and Carrollian holography, and connections to the flat-space limit have been discussed, for example, in 
\cite{Lam:2017ofc, Casali:2022fro,deGioia:2022fcn, Iacobacci:2022yjo,Banerjee:2022oll,Sleight:2023ojm,deGioia:2023cbd,Bagchi:2023fbj,Mason:2023mti,deGioia:2024yne}. We are not going to discuss these interesting proposals in this paper\footnote{We will also not  discuss whether the limit exists as a limit of the dual CFT as a theory ({\it c.f.} footnote \ref{ft:limit}) or as a limit of the bulk geometry ({\it c.f.}, for example, \cite{Barnich:2012aw}).}, but we note that a minimal possibility for flat space holography is that {\it it is the flat-space limit of the AdS/CFT}, with the flat space results emerging from correlators of standard relativistic CFT in a suitable limit.  

Many of the prior works focused on special cases ({\it e.g.} scalar 4-point functions computed by Witten diagrams, bulk massless fields, {\it etc.}). In this work we aim to provide a formulation that would apply in generality:
any $n$-point function of massless and massive spinning fields with general interactions. We will focus our analysis in the simplest setup that involves most of these ingredients while it is also physically interesting: the 3-point function of an abelian gauge field with a massive spin-1 complex Proca field.
Our aim is to obtain the scattering of the photon off a massive vector field (Figure \ref{diag:Mink}) by taking a limit of the corresponding process in AdS (Figure \ref{diag:EAdS}). In flat space this scattering process captures the electromagnetic properties of the massive particle (charge, magnetic and quadrupole moments for a spin one particle) and as such it is interesting on its own right. In particular, our analysis may pave a way to obtain non-perturbative results about electromagnetic form factors of higher-spin (hadronic) states using holography and CFT results.

3-point functions in CFT are fixed by conformal invariance, up to constants, so this is a case where the results is known non-perturbatively, and it would allow us to directly take the limit on the CFT side.
On the other hand, to understand what is the precise limit to be taken, it is useful to have a bulk realization in AdS. We will work with Euclidean signature in AdS with flat boundary (AdS in Poincar\'e coordinates, or more accurately with the boundary conformal structure of AdS  represented by a flat metric). We will Fourier transform along the boundary directions and, correspondingly, we will consider the CFT in momentum space. 

In AdS/CFT correspondence, the massive field is dual to a non conserved operator whereas the gauge field is dual to a conserved current in the boundary theory, so the relevant CFT 3-point function is that of a conserved current with a non-conserved vector operator and its complex conjugate. This 3-point function (in momentum space) was determined in our earlier work \cite{Marotta:2022jrp} by solving the conformal Ward identities, following \cite{Bzowski:2013sza,1510.08442,Bzowski:2017poo,Bzowski:2018fql}, and it depends on the conformal dimension $\Delta$ of the non-conserved operator, the spacetime dimension $d$ and three parameters, whose values are theory-specific.  

In AdS, we work with the most general effective action of the Proca field coupled of an abelian  gauge field, including  up to three derivative terms. This action involves three coupling constants: the minimal coupling, and two more couplings that may be associated with the magnetic and quadrupole moments of the massive spin-one field.
This action might be thought as arising from a compactification of ten or eleven dimensional supergravity, where the massive vectors correspond to Kaluza-Klein modes of some higher-dimensional field. The boundary values of the bulk fields act as the sources of the corresponding boundary operators and the holographically renormalized bulk partition function  provides the generating functional of the boundary CFT correlators. We work out the 3-point function using the original GKPW prescription \cite{9802109,9802150} and holographic renormalization \cite{0209067}. 
Comparison of the 3-point function computed using the AdS/CFT correspondence with the general CFT 3-point function shows that there is an 1-1 relation between the three arbitrary parameters that appear in the solution of the conformal Ward identities and the three AdS bulk coupling. This relation depends on the AdS radius $L$ and the conformal dimension $\Delta$ of the non conserved operators and is valid in the regime where the boundary theory is strongly coupled. This explicit matching provides a
non trivial test of AdS/CFT correspondence for the massive spin-1 field described by a higher derivative effective action.

After computing the above 3-point function, we analyse it in the flat limit where we send the AdS radius $L$ to infinity. The flat space amplitudes arise from the bulk region where the AdS metric reduces to the flat metric with the vanishing Ricci tensor and Ricci scalar. In the standard Poincar\'e coordinates (see equation \eqref{stanpoin54}), the Ricci tensor can be expressed in terms of the radial coordinate $z$ as $R_{MN} =-d\, \delta_{MN}/z^2$ ($M, N=0,\dots d$). Therefore, the dominant region in the flat limit corresponds to the deep interior of the AdS background where $z$ is large. 
We parametrized this AdS region as $z=L\,e^{\frac{\uptau}{L}}$.
In the flat limit, $\uptau$ is interpreted as Euclidean time.\footnote{We work in the Euclidean AdS signature and Wick rotate the radial direction to make it time like after taking the flat limit.} Further, in this flat region, the $AdS$ isometry algebra becomes the Poincar\'e algebra through the Inonu Wigner contraction  \cite{InonuWigner}. In particular, the AdS isometries include scaling and special conformal transformation, and we show how in the flat space limit these isometries disappear and instead we obtain  translational invariance in $\uptau$ together with Lorentz transformations that rotate $\uptau$ to the other boundary directions.

\begin{figure}[!tbp]
  \centering
  \begin{minipage}[b]{0.45\textwidth}
   \begin{center}
\begin{tikzpicture}[scale=.8]
\node[above] at (0,3) {$i^+$};
\node[below] at (0,-3) {$i^-$};
\node[right] at (1.5,1.7) {$\mathscr{I}^+$};
\node[right] at (3,0) {$i^0$};
\node[right] at (1.5,-1.7) {$\mathscr{I}^-$};
\node[right] at (0.3,-1.0) {$W$};
\node[right] at (-0.4,1.3) {$W$};
\node[right] at (0.7,0.5) {$\gamma$};
\draw [-, snake=coil] [ thick](0,0) -- (1.5,1.5);
\draw [-] [ thick](-3,0) -- (0,-3);
\draw [-] [ thick](3,0) -- (0,-3);
\draw [-] [ thick](0,3) -- (3,0);
\draw [-] [ thick](-3,0) -- (0,3);
\draw plot [smooth] coordinates {(0,-3) (.5,-1.5) (0,0) (0-.5,1.5) (0,3)};
\end{tikzpicture}
\end{center}
\caption{Scattering of a photon $\gamma$ off a massive spin-1 particle $W$ in Minkowski spacetime.}
\label{diag:Mink}
  \end{minipage}
  \hfill
  \begin{minipage}[b]{0.45\textwidth}
   \begin{center}
\begin{tikzpicture}[scale=.8]
\draw (0,0) circle (3cm);
\draw [-, snake=coil] [ thick](2.1,2.1) -- (0,0);
\draw [-] [ thick](-2.1,2.1) -- (0,0);
\draw [-] [ thick](0,0) -- (0,-3);
\node[right] at (0.0,-1.5) {$W$};
\node[right] at (-1.1,1.3) {$W$};
\node[right] at (0.7,0.5) {$\gamma$};
\end{tikzpicture}
\end{center}
\caption{Same process as in Fig. \ref{diag:Mink} but now in Euclidean AdS.\newline } 
\label{diag:EAdS}
  \end{minipage}
\end{figure}

We would like to take the flat space limit in a way that keeps the physics we want to probe. Suppose we want to compute the scattering amplitudes for a theory described by flat space by a Lagrangian $L_{\rm flat}[m^2_i, g_j]$ that depends on set of massless fields, massive fields with masses $m_i^2$ and coupling $g_j$ via a flat-space limit from AdS. Then the proposal is to start with the same action now in AdS (with AdS radius $L$) and then consider the flat space limit $L \to \infty$ keeping fixed the masses $m_i^2$ and coupling $g_j$ (in Planck units). Given the standard relation between masses and conformal dimensions, for example $m^2 L^2 = \Delta (\Delta-d)$ for scalar fields (or equation \eqref{3.13} for the case we consider), keeping fixed the mass implies that the conformal dimension must tend to infinity, $\Delta \to \infty, $ as $L \to \infty$. The crucial question is then whether AdS amplitudes, or more generally CFT correlators, admit such a limit. 

The main building blocks for momentum space CFT 3-point functions are the so-called $triple$-$K$ integrals \cite{Bzowski:2013sza},
\begin{eqnarray} \label{Intro_3K}
J_{N\{k_1,k_2,k_3\}}(p_1,p_2,p_3)&\equiv&\int_0^\infty dx\,x^{\frac{d}{2}+N-1} \prod_{i=1}^3 p_i^{\Delta_i-\frac{d}{2}+k_i}\,K_{ \Delta_i-\frac{d}{2}+k_i}(x p_i)\, .	
\end{eqnarray}
where $p_i$ are the magnitudes of momenta, $p_i = \sqrt{{\bf p}_i^2}$,
$K_{ \Delta_i-\frac{d}{2}+k_i}(x p_i)$ are modified Bessel functions of the second kind and $N$ and $k_i$ are parameters (which are integers in the cases we discuss). In this integral, the $x=0$ region is the UV part of the integral, while the $x \to \infty$ corresponds to the IR part of the integral. In the AdS computation these integrals arise from the corresponding Witten diagrams with the Bessel functions being the (momentum-space) bulk-to-boundary propagators and the integral over $x$ originating from the integral over the bulk vertex, with $x$ identified with the AdS radial coordinate. The flat-space limit corresponds to considering the deep interior of AdS, $z \to \infty$, and thus the IR region of the $triple$-$K$ integral.
In the flat-space limit the momenta along the boundary directions become the spatial momenta of the flat-space scattering amplitude, and thus we want to keep fixed ${\bf p}_i$ as $x \to \infty$. In addition, we need to send $\Delta \to \infty$ when the corresponding bulk field is massive. 

Thus, the flat-space limit rests (in part) in our ability to take the limit of the $triple$-$K$ integrals. 
For massless fields this involves taking the large argument limit of a modified Bessel function, while for massive fields we need to take a limit where both the argument and the order and the argument of the Bessel function tends to infinity. This former limit is well known, but the latter (called uniform expansion in the mathematics literature) is less known and we review it in detail in appendix \ref{UnEx}.
The limits of the Bessel function also tell us how the AdS bulk-to-boundary propagators behave in this limit, and after Wick-rotating to Minkowski spacetime, the answer is that they tend to plane waves,
\begin{equation} \label{K_lim}
    {K}_{\Delta-\frac{d}{2}+\ell}(z\,k) \to \frac{1}{\sqrt{Z_\Delta}} e^{-i E t}
\end{equation}
where $t=-i \uptau$ is Minkowski time (with $\uptau=L \log (z/L)$).
$E=\pm \sqrt{k^2+m^2}$ is the energy variable of the flat-space $(d+1)$-momentum vector, 
$(E, {\bf k})$, where ${\bf k}$ is the momentum vector in the CFT. In other words, the momentum variable of the CFT directly becomes the spatial part of the momentum variable in flat space and the energy variable is what is dictated by the on-shell condition. Note that the correct on-shell relation for $E$ automatically emerges from the limit. The two signs correspond to whether after Wick-rotation the plane wave corresponds to in- or out-state. The factor $Z_\Delta$ is a renormalization factor. In the cases we discuss, the $Z$-factor tends to infinity for the massless photon and to zero for the massive vector. One would need to renormalize the CFT operators by precisely these factors in order for the flat-space limit to exist.  Using (\ref{K_lim}) in (\ref{Intro_3K}) we find that the $triple$-$K$ integral becomes (proportional) to the energy-preserving delta function,
\begin{eqnarray}
\lim_{L \to \infty}  J_{N\{k_1,k_2,k_3\}}(p_1,p_2,p_3) \sim \delta(E_1+E_2+E_3)   
\end{eqnarray}
where the limit is taken with $\Delta_i/L=m_i$ fixed. Note that the conservation of the spatial momentum is automatic since the momentum space CFT 3-point functions already contain the momentum-preserving delta function, $\delta({\bf k}_1 + {\bf k}_2 + {\bf k}_3)$. 
To complete the flat-space limit of the 3-point function one needs to take the limit of the form factors (introduced in equation \eqref{2.5}) and these involve factors of $\Delta$ (which follow from the solution of the conformal Ward identities). These factors are crucial in order to obtain the correct flat space result, 
 \begin{eqnarray}
\lim_{L \to \infty} \sqrt{Z_{W_1} Z_A Z_{W_3}} \,A_3^{\mu_1\mu_2\mu_3} \;=\; -2 \pi i
\delta(E_1+E_2+E_3)\, {\cal M}_3^{\mu_1\mu_2\mu_3}\, ,
\end{eqnarray}
where $A_3^{\mu_1\mu_2\mu_3}$ is the momentum-space CFT 3-point function and ${\cal M}_3^{\mu_1\mu_2\mu_3}$ is the flat space scattering amplitude.
 
Together with the 3-point function we also analyse the flat limit of the AdS propagators, with the boundary directions Fourier transformed to momentum space. Again, the flat limit of these propagators corresponds to sending $L$ and $\Delta$ to infinity. An important role is played by the bulk to boundary (Btb) propagators of the gauge and Proca fields. These dictate the external leg factors of the fields in the flat limit which turns out to be very crucial for matching the flat space 3-point amplitude with the CFT 3-point function. More generally, the solution of the field equations in AdS properly limit into corresponding solutions in flat space. The AdS solutions depend on the fields that parametrize their boundary conditions (which play the role of sources in AdS/CFT) and these morph into polarization vectors in the flat space limit. 

We also consider the bulk-to-bulk (BtB) propagator of the gauge field. Even though we only need its near boundary behaviour in computing the 3-point function via holographic renormalisation, we have analysed the flat limit of the full BtB propagator in  momentum space. Since this propagator plays the role of Green's function in AdS, we expect it to limit to the Feynman propagator since the latter also plays the role of Green's function in flat space. We find that this is indeed the case, as expected. However, this analysis gives an 
interesting insight about the longitudinal part of the propagator. As is common in AdS/CFT, we used the radial/axial gauge where $A_0=0$. In the flat space limit, the transverse part of the gauge BtB propagator matches exactly with the transverse part of the Feynman propagator in the flat space limit, while the longitudinal part divergences. This divergence is precisely linked with an additional singularity (an unphysical double pole) that is present in the  Feynman propagator in the axial gauge in flat space \cite{Caracciolo:1982dp, Leibbrandt:1987}, and our results match these earlier results.

The rest of the paper is organised as follows. In section \ref{sec2review3er}, we review results obtained in previous literature: we summarise the expression of the momentum-space CFT 3-point function involving a conserved current and two generic non conserved operators having the same conformal dimension, and we also review results about the flat limit of AdS at the geometric and group algebra level. In section \ref{secpoincareads}, we explicitly show how the AdS isometries limit to the Poincar\'e isometries and how the scaling and special conformal symmetry of the CFT correlators recombine to Poincar\'{e} transformations in the large $L$ limit. In section \ref{s3a}, we shall introduce the bulk theory involving a gauge field and two charged massive spin-1 fields and derive the boundary CFT 3-point function using this bulk theory and the procedure of holographic renormalisation. This fixes the coefficients appearing in the CFT 3-point function in terms of bulk quantities. In section \ref{flat}, we analyse the flat limit of the BtB propagator of the gauge field and Btb propagators of the gauge and Proca fields. In section \ref{sec6flatdr}, we consider the flat space limit of the 3-point function and show that it matches with the expected result in the flat space. We end with some discussion in section \ref{s4}. 

The papers contains a number of technical computations, which require dealing with many subtle issues. While the techniques and subtleties are all known by the experts, detailed expositions are rare in the literature and we present a comprehensive analysis in a series of appendices. 
Appendix \ref{geo768} contains our conventions, and in appendix \ref{UnEx} we discuss the limiting behaviour of the modified Bessel functions. In particular, we present a self-contained discussion of the uniform expansion of the Bessel function when both the argument and the order of the Bessel function goes to infinity. Appendix \ref{appen:D} contains a derivation of the most general form of effective action in AdS, which contains up to cubic terms in the gauge and Proca fields, and up to the three derivative  terms. This is the starting point for our holographic computation in section \ref{s3a}.
In appendix \ref{Classical} we compute the bulk-to-boundary and the bulk-to-bulk propagators of the gauge field in axial gauge, and the bulk-to-boundary propagator for the Proca field. Appendix 
\ref{app:lorenz} contains the computation of the gauge field bulk-to-boundary propagator in Lorenz gauge.
In appendices \ref{s3proca} and \ref{s3} we work out holographic renormalization for the Proca and gauge field, respectively. The massive spin-1 field corresponds to an irrelevant operator and this requires special attention. Appendix \ref{exact} contains the computation of the corresponding flat space scattering amplitude. Finally, in appendix \ref{app: multipoles} we present a self-contained summary of the relation between electromagnetic form factors and couplings in the effective action.

\section{Review of CFT results}
\label{sec2review3er}
In this section, we summarise the CFT 3-point function involving a conserved current and two non conserved spin 1 fields in momentum space following \cite{Marotta:2022jrp}. This will be needed later to compare with the bulk 3-point function of a gauge field and two massive spin-1 Proca fields. The results in \cite{Marotta:2022jrp} are given in an index free notation where Lorentz indices have been contracted with auxiliary vectors. Here, we state the result in terms of explicit indices which will be more useful for our purposes.

The desired 3-point correlator  was determined from the CFT Ward identities. Extracting the momentum conserving delta function, it can be expressed as
\be
\mathcal{A}_3^{\mu_1\mu_2\mu_3}&=& (2\pi)^d \delta^d({\bf p}_1+{\bf p}_2+{\bf p}_3) \;\llangle[\Bigl]\mathcal{O}_{1}^{\mu_1}({ \bf p}_1) {\cal J}^{\mu_2}({\bf p}_2) \mathcal{O}_{3}^{\mu_3}({\bf p}_3)   \rrangle[\Bigl]  \label{2.1we}
\ee
The operators $\mathcal{O}_1$ and $\mathcal{O}_3$ can have different conformal dimensions, say $\Delta_1$ and $\Delta_3$ respectively. However, in our case, they will correspond to bulk fields with the same mass, hence, we shall take $\Delta_1=\Delta_3=\Delta$. The reduced correlator in \eqref{2.1we} can be decomposed in a transverse and longitudinal part as 
\be
&&\hspace*{-1.39cm}\llangle[\Bigl] \mathcal{O}_{1}^{\mu_1}({\bf p}_1)\, {\cal J}^{\mu_2}({\bf p}_2) \,\mathcal{O}_{3}^{\mu_3}({\bf p}_3)    \rrangle[\Bigl]\non\\[.2cm]
&=&  \llangle[\Bigl]\mathcal{O}_{1}^{\mu_1}({\bf p}_1)\, { j}^{\mu_2}({\bf p}_2) \,\mathcal{O}_{3}^{\mu_3}({\bf p}_3) \rrangle[\Bigl]\;\;{+}\;\;\f{p_2^{\mu_2}}{p_2^2}\llangle[\Bigl]  \mathcal{O}_{1}^{\mu_1}({\bf p}_1)\, p_{2\nu}{\cal J}^{\nu}({\bf p}_2) \,\mathcal{O}_{3}^{\mu_3}({\bf p}_3) \rrangle[\Bigl]\, ,  \label{decnb4}
\ee
where $j^\mu$ denotes the transverse part of the conserved current
\be
j^\mu({\bf p}_2) = \pi^{\mu}_{\;\;\nu}({\bf p}_2) {\cal J}^\nu({\bf p}_2), \qquad\quad \pi^{\mu\nu}({\bf p}_2) =\delta^{\mu\nu}-\f{p_2^\mu p_2^\nu}{p_2^2}, \qquad \qquad p_2^\mu\; \pi_{\mu\nu}({\bf p}_2) =0\, .
\ee
The second term on the right hand side of \eqref{decnb4} is the longitudinal contribution and the conservation Ward identity for the symmetry current relates it to the 2-point function of the operators $\mathcal{O}^\mu$. 
This relates one of the coefficients of the 3-point function with  
the normalization of the 2-point function of  $\mathcal{O}^\mu$, as we discuss below. 
Focusing on the transverse part, we decompose it in form factors,
\be
\llangle[\Bigl]  \mathcal{O}_{1}^{\mu_1}({\bf p}_1)\, j^{\mu_2}({\bf p}_2) \mathcal{O}_{3}^{\mu_3}({\bf p}_3)   \rrangle[\Bigl]&=&(\pi\cdot p_1)^{\mu_2}A^{\mu_1\mu_3}+ \pi^{\mu_2\mu_1}B^{\mu_3} +\pi^{\mu_2\mu_3}C^{\mu_1}\,  ,\label{3ptrf}
\ee
where
\be
A^{\mu_1\mu_3}&=& A_1\;\delta^{\mu_1\mu_3} +A_2\;p_1^{\mu_1}\,(p_1+p_2)^{\mu_3}+A_3\;p_2^{\mu_1}\,(p_1+p_2)^{\mu_3}+A_4\;p_1^{\mu_1}p_2^{\mu_3}+A_5\;p_2^{\mu_1} p_2^{\mu_3}\, ;\non\\[.3cm]
B^{\mu_3}&=&B_1\;(p_1+p_2)^{\mu_3}+B_2\;p_2^{\mu_3}\, ;\non\\[.3cm]
C^{\mu_1}&=&C_1\;p_1^{\mu_1}+C_2\;p_2^{\mu_1}\, .\label{2.5}
\ee
The form factors $A_i , B_k, C_k\  (i=1,...,5, k=1,2)$ depend on the magnitudes of the momenta, $p_j=|{\bf p}_j| = \sqrt{{\bf p}_j^2}\  (j=1, 2, 3)$. In the above expressions
we used the momentum conserving delta function to express $p_3^\mu=-p_1^\mu-p_2^\mu$.\footnote{In \cite{Bzowski:2013sza} the momentum conserving delta function was solved differently for different indices, 
$ \mu_1 \rightarrow {\bf p}_1,  {\bf p}_2, \ 
 \mu_2 \rightarrow {\bf p}_2, {\bf p}_3, \ 
\mu_3 \rightarrow {\bf p}_3, {\bf p}_1$. This results in form factors $\tilde{A}, \tilde{B}, \tilde{C}$ that relate to the ones we use here by
\be
A_1 &=& -\tilde A_1\quad,\quad A_2= \tilde A_4-\tilde A_2\quad,\quad A_3= \tilde A_5-\tilde A_3\quad,\quad A_4= \tilde A_4\quad,\quad A_5=\tilde A_5\non\\[.3cm]
B_1&=&\tilde B_2-\tilde B_1\quad,\quad  B_2 = -\tilde B_2\quad,\quad  C_1 = \tilde C_1\quad,\quad  C_2 = \tilde C_2\, . \non
\ee}

 As discussed in section 3.5 of \cite{Marotta:2022jrp}, the correlator is antisymmetric under exchange of $(\mu_1, p_1)$ and $(\mu_3, p_3)$ that this implies,
\begin{align} \label{symm}
A_i(p_1, p_2, p_3) &= A_i(p_3, p_2, p_1), \quad i=1, 2, 5, \qquad A_3(p_1, p_2, p_3)  = -A_4 (p_3, p_2, p_1) \\
B_1(p_1, p_2, p_3) & =C_1(p_3, p_2, p_1), \qquad
B_2(p_1, p_2, p_3) =-C_2(p_3, p_2, p_1)\, .\non
\end{align}

The functions $A_i, B_i$ and $C_i$ are determined by solving the Ward identities, and they are given in terms of triple-$K$ integrals:
\be
 A_1 &=& -a_5 J_{2\{0,1,0\}}+ a_1J_{1\{0,0,0\}}\, ;  \non\\[.2cm]
  A_2 &=& -a_5 J_{3\{-1,2,-1\}}+ a_2J_{1\{-1,0,-1\}} 
  +2 a_4 J_{2\{-1,1,-1\}} \, ;   \non\\[.2cm]
  A_3 &=& -A_4= a_5 J_{3\{0,1,-1\}}- a_4 J_{2\{0,0,-1\}}\, ;  \non\\[.2cm]
 A_5 &=& a_5 J_{3\{0,0,0\}}\, ;\non\\[.2cm]
   B_1 &=& C_1= -a_5 J_{2\{0,1,0\}}+ b_1J_{1\{0,1,-1\}}+(b_1-b_2)J_{1\{1,0,-1\}} +(b_1-b_2 + a_4)J_{1\{0,0,0\}}\, ; \non\\[.3cm]
       B_2&=& -C_2=-a_5 J_{2\{0,0,1\}}+ b_2J_{1\{0,0,0\}}\, ;,\label{2316a}
\ee
where $J_{N\{k_1,k_2,k_3\}}$ denote the triple K integrals and are defined by
\begin{eqnarray}
J_{N\{k_1,k_2,k_3\}}(p_1,p_2,p_3)&\equiv&\int_0^\infty dx\,x^{\frac{d}{2}+N-1} \prod_{i=1}^3 p_i^{\Delta_i-\frac{d}{2}+k_i}\,K_{ \Delta_i-\frac{d}{2}+k_i}(x p_i)\, .	\label{B.51}
\end{eqnarray}
For more details and useful properties of these integrals, see \cite{Bzowski:2013sza, 1510.08442, Bzowski:2015yxv}.
Note that \eqref{2316a} already satisfy the symmetry constraints \eqref{symm}.

The 3 point function of a conserved current and two arbitrary spin 1 operators with the same conformal dimension $\Delta_1=\Delta_3=\Delta$ is given in terms of only 3 independent parameters. This means that not all the parameters $a_i, b_i$ 
in \eqref{2316a} are independent. There are relations among different constants and three of the constants are fixed in terms of the remaining three  as 
\be
 a_1&=& (d-2)\Delta a_5-(\Delta-1)a_4 +b_2 \quad;\quad
     a_2 =2(d-2)\Delta a_5  -(2\Delta+d-4)a_4   +\f{(2\Delta-d)}{(\Delta-1)}b_2\non\\[.3cm]
    b_1&=&\f{(2\Delta-d)}{2(\Delta-1)}b_2\, .
\label{2349afnewd}
\ee
Thus, the 3-point function is parametrised by three independent parameters as expected, and we have chosen $a_4, a_5$ and $b_2$  to be the three independent parameters. 
One of these parameters is fixed in terms of the normalisation of the non-conserved operator. Indeed,
the 2-point function of operators $\mathcal{O}_1$ and $\mathcal{O}_3$ is given by \cite{Marotta:2022jrp} 
\be
\llangle[\Bigl]  \mathcal{O}^*_{\mu}(p) \mathcal{O}_{\nu}(-p)   \rrangle[\Bigl]&=&a_0\left[ \delta_{\mu\nu} -\left(\f{2\Delta-d}{\Delta-1}\right)\f{p_\mu p_\nu}{p^2} \right]p^{2\Delta-d}\,  ,\label{3ptrfdrt}
\ee
Now, the generating functional of the CFT correlators is given by
\be
Z[{A}^{}_{(0) \mu}, \mathcal{W}^{}_{(0) \mu},\mathcal{W}^{*}_{(0) \mu}]= \int \mathcal{D}\Phi\;\exp\biggl[-S_{CFT}\; -\int d^dx \Bigl(\mathcal{J}^\mu {A}^{}_{(0)\mu} + \mathcal{O}^{*\mu}\mathcal{W}^{}_{(0) \mu}+ \mathcal{O}^{\mu}\mathcal{W}^{*}_{(0) \mu}\Bigl)\biggl]
\ee
where ${A}^{}_{(0) \mu}, \mathcal{W}^{}_{(0) \mu}$ and $\mathcal{W}^{*}_{(0) \mu}$ are the sources for the CFT operators $\mathcal{J}^\mu, \mathcal{O}^{*\mu} $ and $\mathcal{O}^{\mu}$, respectively. In the AdS/CFT correspondence, these sources are the fields that determine the boundary conditions of the corresponding bulk fields. Demanding invariance of the generating functional under the $U(1)$ transformation, namely
\be
\delta A_{(0) \mu}(x) = \p_\mu\lambda(x)\;;\quad \delta \mathcal{W}^{}_{(0)\mu} = i g\lambda(x){\mathcal{W}^{}_{(0) \mu}} \;;\quad \delta \mathcal{W}^{*}_{(0)\mu} = -i g\lambda(x){\mathcal{W}^{*}_{(0) \mu}}
\ee
we find the conservation ward identity
\begin{equation}
    \partial^\mu \langle \mathcal{J}^\mu(x) \rangle_s 
    = i g \left(
    \mathcal{W}^{}_{(0) \mu}(x) \langle \mathcal{O}^{*\mu}(x) \rangle_s
    -\mathcal{W}^{*}_{(0) \mu}(x) \langle \mathcal{O}^{\mu}(x) \rangle_s
    \right)\, ,
\end{equation}
where the subscript $s$ indicates that these are identities for expectation values in the presence of sources. Differentiating w.r.t. $\mathcal{W}^{}_{(0) {\mu_1}}(x_1), \mathcal{W}^{}_{(0) \mu_3}(x_3)$, (and renaming $x, \mu \to x_2, \mu_2)$, and Fourier transforming to momentum space yields,  \be
\hspace*{-.8cm}\llangle[\Bigl] \mathcal{O}^{*\mu_1}({\bf p}_1)\, {\bf p}_{2\mu_2}{\cal J}^{\mu_2}({\bf p}_2) \,\mathcal{O}^{\mu_3}({\bf p}_3)    \rrangle[\Bigl]
\;=\;   \left( g
\llangle[\Bigl]  \mathcal{O}^{*\mu_1}(-{\bf p}_3) \,\mathcal{O}^{\mu_3}({\bf p}_3) \rrangle[\Bigl]\;-\;g\llangle[\Bigl]\mathcal{O}^{*\mu_1}({\bf p}_1) \,\mathcal{O}^{\mu_3}(-{\bf p}_1) \rrangle[\Bigl] \right)\,  \label{decnb4e}
\ee
Using this, we find 
\cite{Marotta:2022jrp}
  \begin{eqnarray}
 a_0\;=\;  2^{\frac{d}{2} -4}\, \frac{(d-2\Delta)}{g(d-2)}\,\Gamma \left(\frac{ d-2\Delta}{2}\right)\Gamma \left(\frac{ 2\Delta-d}{2}\right)\,\Gamma\left(\frac{d}{2}\right)\Bigl[(\Delta-1)\left(-a_4+(d-2) a_5\right)+b_2
 \Bigl]\label{gtr5d}
\end{eqnarray}
Note that this relation involves a new parameter, namely $g$, which enters via the Ward identity. Altogether, the Ward identity introduces one relation between the parameters in the 3-point function and the normalization of the 2-point, but it also contains an additional parameter (the gauge coupling).
Thus, up to 3-point functions we need a total of three parameters.

Finally, we comment about the divergences appearing in the 3-point function. For integer values of $\Delta$, many of the triple-$K$ integrals appearing in \eqref{2316a} diverge and hence regularisation is required and renormalization may be needed. However, in this paper we consider $\Delta$ to be non-integer. In this case also some of the triple K integrals, namely $J_{1\{0,1,-1\}},J_{1\{1,0,-1\}}, J_{1\{-1,1,0\}} $ and $J_{1\{-1,0,1\}}$ are individually divergent. However, the divergences cancel for the combination in which they appear in the 3-point function. The details of this analysis can be found in \cite{Marotta:2022jrp}.

\section{Poincar\'{e} symmetry from AdS isometries}
\label{secpoincareads}

\subsection{Flat space limit of AdS}
\label{FlatAds}
At the geometric level, taking the flat space limit of AdS corresponds to sending $L$ to infinity. The AdS metric in the Poincar\'{e} coordinates is given by
\be
ds^2 =\f{L^2}{z^2}\Bigl(dz^2+\delta_{\mu\nu}dx^\mu dx^\nu\Bigl) \qquad;\qquad x^a =(z, x^\mu)\label{adscvgt}
\ee
In the limit $L\rightarrow\infty$, taken such that the metric $G_{MN}$ has a (finite) limit,
the Riemann, Ricci and scalar curvatures vanish and one gets a flat geometry (see equation \eqref{geomet56}). To analyse this limit efficiently, it is convenient to parametrise the radial coordinate $z$ as \cite{2106.04606}
\begin{eqnarray}
\frac{z}{L}=e^{\frac{\uptau}{L}}\qquad;\qquad \uptau\in \; (-\infty,\,\infty)\label{5.43}
\end{eqnarray}  
In the large $L$ limit, $\uptau$ becomes $(d+1)^{\mbox{th}}$ flat space direction. Indeed, in this limit, the AdS metric \eqref{adscvgt} becomes the flat space metric as 
\begin{eqnarray}
ds^2\;=\;  (d\uptau)^2+   e^{-2\frac{\uptau}{L}\delta_{\mu\nu} dx^\mu dx^\nu}\;=\; \delta_{ab}dx^adx^b+{\cal O}\left(\frac{1}{L}\right) \label{flatmetricfgtr}
\end{eqnarray}
where $a,b=1,\cdots,d+1$ and we have denoted $\uptau$ by $x^{d+1}$ in the second equality. To get to Minkowski space one may additionally Wick rotate $\uptau = -i t$ {}\footnote{Note that the analogous flat space limit of the de Sitter metric directly leads to Minkowski space.}.

It is also instructive to see how the Poincar\'{e} algebra emerges from the AdS isometry algebra in the flat limit. The isometry algebra of Euclidean AdS$_{d+1}$ is so$(d+1,1)$ , which is also the conformal algebra on $R^d$, 
is given by 
\be
[M_{AB}, M_{CD}] \;=\; \eta_{BC}M_{AD} -  \eta_{AC}M_{BD}+ \eta_{AD}M_{BC}- \eta_{BD}M_{AC}\label{isometryalgebraads}
\ee
where, $\eta_{AB}=(+,\dots,+,\,-)$ and
$$A,B,C,D =1,2,\cdots, d+1,d+2\;\;\equiv\;\; \{a,d+2\}\;\; \equiv\;\; \{\mu, d+1,d+2\}$$
To recast \eqref{isometryalgebraads} in the conformal algebra, we need to make the following redefinitions \cite{DiFrancesco}
\be
M_{\mu\nu} \;=\; L_{\mu\nu}\;;\quad 
  M_{d+1,\mu}\;=\; \f{1}{2} (P_\mu +K_\mu)\;\;;\quad 
   M_{d+2,\mu}\;=\; \f{1}{2} (P_\mu -K_\mu) \;\;;\quad M_{d+2,d+1}\;=\; D\quad \label{e62w}
\ee
With this, the algebra \eqref{isometryalgebraads} reduces to 
\be
[L_{\mu \nu}, L_{\rho \sigma}] &=& \delta_{\nu \rho }L_{\mu \sigma} -  \delta_{\mu \rho}L_{\nu \sigma}+ \delta_{\mu \sigma}L_{\nu \rho}- \delta_{\nu \sigma}L_{\mu \rho}\non\\[.3cm]
\;[L_{\mu\nu}, P_\rho]&=& \delta_{\nu \rho }P_{\mu}-\delta_{\mu \rho }P_{\nu}\quad;\quad
[L_{\mu\nu}, K_\rho]\;=\; \delta_{\nu \rho }K_{\mu}-\delta_{\mu \rho }K_{\nu} \non\\[.3cm]
[K_\mu, P_\nu]&=& 2\delta_{\mu\nu} D - 2L_{\mu\nu}\;;\quad
[D, P_\mu]\; =\; P_\mu\;\;;\qquad [D, K_\mu] \;=\; -K_\mu
\ee
This is the standard conformal algebra: $L_{\mu\nu}, P_\mu, K_\mu, D$ represent the rotation, translation, special conformal transformation and the dilatation generator, respectively.  

The  (Euclidean) AdS isometry algebra \eqref{isometryalgebraads} reduces to the algebra of the Euclidean group in the flat space limit via the Inonu Wigner contraction \cite{InonuWigner}. Upon Wick rotation this becomes the Poincar\'e algebra, and we will loosely use this terminology even when we work with Euclidean signature.
To see this, we note that upon splitting the $(d+2)^{th}$ component  the algebra \eqref{isometryalgebraads} can be written as 
\be
[M_{ab}, M_{ce}] &=& \delta_{bc}M_{ae} -  \delta_{ac}M_{be}+ \delta_{ae}M_{bc}- \delta_{be}M_{ac}\non\\[.2cm]
\;[M_{ab}, M_{c, d+2}] &=& \delta_{bc}M_{a,d+2} -  \delta_{ac}M_{b,d+2}\quad;\qquad 
\;[M_{a,d+2}, M_{b,d+2}] \;=\;   M_{ab}\label{jythgr}
\ee
Now, writing $M_{a,d+2}\equiv L \,\textbf{P}_a$ and taking the limit $L\rightarrow\infty$, the algebra \eqref{jythgr} reduces to 
\be
[M_{ab}, M_{ce}] &=& \delta_{bc}M_{ae} -  \delta_{ac}M_{be}+ \delta_{ae}M_{bc}- \delta_{be}M_{ac}\non\\[.2cm]
\;[M_{ab}, \textbf{P}_c]&=& \delta_{bc  }\textbf{P}_{a}-\delta_{ac }\textbf{P}_{b}
\quad;\qquad 
\;[\textbf{P}_{a}, \textbf{P}_{b}] \;=\;  0
\ee
 This is the standard algebra of the Euclidean group in flat $d+1$ dimensional space.

\subsection{From AdS to Poincar\'{e}}

It was mentioned in the introduction that CFT correlators are expected to turn into S-matrix in the flat limit. This means that the conformal symmetry should morph into the Poincar\'e symmetries in the flat limit. 
In this subsection, we explicitly show how this happens. 

We begin by noting that the generator $M_{a,d+2}$ introduced in the previous subsection 
becomes the momentum generator in $d+1$ dimensional flat space, up to a rescaling by the AdS radius. From equation \eqref{e62w}, this implies that the combination $P_\mu-K_\mu$ of the CFT algebra becomes the momentum component $\textbf{P}_\mu$ ( with $\mu =1,2,\cdots,d$) whereas the CFT generator $D$ becomes the momentum component $\textbf{P}_{d+1}$ in the flat limit. Together, they form the flat space momentum in $d+1$ dimensions
\be
\textbf{P}_{a}= (\textbf{P}_{\mu},\textbf{P}_{d+1}) \sim (P_\mu- K_\mu, D)\label{e615t}
\ee
On the other hand, the combination $P_\mu+K_\mu$ of the CFT generators provides $M_{d+1,\mu}$ components of the Lorentz generator in the flat limit, i.e.
\be
M_{ab} = (M_{\mu\nu},M_{d+1,\mu} ) \sim (L_{\mu\nu}, P_\mu+K_\mu)\label{e616t}
\ee
To see these relations more explicitly at the level of symmetry transformations, we note that the AdS isometry transformations in $(\uptau,x^\mu)$ coordinates are given by\cite{0201253}
\begin{enumerate}
\item $\mbox{Rotations and translation of } x^\mu$.
\be
\delta x^\mu\;\;=\;\; \alpha^\mu_{\;\;\nu}x^\nu\;+\;b^\mu\label{tauxtrans}
\ee
\item Scaling of $x^\mu$ and translation of $\uptau$
\be
\delta x^\mu \;\;=\;\;\lambda x^\mu\qquad;\qquad 
\delta \uptau\;\;=\;\;L \lambda\label{tau1}
\ee
 \item Special conformal transformation of $(\uptau, x^\mu)$
\be
\delta x^\mu &=& 2 (\delta_{\sigma\nu}c^\sigma x^\nu)x^\mu -x^2 c^\mu\quad;\qquad
\delta\uptau \;\;=\;\; 2 L(\delta_{\mu\nu}c^\mu x^\nu)\label{taulorentz}
\ee
where, $x^2 \equiv L^2 e^{\f{2\uptau}{L}}+\delta_{\mu\nu}x^\mu x^\nu$.
\end{enumerate}
On the other hand, we have the following isometries in flat space
\be
\delta x^\mu &=& \omega^\mu_{\;\;M} x^M  +a^\mu
\;=\;  \omega^\mu_{\;\;\nu} x^\nu + \omega^\mu_{\;\;\uptau} \uptau +a^\mu\label{lorentflatr45}\\[.3cm]
\delta\uptau &=& \omega^\uptau_{\;\;M} x^M  +\beta\;\;=\;\; \omega^\uptau_{\;\;\mu} x^\mu  +\beta
\ee
We shall now show how to recover these isometries from the flat limit of AdS and relate the flat space parameters $\omega^\mu_{\;\nu}, \omega^\uptau_{\;\mu}, a^\mu,\beta$ in terms of the AdS isometry parameters $\alpha^\mu_{\;\nu}, b^\mu, c^\mu $ and $\lambda$. We start with the transformation of $\uptau$. From \eqref{tau1}, we find that it has the structure of translation in the limit $L\rightarrow\infty$ if we simultaneously send $\lambda$ to 0, i.e.
\be \label{eq:time_transl}
\beta = \lim_{L\rightarrow\infty\atop \lambda \rightarrow 0} L \lambda\qquad\implies\qquad \delta\uptau = \beta
\ee
We also see that in this limit the 
scaling transformation of $x^\mu$ disappears. We  now consider the rotation of $\uptau$. From equation \eqref{taulorentz}, we see that it has the correct flat space structure if we define
\be
\omega^\tau_{\;\;\nu}\;\;\equiv\;\;\lim_{L\rightarrow\infty\atop c_\nu\rightarrow 0}2Lc_\nu \qquad\implies\qquad \delta\uptau = \omega^\uptau_{\;\;\nu}x^\nu\label{defomega}
\ee
This completes the analysis for the transformations of $\uptau$. Next, we consider the transformations of $x^\mu$. In the limit $L\rightarrow\infty$ and $c_\nu\rightarrow0$, equation \eqref{taulorentz} gives
\be
\delta x^\mu 
&=&\lim_{L\rightarrow\infty\atop c_\nu\rightarrow 0}2 (\delta_{\sigma\nu}c^\sigma x^\nu)x^\mu -\Bigl[ L^2 \Bigl(1+\f{2\uptau}{L}+\f{4\uptau^2}{L^2}+\cdots\Bigl)+\delta_{\sigma\nu}x^\sigma x^\nu      \Bigl] c^\mu\non\\
&=&\omega^\mu_{\;\;\uptau}\uptau -\lim_{L\rightarrow\infty\atop c_\nu\rightarrow 0} L^2 c^\mu
\ee
where, we have ignored the terms which vanish when $L\rightarrow \infty$ or $c_\mu\rightarrow0$. 
In writing the last equality, we have used equation \eqref{defomega} and $\omega^\mu_{\;\;\uptau}= -\omega^{\;\;\mu}_\uptau$. Combining the above equation with \eqref{tauxtrans}, we find
\be
\delta x^\mu = \alpha^\mu_{\;\;\nu}x^\nu +\omega^\mu_{\;\;\uptau}\uptau +b^\mu-\lim_{L\rightarrow\infty\atop c_\nu\rightarrow 0} L^2 c^\mu
\ee
Finally we consider $b^\mu = L^2 c^\mu + a^\mu$, where $a^\mu$ is independent of $L$, so that 
 the combination $ (b^\mu-  L^2 c^\mu) = a^\mu$ has a finite limit giving a finite translation and we recover the expected Poincar\'e transformation of $x^\mu$, as given in \eqref{lorentflatr45},  in the flat limit. From the above derivation, we also see that the translation of $x^\mu$ in the flat limit comes from a combination of original translation and special conformal transformation as indicated in \eqref{e615t}. Similarly, the rotation of $x^a$ comes from a linear combination of the original rotation and translation of $x^\mu$ and the special conformal transformation of $(x^\mu, \uptau)$ as suggested by equation \eqref{e616t}.

\section{ 3-point function from bulk theory} 
\label{s3a}

\subsection{Bulk theory}
\label{Bulk theory}

In this section we derive the CFT 3-point function $\Bigl\langle {\cal O}^{*\mu}\mathcal{J}^\tau{\cal O}^\nu \Bigl\rangle $  
of a U(1) conserved current $\mathcal{J}^\mu$ with a   vector operator ${\cal O}^\nu$ charged under the U(1) using AdS/CFT.   For this purpose we need a bulk action in AdS, whose cubic terms are linear in the gauge field $A_M$ and quadratic in massive vector fields, $W_N$.  As shown in appendix \ref{Classical}, the most general such action in Euclidean signature describing the interaction between a $U(1)$ gauge field and a complex massive spin one field in $d+1$ dimensional curved spacetime up to 3 derivative terms is given by
\begin{eqnarray}
S&=&\!\!\!\!\!\int d^{d+1}x\sqrt{G} \Bigl[-\f{1}{16\pi G_N}(R-2\Lambda)+\frac{1}{4} F^{MN}F_{MN}+\frac{1}{2}W^{*}_{MN} W^{MN} +m^2 W^{*}_M W^M  \nonumber\\
&&-ig\,\alpha F^{MN}W^*_MW_N+\,ig\beta F^{MN}\,\Big(  \nabla_{M} W^*_P\nabla^PW_{N} -\nabla_{M} W_P\nabla^PW_{N}^*\Big)
\Bigl]\, ,
\label{5.6}
\end{eqnarray}
where $M,\,N,P$ run from $0$ to $d$, $\Lambda$ is the cosmological constant  and $F_{MN}=\partial_M A_N - \partial_N A_M$ is the field strength of the gauge field. We have also introduced the field strength of the massive spin 1 field as
 \begin{eqnarray}
 W_{MN}= D_MW_N-D_NW_M, \qquad D_M= \nabla_M +i g A_M\, ,
 \end{eqnarray}
 with $\nabla_M$ being the diffeomorphism covariant derivative.  
 The cubic terms are parametrized by three independent parameters, $g, \alpha, \beta$,  matching the number of independent parameters that we found in the CFT analysis. One of them is the gauge coupling constant $g$ and it multiples the terms introduced  by minimal coupling. The other two, $\alpha$ and $\beta$, were first introduced in the context of zero cosmological constant and their physical meaning is as follows: $\alpha$ is the gyromagnetic coupling which is related to the magnetic moment of the massive vector field $W_M$ and $\beta$ is related to its quadrupole moment, see, {\it e.g.}, \cite{Aronson:1969ltq, Kim:1973ee, Brodsky:1992px, comptonsum, comptonsum1} and the discussion in appendix \ref{app: multipoles}.

We shall use the above action in an AdS background. Einstein equations imply that the matter fields $A_M$ and $ W_M$ couple to the metric through their energy momentum tensor. This back-reaction can modify the AdS background. However, we shall ignore such back-reaction. The reason is that we are interested in computing the 3-point function in the CFT, so the corresponding sources are only turned on infinitesimally (to implement the operator insertion) and then are turned off. As the bulk energy momentum tensor is quadratic in the fields, one may then neglect the backreaction.
 The gauge field equation derived from \eqref{5.6} in the AdS background is given by
\begin{eqnarray}
\nabla_M\,F^{MN}=J^N\qquad\implies\qquad  \Bigl(\nabla_M\nabla^M +\frac{d}{L^2} \Bigl)A_N-\nabla_N\nabla_MA^M=J_N\label{ytr5a}
\end{eqnarray}
with the source current given by 
\begin{eqnarray}
J^N&=&2i g\Big( W^*_M\nabla^{[M} W^{N]}-W_M\nabla^{[M} W^{*N]}\Big)+2ig\,\alpha\nabla_M\Big( W^{*[M}\,W^{N]}\Big)\nonumber\\
&&
-2ig\,\beta \nabla_M\Big( \, \nabla^{[M|} W^*_P\,\nabla^PW^{|N]}-\nabla^{[M|} W_P\,\nabla^PW^{*|N]}\Big)\, ,
     \label{B.26aa}
\end{eqnarray}
where the antisymmetrization on right hand side is only over the indices $M$ and $N$. In writing (\ref{B.26aa}) we have neglected terms with higher powers in the gauge coupling $g$ since we shall be only interested in the cubic interactions, which are linear in the gauge field, in what follows. Taking the covariant derivative of both sides of \eqref{ytr5a}, we find that the left hand side vanishes identically giving the conservation equation $\nabla_MJ^M=0$. It is easy to check that the current given in \eqref{B.26aa} satisfies this conservation condition on-shell. For doing calculations, we shall Fourier transform the boundary directions as
\begin{eqnarray}
T_M(z,\,k)=\int d^d x 
\;e^{-i k\cdot x}\; T_M(z,\,x),
\end{eqnarray}  
where $T_M$ can be any bulk quantity. From now on, we shall work in this Fourier basis. 

To proceed further, we need to gauge fix $A_M(z,k)$. We shall work in the axial gauge and in Euclidean signature, setting $A_0(z,k)=0$. For the 3-point function, we shall need the perturbative solution of the gauge field up to first order in the coupling $g$. It is given by (see appendix \ref{renormalised} for details)
\begin{eqnarray}
A_\mu(z,\,k)=\mathbb{K}_{\mu}^{\;\;\nu}(z,k) A_{(0)\nu}(k) +\,\int dw \sqrt{G} \;{\cal G}_{\mu\nu}(z,\,w;\,k)\,J^\nu(w,\,k)\, , \label{ftr5a}
\end{eqnarray} 
where $\mathbb{K}_\mu^{\;\;\nu}(z,k)$ and $\mathcal{G}_{\mu\nu}(z,w;k)$ denote the bulk-to-boundary and bulk-to-bulk propagators of the gauge field, respectively. Their expressions are given in equations \eqref{C.70} and \eqref{C.85}. The field $A_{(0)\mu}(k)$ denotes the boundary value of the gauge field and $J^\nu(w,k)$ can be obtained from \eqref{B.26aa} by specialising $N$ to the boundary index $\nu$. 

Next, we consider the massive fields. For the 3-point function we are interested in, we only need the free field classical solution for these massive fields. The reason is that we will determine the 3-point function through the back reaction of the massive fields to $A_\mu$, using \eqref{ftr5a}, and since the massive field enters quadratically there, higher-order corrections to the massive field will not contribute to the 3-point function of interest.
These can be expressed in terms of the massive spin-1 bulk-to-boundary propagators $\mathcal{K}_{M}^{\;\;\mu}(z;k)$ and $\bar{\mathcal{K}}_{M}^{\;\;\mu}(z;k)$ as
\begin{eqnarray}
W_M(z,\,k)=\mathcal{K}_{M}^{\;\;\mu}(z,\,k)\, w_\mu(k)\quad;\quad W^*_M(z,\,k)=\bar{\mathcal{K}}_{M}^{\;\;\mu}(z,\,k)\, w^*_\mu(k)\label{mass56t}
\end{eqnarray}
The propagators $\mathcal{K}_{M}^{\;\;\mu}(z;k)$ and $\bar{\mathcal{K}}_{M}^{\;\;\mu}(z;k)$ are given in equations \eqref{C.64} and \eqref{C.66}, respectively. The $w_\mu$ and $w^*_\nu$ are related to the boundary values of $W_\mu$ and $W^*_\nu$, respectively. Note that we only need to specify the boundary component of the massive fields. The radial component $W_z$ gets fixed in terms of the boundary components.  

The bulk fields $W_M$ and $W^*_M$ are dual to the non conserved CFT operators of section \ref{sec2review3er}. Their mass $m$ is related to the conformal dimension $\Delta$ of the boundary operators by the relation
\begin{eqnarray}
L^2\,m^2=(\Delta-1)(\Delta +1-d)\label{3.13}
\end{eqnarray}
which follows from equation \eqref{B.45} of appendix \ref{C.4}.

\subsection{Three-point function}
\label{s3an}
In this subsection, we use the AdS/CFT correspondence to obtain the 3-point function involving two spin-1 operators and a conserved current in the CFT dual to the bulk theory described above. This 3-point function will be a special case of the 3-point function given in section \ref{sec2review3er}. The 3 arbitrary parameters appearing in the CFT result \eqref{3ptrf} will be fixed in terms of the bulk parameters. 

According to the AdS/CFT correspondence, 
the on-shell bulk partition function $Z_{\rm onshell}$   with given boundary behaviour of the bulk fields is identified with the generating functional of the dual  CFT-correlation functions\cite{9802109, 9802150} ,
\begin{eqnarray}
Z_{\rm onshell}[\Phi_{(0)}]&=& \Bigl\langle e^{-\int d^d x\, \Phi_{(0)}(x)\,{\cal O}(x)}\Bigl\rangle
\end{eqnarray}
where $\Phi_{(0)}$ denotes the field parametrizing the Dirichlet boundary conditions of the bulk field $\Phi$ which is dual to the CFT operator ${\cal O}$.

In the saddle point approximation, the generator of the connected QFT correlators, denoted by $W[\Phi_{(0)}]$, is given by the on-shell value of the action, namely, 
\begin{eqnarray}
W[\Phi_{(0)}] &=&-S_{\rm onshell}
\end{eqnarray}
This is the main ingredient to compute the correlation functions of boundary CFT operators from the bulk action. 
To obtain renormalized correlators we still need to holographically renormalize \cite{0209067}. We regulate the theory by putting the boundary at $z=\epsilon$ and add counterterms to cancel the infinities.
The full renormalized action is obtained by 
\begin{eqnarray}
S_{\rm ren}&=&\lim_{\epsilon\rightarrow 0}\;\Bigl( S_{\rm reg}+S_{\rm ct}\Bigl)
\end{eqnarray}
where $S_{\rm reg}$ denotes the regularised action and $S_{\rm ct}$ denotes the counterterms. 

The details of the holographic renormalisation for the bulk theory described by action \eqref{5.6} is given in appendix \ref{s3}. Given the renormalized on-shell action, we can now evaluate the desired 3-point function. The first step for this is to obtain the exact renormalized 1-point function of the conserved current. It is given by (for details, see appendix \ref{s3})
\be
\langle \mathcal{J}^\mu(k)\rangle &=&\lim_{\epsilon\rightarrow0}\; \f{1}{\epsilon^{\f{d}{2}}\sqrt{\gamma}}\;\f{\delta S_{\rm ren}}{\delta {A}_{\mu}(k,\epsilon)}\label{pi314}
\ee
where we have used the Fefferman Graham coordinates,
\be
ds^2 =L^2\f{d\rho^2}{4\rho^2}+\gamma_{\mu\nu}dx^\mu dx^\nu,\quad \gamma_{\mu\nu} =\f{L}{\rho}\delta_{\mu\nu}\, .
\ee
Here $\gamma_{\mu\nu}$ is the induced metric at $\rho=$constant and the IR regulating boundary is at $\rho=L\epsilon$.  

The CFT 3-point function of the conserved current and two spin-1 operators is obtained by differentiating \eqref{pi314} with respect to the sources of the spin-1 operators. The final result is given by 
\be
\Bigl\langle {\cal O}^{*\mu}(p_1)\mathcal{J}^\tau(p_2){\cal O}^\nu(p_3) \Bigl\rangle 
=\delta^{\tau\lambda}\f{(2\pi)^{d}(2\pi)^{d}\;\;\delta^2}{
\delta \mathcal{W}_{(0) \mu} (-p_1)
\delta \mathcal{W}^{*}_{(0) \nu}(-p_3)}\int_0^\infty d\sigma\sqrt{G}\;\mathbb{K}_{\lambda\kappa}(\sigma;p_2)J_{(0)}^\kappa(\sigma,p_2)\, ,\label{4.42nju}
\ee
where $\mathcal{W}_{(0) \mu}$ and $\mathcal{W}_{(0) \mu}^*$
denote the fields associated with the boundary conditions of the bulk fields $W_M$ and $W^*_M$, respectively (see \eqref{eq:sub}, \eqref{wmu789nj}). These are the sources of the boundary operators ${\cal O}^*_\mu$ and ${\cal O}_\mu$ respectively.  The $J^\kappa_{(0)}$ denotes the boundary component of the current \eqref{B.26aa} but with terms only up to $O(g)$ in the coupling constant. Terms with higher orders in $g$ are relevant for bulk calculations of four and higher point correlation functions but  do not contribute to the 3-point function considered in this section. The source current $J^\kappa_{(0)}$ is a function of the massive fields $W_\mu$ and $W^*_\mu$ whose classical solutions are given in equation \eqref{mass56t}.  

After a long but straightforward calculation and using the definition of triple-K integrals given in \eqref{B.51}, the transverse contribution to the 3-point function is obtained to be
\begin{eqnarray}
\llangle[\Bigl] {\cal O}^*_\nu(p_1)\, {\cal J}_\mu(p_2)\,{\cal O}_\rho(p_3)\rrangle[\Bigl]\;\;=\;\; (\pi_2\cdot p_1)_\mu\,{\cal A}_{\nu\rho}+(\pi_2)_{\mu\nu}\,{\cal B}_\rho+ (\pi_2)_{\mu\rho}\,{\cal C}_\nu\label{4.40}
\end{eqnarray}
The form factors ${\cal A}$, ${\cal B}$ and ${\cal C}$, have the same structure as in equations  \eqref{2.5} and \eqref{2316a}. However, the 
coefficients $a_i$ and $b_i$ appearing in \eqref{2316a} are now given in terms of the AdS bulk parameters as 
\begin{eqnarray}
 a_1&=& gC_0\left[-2+\f{2(d-2)}{L^2}\beta\right]\nonumber\\[.2cm]
 a_2&=&gC_0\left[-4+\,\frac{2(d-2)}{\Delta-1}\alpha+\frac{1}{L^2}\,\frac{2(d-2)(2(2-\Delta)+d(\Delta-1))}{(1-\Delta)}\beta\right] \nonumber\\[.2cm]
 a_4&=&gC_0\left[\frac{1-\alpha}{\Delta-1}+\frac{{1}}{L^2}\, \frac{2(d-2+\Delta(1-d))}{1-\Delta}\beta\right]\non\\[.2cm]
 a_5 &=&gC_0\left[ \f{2\beta}{L^2}\right]\non\\[.2cm]
 b_1&=&gC_0\left[\frac{d-2\Delta}{2(\Delta-1)}\Bigl( 1+\alpha-\frac{2\,{\Delta}}{L^2}\,\beta\Bigl)\right]\nonumber\\[.2cm]
 b_2&=&gC_0\left[-(1+\alpha)+\,\frac{{2\Delta}}{L^2}\beta \right]
\label{4.42fg}
\end{eqnarray}
where, we have defined\footnote{The AdS-radius $L^{2\Delta-d-1}$ that appears in the definition of $C_0$ has been extracted from the metric factors involved in the integral of three Btb-propagators. All  the other factors appearing in the definition of $C_0$ collect the overall constants  present in equations \eqref{C.131} and \eqref{C.64}.}
\begin{eqnarray}
C_0= -\frac{2^{2-\frac{d}{2}}}{\Gamma\left(\frac{d}{2}-1\right)}\,\left[\frac{2^{\frac{d}{2}+1-\Delta}}{\Gamma\left(\Delta -\frac{d}{2}\right)}\right]^2\,L^{2\Delta-d-1}\label{4.42}
\end{eqnarray}
The relations given in equation \eqref{2349afnewd} can be easily seen to be satisfied for the values of $a_4,a_5$ and $b_2$ given above. The AdS/CFT correspondence has fixed the 3 arbitrary parameters in the boundary CFT 3-point function in terms of the bulk coupling parameters. 

The CFT 3-point function, reviewed in section \ref{sec2review3er}, of one conserved current and two non conserved operators ( with same conformal dimensions ) spans a 3-dimensional space. In the bulk effective theory also, we have 3 parameters $g, \alpha$ and $\beta$ which also span a 3-dimensional space. The 3 independent parameters in the CFT side were chosen to be $a_4, a_5$ and $b_2$. Their expression in terms of the bulk parameters is given above. We can also invert these relations to express the bulk parameters in terms of the independent boundary CFT parameters as 
\begin{align}
    g&=-\frac{(\Delta-1)\left(-a_4+(d-2) a_5\right)+b_2}{2 C_0} \label{eq:g}\\
    \alpha&=-\frac{(\Delta-1)(-a_4 + d a_5) + 2 a_5 -b_2}{(\Delta-1)\left(-a_4+(d-2) a_5\right)+b_2} \\
    \beta=& -\frac{a_5 L^2}{(\Delta-1)\left(-a_4+(d-2) a_5\right)+b_2}
\end{align}
If we had less than 3 parameters in the bulk, then they would not span the full 3-dimensional CFT space mentioned above. Similarly, if we had more than 3-parameters in the bulk, say coming from the higher derivative terms, then the relation between the CFT and bulk parameters would be degenerate.

One important point to note is that each coupling in the bulk (minimal, gyromagnetic, quadrupole) is consistent with the boundary CFT 3-point function by itself. 
This follows from the fact that the bulk action is AdS invariant for any value of the couplings, and the AdS isometries imply that the contribution of each term in the boundary correlator is a CFT correlator on its own.
Moreover, the matching happens for arbitrary values of these parameters. 
The matching of the 3-point function considered here is a non-trivial confirmation of the gauge/gravity correspondence for an effective field theory of charged massive spin-1 and gauge field up to three derivative terms.

\subsection{Conservation Ward identity from the bulk}
\label{long1}
The transverse ward identity \eqref{decnb4e} relates the 2-point function with the longitudinal part of the 3-point function involving the divergence of the conserved current. We shall now show that it is consistent with our bulk analysis. The Ward identity \eqref{decnb4e} is easiest to derive by the procedure of holographic renormalisation. Using \eqref{8.8}, we find the 1-point function of the divergence of the current to be (focusing on odd $d$ for now)
\be
\llangle {\bf p}_{2\mu}\mathcal{J}^\mu({\bf p}_2)\rrangle =  -\frac{2}{L}\,
\,\delta^{\mu\nu}\ \left(\frac{d}{2} -1\right) p_{2\mu} A_\nu^{(d -2)}\label{8.7yt}
\ee
where $ A_\nu^{(d -2)}$ appears in the asymptotic expansion of the gauge field (see equation \eqref{5.11gft}). 

Now, up to $O(g)$, the RHS of the above equation in momentum space takes the form (see equation \eqref{fgtyho})
\be
&&\hspace*{-1cm}(d-2)\delta^{\mu\nu} {\bf p}_\mu A_\nu^{(d-2)}({\bf p})\non\\
&=& 
g(2\Delta-d)\int \f{d^dk}{(2\pi)^d} \delta^{\mu\nu}\biggl[ \mathcal{W}_\mu^{*(0)}({\bf k}) \mathcal{W}_\nu^{(2\Delta-d)}({\bf p}-{\bf k})-\mathcal{W}_\mu^{(0)}({\bf k}) \mathcal{W}_\nu^{*(2\Delta-d)}({\bf p}-{\bf k})    \biggl]\label{fgtyhoo}
\ee
where $ \mathcal{W}_\mu^{(0)}$ and $\mathcal{W}_\nu^{(2\Delta-d)}$ (and their complex conjugates) are the source and vev part  of the near boundary expansion of the Proca field as given in equations \eqref{eq:sub} and \eqref{wmu789nj} respectively.

Now, using the 1-point function \eqref{8.7yt} and the expressions of $\mathcal{W}_\mu^{(2\Delta-d)}$ (and its complex conjugate) given in \eqref{wnutyui}, we find 
\be
&&\Bigl\langle \mathcal{O}^{*\nu}({\bf p}_1){\bf p}_{2\mu}\mathcal{J}^\mu({\bf p}_2) \mathcal{O}^{\sigma}({\bf p}_3)\Bigl\rangle \non\\[.2cm]
&=&-\f{1}{L}(d-2)\delta^{\mu\tau}{\bf p}_{2\mu}\f{\delta^2 {A}^{(d-2)}_\tau({\bf p}_2)}{\delta\mathcal{W}^{(0)}_\nu(-{\bf p}_1)\mathcal{W}^{*(0)}_\sigma(-{\bf p}_3)}(2\pi)^d(2\pi)^d\non\\[.2cm]
&=& -\f{1}{L} g(2\Delta-d)\biggl[  \left(\f{p_1}{2}\right)^{2\Delta-d}   \f{\Gamma\Bigl(\f{d}{2}-\Delta\Bigl)\; L^{2\Delta-d}}{\Gamma\Bigl(\Delta-\f{d}{2}\Bigl)} \Bigl(\delta^{\nu\sigma}+\f{ {\bf p}_1^\nu {\bf p}^\sigma_{1}(d-2\Delta)}{p_1^2(\Delta-1)}\Bigl) \non\\
&&- \left(\f{p_3}{2}\right)^{2\Delta-d}   \f{\Gamma\Bigl(\f{d}{2}-\Delta\Bigl)\; L^{2\Delta-d}}{\Gamma\Bigl(\Delta-\f{d}{2}\Bigl)} \Bigl(\delta^{\sigma\nu}+\f{ {\bf p}_3^\nu {\bf p}^\sigma_{3}(d-2\Delta)}{p_3^2(\Delta-1)}\Bigl) \biggl] (2\pi)^d\delta^d({\bf p}_1+{\bf p}_2+{\bf p}_3)  \non\\[.3cm]
&=&g \biggl[\llangle[\Bigl] \mathcal{O}^{*\nu}(-{\bf p}_3)\mathcal{O}^\sigma({\bf p}_3)\rrangle[\Bigl] - \llangle[\Bigl] \mathcal{O}^{*\nu}({\bf p}_1)\mathcal{O}^\sigma(-{\bf p}_1)\rrangle[\Bigl] \biggl] (2\pi)^d \delta^d({\bf p}_1+{\bf p}_2+{\bf p}_3)\label{11220}
\ee
In going to the last equality, we have used the expression of two point function \eqref{2pty786} obtained using holographic renormalization. The above result \eqref{11220} is exactly the transverse ward identity \eqref{decnb4e} we wanted to show. We can also verify the above Ward identity by directly using \eqref{4.42nju} and contracting it with the momenta of the current $\mathcal{J}^\mu$. In this derivation, we considered the case of odd dimensions and arbitrary $\Delta$. The analysis for even dimension and arbitrary $\Delta$ is similar and yields the same final result \eqref{11220}. 

The Ward identity \eqref{decnb4e} also gives the relation \eqref{gtr5d} between the CFT 2- and 3-point function coefficients. Using \eqref{eq:g}
we see that the CFT 2-point function coefficient $a_0$ becomes
\be
a_0&=& 2^{d-2\Delta}(2\Delta-d) \f{\Gamma\Bigl(\f{d}{2}-\Delta\Bigl)\; L^{2\Delta-d-1}}{\Gamma\Bigl(\Delta-\f{d}{2}\Bigl)}\label{2and3}
\ee
This agrees exactly with the two point function coefficient appearing in the 2-point function of Proca field in equation \eqref{2pty786} obtained using holographic renormalisation.

\section{Flat space limit of Propagators} 
\label{flat}
In this section, we consider the AdS propagators for the gauge and Proca fields and analyse them in the flat space limit. More specifically, we shall consider the bulk-to-bulk (BtB) propagator of the gauge field and the bulk-to-boundary (Btb) propagators of both gauge and Proca fields. We shall show how the BtB propagator of gauge field turns into the momentum representation of the gauge Feynman propagator in the limit to flat space. On the other hand, the Btb propagators will turn out to be related to the external leg factors of the corresponding fields in the flat limit. 

In section \ref{FlatAds}, we reviewed how the AdS geometry locally reduces to the flat space geometry when the AdS radius $L$ is taken to be large. We introduced the bulk coordinate $\uptau$ via the relation $\frac{z}{L}=e^{\frac{\uptau}{L}}$. The flat metric corresponds to keeping $\f{z}{L}$ to be $\mathcal{O}(1)$ and neglecting the $\mathcal{O}(\f{1}{L})$ terms in the AdS metric (see equation \eqref{flatmetricfgtr}). It is clear that in this limit, the radial coordinate $z$  is very large. It is consistent with the bulk kinematic region $z\,k>>1$ taken in \cite{1912.10046} as the bulk region relevant for reproducing the flat space S matrix in the flat limit. This will also be the limit that we shall consider in this and next section for the BtB and Btb propagators and on the  three point correlator for getting the corresponding quantities in flat space.     
\subsection{Gauge bulk-to-bulk propagator} 
The  derivation of  the bulk-to-bulk propagator of an abelian gauge field in momentum space in the radial/axial gauge $A_0=0$ has been reviewed in appendix \ref{btbappx} and is given by
\be
 \mathcal{G}_{\mu\nu}(z,w;k) 
         &=& -\f{1}{L^{d-3}}
    \begin{cases}
      ({z}{w})^{\f{d}{2}-1}I_{\f{d}{2}-1}(k z)K_{\f{d}{2}-1}(k w)\pi_{\mu\nu}+\f{{z}^{d-2}}{d-2}\f{k_\mu k_\nu}{k^2},& \text{if } z< w\\[.4cm]
     ({z}{w})^{\f{d}{2}-1}I_{\f{d}{2}-1}(k w)K_{\f{d}{2}-1}(k z)\pi_{\mu\nu}+\f{{w}^{d-2}}{d-2}\f{k_\mu k_\nu}{k^2},              & \text{if } z > w
         \end{cases}\label{C.85f}
\ee
For taking the flat space limit, we shall work in the $\uptau$ coordinate introduced in \eqref{5.43} and write
\begin{eqnarray} \label{eq:K_exp}
K_{d-1}(z\,k)=K_{d-1}(e^{\frac{\uptau_z}{L}}\,k\,L)\qquad;\qquad I_{d-1}(w\,k)=I_{d-1}(e^{\frac{\uptau_w}{L}}\,k\,L)
\end{eqnarray} 
Using the asymptotic expansion of the Bessel function for the large argument given in \eqref{5.49b}, we find
\begin{eqnarray}
&&K_{d-1}(z\,k)= \left(\frac{\pi}{2\,L \,k}\right)^{\frac{1}{2}} e^{-k\,\left(1+\frac{\uptau_z}{L}\right)\,L}\,+{\cal O}\Bigl(\f{1}{L}\Bigl)\;\;;\quad
I_{d-1}(w\,k)= \frac{1}{\sqrt{2\,\pi\,L\,k}}e^{k\,\left(1+\frac{\uptau_w}{L}\right)\,L}+{\cal O}\Bigl(\f{1}{L}\Bigl)\non
\label{5.50}
\end{eqnarray}
With these results, the bulk-to-bulk propagator takes the form
\be
\mathcal{G}_{\mu\nu}(z,w;k)\Big|_{L\rightarrow\infty} &=& -
    \begin{cases}
     \f{e^{-k(\uptau_w -\uptau_z )}}{2k} \pi_{\mu\nu}+\left(\f{L}{d-2}+\uptau_z\right)\f{k_\mu k_\nu}{k^2}+{\cal O}\Bigl(\f{1}{L}\Bigl),& \text{if } \uptau_z< \uptau_w\\[.4cm]
   \f{e^{-k(\uptau_z-\uptau_w)}}{2k}\pi_{\mu\nu}+\left(\f{L}{d-2}+\uptau_w\right)\f{k_\mu k_\nu}{k^2}+{\cal O}\Bigl(\f{1}{L}\Bigl),              & \text{if } \uptau_z > \uptau_w
         \end{cases}\label{C.85ff}
\ee 
To proceed further, we observe that the longitudinal part of the bulk-to-bulk propagator in the flat space limit can be manipulated as\footnote{ The same result can be  obtained by writing in equation \eqref{C.85ff} $\uptau_z=\frac{1}{2} (\uptau_z+\uptau_w) +\frac{1}{2} (\uptau_z-\uptau_w)$ and similarly for $\uptau_w$.}
\be
&&- \frac{L}{d-2}\biggl[\left(\frac{z}{L}\right)^{d-2}\Theta\left(w-z\right)+\left(\frac{w}{L}\right)^{d-2}\Theta\left(z-w\right)\biggl]\non\\[.2cm]
&=& \frac{L}{2-d}\biggl[ e^{(d-2)\left(\frac{\uptau_z+\uptau_w}{2L}+\frac{\uptau_z-\uptau_w}{2L}\right)} \Theta\left(L\,e^{\frac{\uptau_w}{L}}-L\,e^{\frac{\uptau_z}{L}}\right)+z\leftrightarrow w\biggl]\nonumber\\[.3cm]
&=&\left(\frac{L}{2-d} -\frac{\uptau_w+\uptau_z}{2} \right)-\frac{(\uptau_z-\uptau_w)}{2}\Theta(\uptau_w-\uptau_z)-\frac{(\uptau_w-\uptau_z)}{2}\Theta(\uptau_z-\uptau_w)+{\cal O}\Bigl(\f{1}{L}\Bigl)\label{poiuy3}
\ee 
 where we have kept only the leading order terms in $L$. In the limit $L\rightarrow\infty$, the first term in the above expression diverges. We shall shortly connect this divergence with the singularity of the axial gauge propagator in flat space. 
 
 The non-translational invariant piece is a consequence of the divergence. To see this, recall that time translations originate from scaling, $x^{\mu}{}' =e^\lambda x^\mu, \ z' = e^\lambda z$ in the limit $\lambda \to 0, \Lambda \to \infty$, with $\beta = \lambda L$ fixed, see \eqref{eq:time_transl}. In momentum space, $q^\mu{}' = e^{-\lambda} q^\mu$, and the arguments of the Bessel function, $k z$ and $k w$ are invariant under such rescaling. It follows that  
 \begin{equation}
    \mathcal{G}_{\mu\nu}(e^\lambda z, e^\lambda w;e^{-\lambda} k)
    = e^{(d-2) \lambda} \mathcal{G}_{\mu\nu}(z,w;k) \ \Rightarrow \
    \delta_\lambda \mathcal{G}_{\mu\nu}(z,w;k) = (d-2) \lambda \mathcal{G}_{\mu\nu}(z,w;k)\, . 
 \end{equation}
Our computation above shows that the transverse part of the correlation is finite as $L \to \infty$ and thus as $\lambda \to 0$ the transverse part is invariant under time translations,
\begin{equation}
 \lim_{L \to\infty, \lambda \to 0} \delta_\lambda \mathcal{G}^{\perp}_{\mu\nu}(z,w;k)
    = 0\, .  
\end{equation}
On the other hand, the longitudinal part diverges linearly in $L$, and thus
\begin{equation}
    \lim_{L \to\infty, \lambda \to 0} \delta_\lambda \mathcal{G}^{||}_{\mu\nu}(z,w;k)
    =- \beta\, ,
\end{equation}
since $\lambda L=\beta$ in this limit. This is precisely how the longitudinal part in \eqref{poiuy3} transforms under $\delta \uptau = \beta$.
Thus, if we remove the divergence, the correlator will also be time-translation invariant.
Ignoring the non-translation invariant part, we have
\be
\mathcal{G}^{\rm TI}_{\mu\nu}(z,w;k)\Big|_{L\rightarrow\infty} &=& -
    \begin{cases}
     \f{e^{-k(\uptau_w -\uptau_z )}}{2k} \pi_{\mu\nu}+\left(\f{L}{d-2}+\frac{\tau_z-\tau_w}{2} \right)\f{k_\mu k_\nu}{k^2}+{\cal O}\Bigl(\f{1}{L}\Bigl),& \text{if } \uptau_z< \uptau_w\\[.4cm]
   \f{e^{-k(\uptau_z-\uptau_w)}}{2k}\pi_{\mu\nu}+\left(\f{L}{d-2}+\frac{\tau_w-\tau_z}{2}\right)\f{k_\mu k_\nu}{k^2}+{\cal O}\Bigl(\f{1}{L}\Bigl),              & \text{if } \uptau_z > \uptau_w
         \end{cases}\label{tranl_inv}
\ee 
where the superscript TI indicates that we kept only the translational invariant part.

 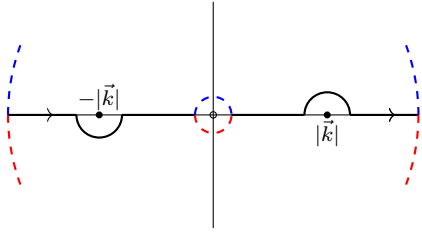
\begin{figure}{}
	\begin{center}
		\begin{tikzpicture}[scale=.30]
		\draw []  (0,-5) -- (0,5);
		\draw[]   (-9,0) -- (9,0);
		\draw[thick]   (-9,0) -- (-6,0);
		\draw[thick]   (-4,0) -- (-0.8,0);
		\draw[thick]   (0.8,0) -- (4,0);
		\draw[thick]   (6,0) -- (9,0);
		
		 \draw[red, thick, dashed] (-0.8,0) arc (-180:0:0.8);
		  \draw[red, thick, dashed] (9,0) arc (0: -20: 9);
		 \draw[red, thick, dashed] (-9,0) arc (180: 200: 9);
		 \draw[blue, thick, dashed] (-0.8,0) arc (180:0:0.8);
		 \draw[blue, thick, dashed] (9,0) arc (0: 20: 9);
		 \draw[blue, thick, dashed] (-9,0) arc (180: 160: 9);
		\draw [-{To[scale width=1]}] (-9,0) -- (-7,0);
		\draw [-{To[scale width=1]}] (7,0) -- (8,0);
	 \filldraw [ thick] (5, 0) circle (3pt);
	 \filldraw [ thick] (-5, 0) circle (3pt);
		\draw (5, -0.8) node { \scriptsize $ |\vec{k}|$};
		\draw (-5, 0.8) node { \scriptsize $ -|\vec{k}|$};
		\draw (0, 0) circle (4pt);
		\draw [-{To[scale width=1]}] (7,0) -- (8,0);
		 \draw[thick] (6,0) arc (0:180:1);
		 \draw[thick] (-4,0) arc (0: -180:1);
		\end{tikzpicture}
	\end{center}
	\caption{ For $x_0<y_0$, we close the contour in the upper half plane and use the blue contour.  For $x_0>y_0$, we close the contour in the lower half plane and use the red contour. }
	\label{figure1}
\end{figure}

To see how to proceed,  let us consider the Feynman propagator of an Abelian gauge field in the axial gauge in flat space. In position space, it is given by \cite{Leibbrandt:1987}
\be
\Delta_{ab}(x-y) = \int \f{d^{d+1}q}{(2\pi)^{d+1}} e^{-iq\cdot(x-y)}D_{ab}(q)\label{deltaabfty}
\ee
with 
\begin{eqnarray}
D_{ab}(q)=\frac{i}{q^2}\Big\{ g_{ab}-\frac{q_a\, n_b+q_b\,n_a}{q\cdot n} +\frac{q_a\,q_b( n^2 +\xi \,q^2)}{(q\cdot n)^2}\Big\}
\end{eqnarray}
where we work with mostly minus Minkowski metric, 
and $n_a$ is a constant four-vector used to impose the gauge condition $n_a\,A^a=0$. The axial temporal  gauge is imposed by taking $n_a\equiv (1,\,0,\dots,\,0)$ and $\xi=0$ which gives
\begin{eqnarray}
D_{\mu\nu}(q)=\frac{-i}{q^2}\Bigl[ \delta_{\mu\nu} -\frac{q_\mu\,q_\nu}{q_0^2}  \Bigl]\qquad;\qquad D_{\mu 0}=D_{00}=0\, . \label{jkutyghfrt}
\end{eqnarray}
\begin{figure}{}
	\begin{center}
		\begin{tikzpicture}[scale=.30]
		\draw []  (0,-5) -- (0,5);
		\draw[]   (-9,0) -- (9,0);
		\draw[thick]   (-9,0) -- (-6,0);
		\draw[thick]   (-4,0) -- (4,0);
		\draw[thick]   (6,0) -- (9,0);
		\draw [-{To[scale width=1]}] (-9,0) -- (-7,0);
		\draw [-{To[scale width=1]}] (7,0) -- (8,0);
	 \filldraw [ thick] (5, 0) circle (3pt);
	 \filldraw [ thick] (-5, 0) circle (3pt);
		\filldraw [ thick] (0, 3) circle (3pt);
		\filldraw [ thick] (0, -3) circle (3pt);
		\draw (5, -0.8) node { \scriptsize $ +E$};
		\draw (-5, 0.8) node { \scriptsize $ -E$};
		\draw (0.9, 3) node { \scriptsize $ +\mu$};
		\draw (0.9, -3) node { \scriptsize $ -\mu$};
		\draw (0, 0) circle (4pt);
		\draw [thick]  (-.5,-.5) -- (0.5,.5);
		\draw [thick]  (.5,-.5) -- (-0.5,.5);
		\draw [-{To[scale width=1]}] (0,0) -- (0,2.8);
		\draw [-{To[scale width=1]}] (0,0) -- (0,-2.8);
		\draw [-{To[scale width=1]}] (7,0) -- (8,0);
		 \draw[thick] (6,0) arc (0:180:1);
		 \draw[thick] (-4,0) arc (0: -180:1);
		\end{tikzpicture}
	\end{center}
	\caption{ The axial gauge propagator in flat space can be regularised by shifting the double poles at the origin along the imaginary axis.  This gives the principle value of the integral. }
	\label{Fig.1}
\end{figure}
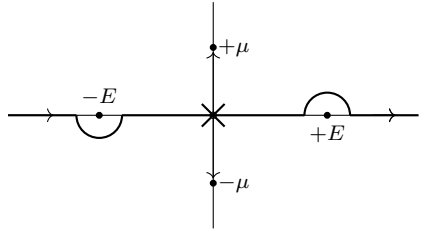
To compare it with the flat space limit result \eqref{tranl_inv}, we need to perform the integration over $q^0$ component in \eqref{deltaabfty}. To perform this integral, we note that the integrand given by \eqref{jkutyghfrt} has the standard single poles of the propagator at the point $q^0= \pm |\vec{q}|=\pm E$ (see figure \ref{Fig.1}), and an unphysical double pole at $q^0=0$. The presence of this double pole makes the integration over $q^0$ divergent. We shall compute this divergent part explicitly. For this, we note that we want to evaluate 
\be
 I&=& -i\int \f{dq^0}{(2\pi)} e^{-iq^0(x^0-y^0)} \f{1}{(q^0)^2-|\vec{q}|^2} \left(\delta_{\mu\nu} - \f{q_\mu q_\nu}{(q^0)^2}\right)\label{axialprop2}
\ee
We can use the standard Feynman prescription for the single poles. However, we need to avoid the double pole at the origin. Thus, for $x_0< y_0$ and $x_0>y_0$, we use the blue and red contours respectively given in Fig. \ref{figure1}. Denoting the radius of the small semi circles around the  origin by $\epsilon$ and following the standard method, we find that the result of the above integral is given by
\be
I&=& 
    \begin{cases}
    - \f{e^{i|\vec{q}|(x_0-y_0)}}{2|\vec{q}|} \pi_{\mu\nu}-\f{iq_\mu q_\nu}{|\vec{q}|^2}\left( \f{1}{\pi\epsilon} +\f{1}{2}(x_0-y_0)  \right),& \text{if } x^0< y^0\label{C.85ffb}\\[.4cm]
  - \f{e^{-i|\vec{q}|(x^0-y^0)}}{2|\vec{q}|}\pi_{\mu\nu}-\f{iq_\mu q_\nu}{|\vec{q}|^2}\left( \f{1}{\pi\epsilon} -\f{1}{2}(x_0-y_0)  \right),              & \text{if } x^0 > y^0
         \end{cases}
\ee
Making use of the step theta function, the longitudinal part can be written as
\be \label{reg_prop}
 \f{1}{\pi\epsilon} -\left[\f{1}{2}(x_0-y_0) \theta(x_0-y_0)-\f{1}{2}(x_0-y_0) \theta(y_0-x_0)\right] 
\ee
This is identical with the longitudinal part of the flat space limit of the bulk-to-bulk propagator of the gauge field given in equation \eqref{tranl_inv}, if we Wick rotate $(x^0,\,y^0)=-i(\uptau_z,\,\uptau_w)$ and
identify $\epsilon \sim 1/L$.

In flat space, a standard approach to regularise the axial gauge propagator is to use the principle value (PV) prescription for the double pole as shown in  Fig. \ref{Fig.1}\cite{Leibbrandt:1987} 
\begin{eqnarray}
{\rm PV}\left(\frac{1}{(q^0)^2}\right)\;\;=\;\; \frac{1}{2} \left[\frac{1}{(q^0+i\mu)^2}+\frac{1}{(q^0-i\mu)^2}\right]\;;\quad\mu>0
\end {eqnarray}
With this prescription, the double pole at $q^0=0$ gets shifted to $q^0=\pm i\mu$ (see Fig. \ref{Fig.1}). We can now use the standard Feynman contour prescription to perform the integration over $q_0$ and then send $\mu \to 0$. This gives the same expression as given in \eqref{reg_prop} except that the terms involving $\f{1}{\epsilon}$ are now absent. Note that different prescriptions to deal with the double pole involve a time-translational non-invariant term in the longitudinal part of the propagator \cite{Caracciolo:1982dp}, as in \eqref{poiuy3}.

Thus, with the understanding that $L \to \infty$ limit is treated in this way, we obtain the final result
 \begin{eqnarray}
\mathcal{G}^{\rm FV}_{\mu\nu}(z,w;k)\Big|_{L\rightarrow\infty}\;\; \simeq\;\;\left\{\begin{array}{lr}
-\frac{1}{2k} e^{-k(\uptau_w-\uptau_z)}\pi_{\mu\nu}-\,\frac{k_\mu\,k_\nu}{k^2}\,\frac{(\uptau_z-\uptau_w)}{2}&\mbox{if}\;\; \uptau_z<\uptau_w\\[.4cm]
-\frac{1}{2k} e^{-k(\uptau_z-\uptau_w)}\pi_{\mu\nu}-\,\frac{k_\mu\,k_\nu}{k^2}\,  \frac{(\uptau_w-\uptau_z)}{2}&\mbox{if}\;\;\uptau_w<\uptau_z
  \end{array}\right.\label{ghtyu}
\end{eqnarray}
where FV stands for "Finite Value". 

\subsection{Bulk-to-Boundary Propagators} 
The bulk-to-boundary propagators dictate the external leg factors of the corresponding field in the flat space limit. We start with the gauge field whose bulk-to-boundary propagator is given in equation \eqref{C.70}. Its flat space limit is easily obtained by using the asymptotic expansion given in equation \eqref{5.49b} 
\begin{eqnarray}
{\mathbb K}_{\mu\nu}(e^{\frac{\uptau}{L}}L,\,k)\Big|_{L\rightarrow\infty}\;\;=\;\; \, L^{\frac{d-3}{2}}\,\Biggl[\left(\frac{\pi}{2}\right)^{\frac{1}{2}}\frac{2^{2-\frac{d}{2}}e^{-L \,k}}{\Gamma\left(\frac{d}{2}-1\right)}\,k^{\frac{d-3}{2}} \pi_{\mu\nu} \, e^{-k\,\uptau}+{\cal{O}}\Bigl(\f{1}{L}\Bigl)\Biggl]\;\;+\;\;\frac{k_\mu k_\nu}{k^2} 
\end{eqnarray} 
Noting \eqref{C.85a}, the gauge field in the flat limit can be written as
\begin{eqnarray}
A_0=0, \qquad A_\mu^\perp(k)\;\;\simeq\;\;
\pi_{\mu\nu} \frac{1}{\sqrt{Z_A}} A_{(0)}^{\nu}(k) e^{-k\,L} e^{-k\,\uptau}, \qquad A_\mu^{||}(k)\;\;\simeq\;\; -i\frac{k_\mu k_\nu A_{(0)}^\nu(k)}{k^2} 
\label{5.59}
\end{eqnarray}
where $A_{(0)}^{\nu}(k)$ is the AdS boundary condition \eqref{C.61}, and we have introduced the normalization functions $Z_A$ which depends on the  AdS radius and the momentum as
\begin{eqnarray}
\frac{1}{\sqrt{Z_A}}=\pi^{\frac{1}{2}}\,k^{\frac{d-3}{2}}\, L^{\frac{d-3}{2}} 
\,\frac{2^{\frac{3-d}{2}}}{\Gamma\left(\frac{d-2}{2}\right)}\, .\label{5.60}
\end{eqnarray}
The factor $e^{-k\,L}$ in \eqref{5.59} may be removed by shifting $\uptau$ by $L$. If we leave this factor in \eqref{5.59} it will cancel out in correlators as a consequence of the time translation invariance of the flat space correlators, or (what is the same) because of the energy-conserving delta function. We will see this explicitly in the next section.
The longitudinal part of the gauge fields $A_\mu$ is independent of $\uptau$, and thus we set it to zero by a gauge transformation that preserves the axial gauge, $A_0=0$. We further define 
\begin{equation} \label{eq:scale_a}
    a_R^\mu = \frac{1}{\sqrt{Z_A}} A_{(0)}^{\mu}(k)\, .
\end{equation}
The factor $1/\sqrt{Z_A}$ tends to infinity as $L \to \infty$, and thus we need to scale the source $A_{(0)\mu}$ to zero in order for $a_R^\mu$ to be finite. 
As the AdS source is arbitrary one may always arrange such that $a_R^\mu$ is finite in the flat-space limit. Thus the flat-space limit of the gauge field becomes
\begin{equation}
A_a(\uptau, k) =  {\cal A}_a e^{-k\,\uptau}, \qquad {\cal A}_a\;\;\equiv \;\;
\,\bigl(0,\,\pi_{\mu\nu} a_R^{\nu}(k)\bigl)\, .\label{ghtyr}  
\end{equation}
 The ${\cal A}_a$ thus defined satisfies the transversality condition $q^a\,{\cal A}_a=0$, where the $(d+1)$ dimensional null momenta is defined as \cite{1201.6449}
 \begin{eqnarray}
 q^a= (q^0,q^\mu)=(\pm ik,\,k^\mu),\qquad  \qquad q^2\equiv \delta_{ab}\,q^a\,q^b=0\, ,\label{5.61}
 \end{eqnarray}  
 with $k$ being the magnitude of $k^\mu$. After Wick rotation to Minkowski spacetime,
 $q_M^a=(\pm k, k^\mu)$ and $\uptau=i t$, the factor $e^{-k \uptau}$ becomes plane waves, $e^{\mp i q_M^0 t}$, and the two signs are related to whether the external leg is associated with an in- or out-state.  The factor ${\cal A}_a$ encodes the $(d-1)$-polarization vectors of the $(d+1)$ vector field in flat space. To see this, let us consider a frame such that the momentum of the photon is along the $d$-direction, $q^a=(\pm i k^d, 0, \ldots, 0, k^q)$, then 
 \begin{equation} \label{eq:A_pol}
     {\cal A}_a(k) = \sum_{\lambda=1}^{d-1} a^{(\lambda)}(k) \epsilon^{(\lambda)}_a\, , 
     \qquad \epsilon^{(\lambda)}_a = (0, \delta^\lambda_i, 0),\quad i=1,\ldots,  d-1\, ,
 \end{equation}
 where $\epsilon^{(\lambda)}_a $ are $(d-1)$ polarisation vectors and $a^{(\lambda)}$
 is determined by the AdS boundary condition by $a^{(\lambda)} = a_R^\lambda$. Upon quantization $a^{(\lambda)}$ become the annihilation or creation operators (depending on the signs $\pm$ in $q^a$) of the mode with polarization vector $\epsilon^{(\lambda)}_a$ \footnote{
 Note that this is similar to what happens in Lorentzian AdS solutions that correspond to CFT excited states. The CFT state may be generated by an Euclidean path integral by turning on a source for a dual operator on the boundary of AdS. Using the real-time AdS formalism of \cite{Skenderis:2008dh,Skenderis:2008dg} one may obtain the bulk Lorentzian solution corresponding to this state and in this solution the annihilation and creation operators are given in terms of the boundary sources \cite{Botta-Cantcheff:2015sav,Christodoulou:2016nej}. It turns out the resulting solution is precisely 
 that of HKLL \cite{Hamilton:2006az}, which is then interpreted as corresponding to a bulk coherent state \cite{Botta-Cantcheff:2015sav,Christodoulou:2016nej}.}. 
 One may check that the polarisation vectors satisfy the expected normalization condition,
 \begin{equation}
     \delta^{a b} \epsilon^{(\lambda)}_a \epsilon^{(\sigma)}_b = \delta^{\lambda \sigma}, \qquad \lambda, \sigma =1, \ldots, d-1\, ,
 \end{equation}
 and the expected completeness relation,
 \begin{equation}
 \sum_{\lambda=1}^{d-1} \epsilon^{(\lambda)}_a \epsilon^{(\lambda)}_b = \delta_{ab}
 + \frac{n^2}{(n\cdot q)^2} q^a q^b -\frac{1}{(n\cdot q)} (n^a q^b + n^b q^a)\, ,
 \end{equation}
 where $n^a=(1,0, \ldots, 0)$ is vector imposing the temporal gauge $n^a A_a=0$.

Next, we consider the massive Proca field whose Btb propagator is given in equation \eqref{C.64}. The extension of the above analysis to the massive Proca field is more involved due to the relation among mass, AdS-radius and the conformal dimension of the dual operator given in equation \eqref{3.13}. Due to this relation, a finite mass in the large AdS radius limit requires that $\Delta$ is also taken to be large keeping  $\Delta/L\simeq m$ finite. This implies that we need to analyse the modified Bessel function appearing in the Btb propagator in the limit of both large argument and large order. This is known as uniform expansion \cite{AsympOlver&74} and is reviewed in appendix \ref{UnEx}. For the modified Bessel functions appearing in the Proca Btb propagator, the uniform expansion gives (see equation \eqref{unibese1})
\begin{eqnarray}
{K}_{\Delta-\frac{d}{2}+\ell}(z\,k)\;=\;\left(\f{\pi}{2\,L}\right)^{\f{1}{2}}(k^2+m^2)^{-\f{1}{4}} \left(\f{k}{m+\sqrt{k^2+m^2}}\right)^{-m\,L-\ell}e^{-\sqrt{k^2+m^2}(L+\uptau)}+{\cal O}\Bigl(\f{1}{L}\Bigl)
\label{ghtry}
\end{eqnarray}
With the expansion \eqref{ghtry}, the flat space limit of the Proca Btb propagator or equivalently the classical solution can be easily worked out. Here, we note the flat space limit of classical solutions given in equations \eqref{3260n1y} and \eqref{C.124}
\begin{align}
W_a(k)&\simeq
{\cal W}_{a}(k) e^{-L\sqrt{k^2+m^2}} \,e^{-\sqrt{k^2+m^2}\uptau}+O\Bigl(\f{1}{L^{\frac{d-5}{2}}}\Bigl),\nonumber\\
w_R^\mu &= \frac{1}{\sqrt{Z_W}} w_\mu, \quad
{\cal W}_a(k)=\left( i\frac{k_\mu w_R^\mu}{m},\,\tilde{\pi}_{\mu \nu} w_R^\nu\right)\, ,\label{ghtyrgt}
\end{align}
where $w^\mu$ is the AdS boundary condition for the Proca field, see \eqref{massive_BC}. The factor of 
$e^{-L\sqrt{k^2+m^2}}$ will cancel out in correlator as a consequence of time translation invariance. The expression of $Z_W$ and $\tilde\pi_{\mu\nu}$ are given by
\begin{eqnarray}
\tilde{\pi}_{\mu\nu}&=&\delta_{\mu\nu} +\frac{k_\mu\,k_\nu}{m(m+\sqrt{k^2+m^2})}\, , \\ 
\frac{1}{\sqrt{Z_W}}&\equiv & 
\frac{L^{\frac{d-3}{2}}}{(k^2+m^2)^{\frac{1}{4}}}
\frac{\left((m+\sqrt{m^2+k^2})/2\right)^{mL}}{(m L)^{mL -\frac{1}{2}}}
e^{m L}
\left(1+ {\cal O}\left(\frac{1}{m L}\right)\right) 
\label{5.67}
\end{eqnarray}
Notice that $1/\sqrt{Z_W}$ goes to zero as $L\to \infty$, opposite to what happens for $1/\sqrt{Z_A}$, so to keep $w_R^\mu$ finite in the flat-space limit we now need to send the AdS source $w_\mu$ to infinity, which is always possible since $w^\mu$ is arbitrary.
The uplifted Euclidean momenta of the Proca field in $(d+1)$ dimensions in the flat-space limit can be written as 
\begin{eqnarray}
 q^a=(\pm i\sqrt{k^2+m^2},\,k^\mu)~~,~~q^2= -m^2 
  \label{5.68}
\end{eqnarray}
After Wick rotation to Minkowski spacetime,
 $q_M^a=(\pm \sqrt{k^2+m^2}, k^\mu)$ and $\uptau=i t$, the factor $e^{-\sqrt{k^2+m^2} \uptau}$ becomes plane waves, $e^{\mp i q_M^0 t}$, and the two signs are related to whether the external leg is associated with an in- or out-state. 

It is easy to check that the subsidiary condition ${\cal W}^a q_a=0$ is satisfied as expected (where the indices in ${\cal W}^a q_a$ are contracted using the $(d+1)$ dimensional Euclidean metric $\delta_{ab}$).  Exactly as in the gauge field case, we can write ${\cal W}^a$ in terms of polarization vectors. Indeed, let $\epsilon_\mu^{(r)}=\delta_\mu^r, r=1, \ldots, d$, the $d$-unit vectors along the boundary directions. Then 
\begin{equation} \label{eq: w_pol}
    w^\mu_R = \sum_{r=1}^d w^{(r)}(k) \epsilon_\mu^{(r)}\, ,  
\end{equation}
{\it i.e} $w^{(r)}(k)$ are Cartesian coordinates of $w^\mu_R$. We now introduce the polarization vectors,
\begin{equation}
    \varepsilon_a^{(r)} = \left( i\frac{k^\rho \epsilon_\rho^{(r)}}{m},\,\tilde{\pi}_{\mu}{}^{\nu} \epsilon^{(r)}_\nu \right)\, 
\end{equation}
One may check that they satisfy the expected normalization condition,
\begin{equation}
     \delta^{a b} \varepsilon^{(r)}_a \varepsilon^{(s)}_b = \delta^{rs}, \qquad r, s =1, \ldots, d\, ,
 \end{equation}
and the expected completeness relation, 
\begin{equation}
 \sum_{r=1}^{d} \varepsilon^{(r)}_a \varepsilon^{(s)}_b = \delta_{ab} + \frac{q_a q_b}{m^2}\, .
\end{equation}
It terms of those,
\begin{equation}
  {\cal W}_a(k) =\sum_{r=1}^d  w^{(r)}(k) \varepsilon^{(r)}_a\, . 
\end{equation}
Exactly as in the gauge field case the field $w_\mu$ that parametrizes the AdS boundary condition has morphed into he creation and annihilation operator (depending on the $\pm$ signs in \eqref{5.68}), which upon quantization give rise to massive modes associated with corresponding polarization vectors, and the AdS radial dependence gave rise to the expected plane wave behavior.

\section{Flat limit of 3-Point Function}
\label{sec6flatdr}
In this section we analyse the flat space limit of the CFT 3-point function between a conserved current and two spin one CFT operators computed in section \ref{s3a} using AdS/CFT correspondence and compare the resulting expression with the 3-point amplitude
 involving a gauge field and two massive spin-1 Proca fields in flat space. As we discussed in the previous section the sources must be scaled in order for the limit to be finite, \eqref{eq:scale_a}, \eqref{ghtyrgt}, thus (using the chain rule) we expect,
 \begin{equation} \label{eq:flat_limit}
\lim_{L \to \infty} \sqrt{Z_{W_1} Z_A Z_{W_3}} \,A_3^{\mu_1\mu_2\mu_3}  \sim \delta(E_1+E_2+E_3) {\cal M}_3^{\mu_1\mu_2\mu_3}   
 \end{equation}
 where $Z_A$ and $Z_W$ are defined in \eqref{5.60} and \eqref{5.67}, respectively, 
 $A_3^{\mu_1\mu_2\mu_3}$ is the AdS 3-point momentum space 3-point amplitude and ${\cal M}_3^{\mu_1\mu_2\mu_3}$ is the corresponding flat space scattering amplitude. As we are working in momentum space, the momentum conserving delta function is already present, but the energy conserving delta function should emerge in the limit.

\subsection{Asymptotic Expression of Triple K Integrals}
\label{triplehgyt}
The 3-point CFT correlator given in \eqref{4.40} in momentum space are expressed in terms of the triple-K integrals. The specific integrals appearing in our correlator take the general form (see equation \eqref{B.51})
\begin{eqnarray}
J_{N\{k_i\}}= \int_0^{\infty} dz \,z^{\frac{d}{2} -1+ N} \,p_1^{\Delta -\frac{d}{2} +k_1} \,K_{\Delta-\frac{d}{2} +k_1} (z\,p_1) \,p_2^{\frac{d}{2} -1+k_2} K_{\frac{d}{2} -1+k_2}(z\,p_2)\,p_3^{\Delta -\frac{d}{2} +k_3} \,K_{\Delta-\frac{d}{2} +k_3} (z\,p_3)\non
\end{eqnarray}
We want to evaluate these integrals in the limit $L,\Delta\rightarrow \infty$ keeping $\f{\Delta}{L}$ fixed. For doing this, we use the asymptotic expansions given in equations \eqref{5.49b} and \eqref{unibese1} to obtain,
\begin{eqnarray}
J_{N\{k_i\}}&\simeq& \left(\frac{\pi}{2}\right)^{\frac{3}{2}}\,L^{\frac{d-5}{2}+N}\,\frac{ (m+\sqrt{ p_1^2+m^2})^{mL+k_1} \,p_2^{\frac{d-3}{2}+k_2}\,(m+\sqrt{p_2^2+m^2})^{m\,L+k_3}}{(p_1^2+m^2)^{1/4}\,(p_3^2+m^2)^{1/4}}\nonumber\\[.4cm] 
&&\,e^{-L (\sqrt{p_1^2+m^2}+p_2+\sqrt{p_3^2+m^2})}\,\int_{-\infty}^\infty d\uptau\,e^{-\uptau (\sqrt{p_1^2+m^2}+p_2+\sqrt{p_3^2+m^2})}\;\;+\;\;\cdots\label{5.74}
\end{eqnarray}
where $\cdots$ terms denote the terms subleading in $L$ and $\Delta$. 

In the flat space limit, $\uptau$ is interpreted as Euclidean time. We want to perform the integration over this variable. To do this, we use equations \eqref{5.61} and \eqref{5.68} and using the convention to treat all momenta as incoming (or choosing the plus sign in \eqref{5.61} and \eqref{5.68} )
we write $p_2=-iq_2^0$ and $-iq_{1,3}^0= \sqrt{p_{1,3}^2+m^2}$. 
Substituting these in the integral in \eqref{5.74} gives 
 \begin{eqnarray}
 \int_{-\infty}^\infty d\uptau\,e^{i\uptau (q_1^0+q_2^0+q_3^0)}&=&2\pi\,\delta(q_1^0+q_2^0+q_3^0)
 \end{eqnarray} 
Thus, we see that the energy conserving delta function, as needed in equation \eqref{eq:flat_limit} to interpret the flat limit of the $d$-dimensional  CFT correlator as an amplitude in the flat space-time with one more dimension, emerges from the integration over the AdS radial direction.
To account for both in-coming and out-going momenta, one may consider either $q^0>0$ and appropriately adjusts the signs in the delta function or use the convention to consider only plus signs in delta function and consider $q^0<0$ for out-going momenta. In the remainder we choose the latter convention. 
With this, the expression in \eqref{5.74} becomes
 \begin{eqnarray}
J_{N\{k_i\}}\simeq (-i)^{\frac{d-5}{2}+k_2} L^{N+\frac{d-5}{2}} \left(\frac{\pi}{2}\right)^{3/2}
\frac{(m-iq_1^0)^{m\,L+k_1}}{\sqrt{q_1^0}}\, (q_2^0)^{\frac{d-3}{2}+k_2} \,\frac{(m-iq_3^0)^{m\,L+k_3}}{\sqrt{q^0_3}}\,(2\pi)\delta(q^0_1+q^0_2 +q^0_3)\nonumber
\end{eqnarray}
 where, on the support of the delta function, the exponential factor $e^{iL(q_1^0+q_2^0+q_3^0)}$ has been set to 1. 
 
For comparing with the flat space result, we need to analytically continue the above result to Lorentzian signature. This is achieved by performing the inverse Wick rotation $-iq^0=E$ with $E$ denoting the energy of the particles. This gives
\begin{eqnarray}
J_{N\{k_i\}}&\simeq & \frac{ L^{N-1}}{C_0}\,\frac{(m+ E_1)^{k_1}\,E_2^{k_2}\,(m+ E_3)^{k_3}}{\sqrt{Z_{W_1}\,Z_A\,Z_{W_3}}}\,
\, 
  \,(2\pi\,i)\delta(E_1+E_2 +E_3)\label{jn567}
\end{eqnarray} 
where $Z_A$ and $Z_W$ are defined in equations \eqref{5.60} and \eqref{5.67} respectively and $C_0$ is defined in equation \eqref{4.42} (with $\Delta$ replaced by $mL+\f{d}{2}$).

As mentioned in section \ref{sec2review3er}, some of the triple K integrals appearing in the 3-point function are divergent. However, one can show that these divergences correspond to the $z\rightarrow 0$ end of the integral. Here, we are concerned with the opposite end $z\rightarrow \infty$. In this region, the integrals are well behaved. Due to this, we do not encounter any issue related to the divergences of triple K integrals in the flat limit.

Scalar 3-point functions of primary operators are also given in terms of triple-K integrals \cite{Bzowski:2013sza}, and our discussion suffices to compute their flat-space limit, yielding the expected answer, i.e. a delta function in energy and momentum. 

\subsection{CFT Correlator in Flat Limit}
\label{flatrety}
We are now ready to take the flat limit of our 3-point function  in \eqref{4.40}. This is easily done by using \eqref{jn567}. Replacing the triple K integrals appearing in the 3-point function by \eqref{jn567} and keeping the leading order terms in $L$, we find after some rearrangement
 \begin{eqnarray}
 A_3^{\mu_1\mu_2\mu_3}\Bigl|_{L\rightarrow\infty}&=&2\pi i\;\delta(E_1+E_2+E_3)\, \frac{g}{\sqrt{ Z_{W_1}\,Z_A\,Z_{W_3}}}\,
 {\cal C}^{\mu_1\mu_2\mu_3}
 \label{5.71}
 \end{eqnarray}
where,
 \begin{eqnarray}
 &&\hspace*{-.9cm}{\cal C}^{\mu_1\mu_2\mu_3}\non\\
 &=& -(1+\alpha) \pi^{\mu_2}_{~\mu}\left[ \left( \eta^{\mu_1\mu} +\frac{p_1^{\mu_1}\,p_1^\mu}{m(E_1+m)}\right)\left(\frac{(p_1+p_2)^{\mu_3}\,p_2}{E_3+m} +p_2^{\mu_3}\right)\right.\nonumber\\[.3cm]
 &&\left. +\left( \eta^{\mu\mu_3}+\frac{p_3^{\mu_3}\,p_3^{\mu}}{m(E_3+m)}\right) \left(\frac{p_1^{\mu_1}\,p_2}{E_1+m} -p_2^{\mu_1} \right)\right] -2 p_{1\,\mu}\pi^{\mu\mu_2}\left[ \eta^{\mu_1\mu_3}  -\frac{ p_1^{\mu_1}\,p_2^{\mu_3}}{m(E_1+m)}\right.\nonumber\\[.3cm]
&&\left. +\;\;\frac{2\,  p_1^{\mu_1} \,(p_1+p_2)^{\mu_3} }{(E_1+m)\,(E_3+m)}\;\;-\;\; \frac{2\,E_2\,   p_1^{\mu_1} \,(p_1+p_2)^{\mu_3} }{m(E_1+m)\,(E_3+m)}\;\;+\;\;\frac{ p_2^{\mu_1}\,(p_1+p_2)^{\mu_3}}{m(E_3+m)}\;\right] \nonumber\\[.3cm]
 &&+2\beta\, p_{1\,\mu}\pi^{\mu\mu_2} \left[\frac{p_1^{\mu_1}\,E_2}{(E_1+m)}\frac{ (p_1+p_2)^{\mu_3}  \,E_2}{(E_3+m)} -\frac{p_2^{\mu_1}\,(p_2+p_1)^{\mu_3}\,E_2}{(E_3+m)}+\frac{ p_1^{\mu_1}\, p_2^{\mu_3}\,p_2}{E_1+m} -p_2^{\mu_1}\,p_2^{\mu_3}\right]
 \end{eqnarray}
 This expression may look complicated, but we shall show in the next subsection that it precisely has the structure to match with the desired flat space 3-point function.

 \subsection{Matching with Flat Space Result}
\label{flatrety1}
 
The expression \eqref{5.71} should be compared with the flat space 3-point amplitude of a $U(1)$ gauge field and two massive spin-1 fields in $d+1$ dimensions at tree level. This has been computed in  appendix \ref{exact} and equation \eqref{F.233} gives the final expression of the flat space amplitude in terms of the $(d+1)$ dimensional polarizations of the external fields. To compare \eqref{F.233} with the result obtained in \eqref{5.71}, we need to use the representation of the polarizations suggested by the flat limit of the Btb propagators as given in equations \eqref{5.59} and \eqref{5.68},  for the gauge and Proca field, respectively. In Minkowski signature, they can be written as
 \begin{eqnarray}
\varepsilon^W_a=\left( \frac{(p\cdot \varepsilon)}{m},\, \varepsilon_\mu +\frac{(p\cdot \varepsilon)}{m(E+m)}\,p_\mu \right)~~;~~\varepsilon^A_a=(0, \,\pi_{\mu\nu}\epsilon^\nu)\, ,\label{11.216a}
 \end{eqnarray}
 where $\varepsilon_\mu$ is any of the vectors $\varepsilon_\mu^{(r)}$ introduced in \eqref{eq: w_pol}
 and $\epsilon^\nu$ is any of the vectors $\epsilon^{(\lambda)}_\nu$ introduced in \eqref{eq:A_pol}. 
 Below we shall denote these vectors by $\epsilon_{1\mu}, \epsilon_{2\mu},\epsilon_{3\mu}$ according to which vector they are associated in the order they appear in the correlator.
It is easy to see that these polarization vectors satisfy the condition $p\cdot \varepsilon(p)=0$ with $p^a=(E,\,p^\mu)$ where the inner product now involves the Minkowski metric $\eta_{ab}$. For the above basis of the transverse polarization vectors, we have
\begin{eqnarray}
  \varepsilon_1^a\,\varepsilon_{3a}&=&\epsilon_{1\mu}\,\epsilon_{3\nu}
\left[  \eta^{\mu\nu}+ \frac {2\,(p_1+p_2)^\nu \,p_1^\mu}{(E_1+m)(E_3+m)} -\frac{2\,p_2\,(p_1+p_2)^\nu\,p_1^\mu}{m(E_1+m)(E_3+m)} +\frac{(p_1+p_2)^\nu\,p_2^\mu}{m(E_3+m)}-\frac{p_1^\mu\,p_2^\nu}{m(E_1+m)}\right]\, ,\nonumber\\[.3cm]
 p_2^a\,\varepsilon_{1a} &=&
 \epsilon_{1\mu}\left[p_2^\mu -\frac{p_2\,p_1^\mu}{E_1+m} \right]\qquad;\qquad   p_2^a\,\varepsilon_{3a} =
  \epsilon_{3\mu}\left[ 
  p_2^\mu+\frac{p_2\,(p_1+p_2)^\mu}{E_3+m}\right] 
  \end{eqnarray}
Using these in equation \eqref{F.233} gives
 \begin{eqnarray}
 {\cal M}_3& =& \hat g \,\epsilon_{1\mu_1}\,\epsilon_{2\mu_2}\,\epsilon_{3\mu_3}\Bigg[2 p_{1\mu}\,\pi^{\mu\mu_2}\Bigg(  \eta^{\mu_1\mu_3}+ \frac {2\,(p_1+p_2)^{\mu_3} \,p_1^{\mu_1}}{(E_1+m)(E_3+m)} -\frac{2\,p_2\,(p_1+p_2)^{\mu_3}\,p_1^{\mu_1}}{m(E_1+m)(E_3+m)} \nonumber\\
 &&+\frac{(p_1+p_2)^{\mu_3}\,p_2^{\mu_1}}{m(E_3+m)}-\frac{p_1^{\mu_1}\,p_2^{\mu_3}}{m(E_1+m)}\Bigg)+2\hat{\beta}\, p_{1\mu}\,\pi^{\mu\mu_2} \left(p_2^{\mu_1} -\frac{p_2\,p_1^{\mu_1}}{E_1+m}\right)\left(p_2^{\mu_3}+\frac{ p_2\,(p_1+p_2)^{\mu_3}}{E_3+m}\right)\nonumber\\
 &&-\left(1+\hat\alpha\right) \pi^{\mu_2}_{~\mu}\Bigg\{ -\tilde{\pi}_1^{\mu_1\mu}\, \left(p_2^{\mu_3} +\frac{p_2\,(p_1+p_2)^{\mu_3}}{E_3+m}\right)+ \tilde{\pi}_3^{\mu_3\mu}\,\left( p_2^{\mu_1} -\frac{ p_2\,p_1^{\mu_1}}{E_1+m}\right)\Bigg\}\Bigg]
 \label{5.83} 
 \end{eqnarray}
 By comparing this  $(d+1)$ dimensional amplitude  with the $d$-dimensional CFT correlator in flat limit
 given in equation \eqref{5.71}, we see that they match exactly provided we identify the flat space gyromagnetic ratio $\hat\alpha$ and quadrupole couplings $\hat\beta$ with their AdS counterparts $\alpha,\beta$, respectively. Doing this, we find
 \begin{eqnarray}
\lim_{L \to \infty} \sqrt{Z_{W_1} Z_A Z_{W_3}} \,A_3^{\mu_1\mu_2\mu_3} \;=\; -2 \pi i
\delta(E_1+E_2+E_3)\, {\cal M}_3^{\mu_1\mu_2\mu_3}\, ,
\end{eqnarray}
Thus the flat space limit of the CFT correlator correctly reproduces the interacting part of the flat-space S-matrix.

\section{Discussion}
\label{s4}

We discussed in this paper the computation of the flat space scattering amplitude of massive 
spin 1 field, its complex conjugate and a $U(1)$ gauge field in $d+1$ dimensions via a flat-space limit of a $d$-dimensional 3-point CFT correlator of a conserved current, a non-conserved vector current and its complex conjugate. This computation may also be formulated as a flat space limit of a corresponding tree-level AdS amplitude, with the bulk interactions involving both minimal and non-minimal couplings, with the latter being the gyromagnetic and the quadrupole couplings.

The bulk AdS computation and the agreement with the CFT result is in itself a new test of the 
AdS/CFT. We computed the boundary 3-point correlation function following the procedure of holographic renormalization. This fixes the three coefficients appearing in the general CFT 3-point function of a conserved current and two non conserved operators in terms of bulk parameters. One feature of this matching is that each bulk coupling is separately consistent with the expected conformal invariance. This is not surprising since each bulk coupling is invariant under the AdS isometries by itself. Further, since the matching occurs for arbitrary values of the bulk couplings, conformal symmetry does not impose any restriction on the bulk couplings at the level of 3-point function, leaving for example, the AdS gyromagnetic ratio  $\alpha$ completely arbitrary. Unitarity and crossing symmetry may impose constraints which may fix or
restrict the allowed values of $\alpha$ but this would require analysing higher point functions. 

The flat-space limit amounts to sending the AdS radius $L$ to infinity while keeping fixed all parameters (masses and coupling constants) that appear in the bulk action. From the CFT perspective, one zooms in on the IR region while sending to infinity the conformal dimension of the operator dual to the massive fields. In this limit, we show that the $d$ dimensional CFT 3-point function matches with a corresponding 3-point scattering amplitude in $d+1$ dimensional flat space. 
The flat-space limit turns AdS isometries into Poincar\'e isometries and classical solutions in AdS to plane wave solutions in flat space, with the fields parametrizing the boundary condition in AdS becoming polarization vectors in flat space. 

We also analysed the flat-space limit of the BtB propagator of the gauge field in the axial gauge and explicitly showed that it matches with the flat space Feynman propagator in the axial gauge. The longitudinal part of the Feynman propagator in the axial gauge is prescription dependent and we show that the principle value prescription in flat space agrees with the translation invariant part of AdS expression in the flat limit (as one may have anticipated based on earlier flat space analyses). The polarisation vectors of the fields in the flat-space limit are also dictated by the Btb propagators. In particular, the matching of the 3-point function requires matching the flat space polarisation vectors to that that emerge from the flat-space limit of AdS. The conservation of the spatial momenta in the flat-space limit is ensured by working with momentum space CFT. On the other hand, the energy conserving delta function emerges from the {\it triple-K} integrals that underlie momentum space CFT 3-point functions. One of the main ingredients for the flat-space limit matching was the uniform expansion of modified Bessel functions in which both the argument as well as the order of the modified Bessel functions were taken to be large. This was crucial for taking the limit of the modified Bessel functions associated with the non conserved operators. 

The bulk AdS computation was done at tree-level, but  the CFT three-point function is fixed non-perturbatively by conformal invariance. This implies that bulk loops in AdS will lead to an AdS amplitudes of the same form as at tree-level but with quantum corrected parameters. Moreover, quantum corrections of the flat space gyromagnetic and quadruple coupling may be directly obtained by the flat-space limit of the corresponding AdS diagrams. The reason is that the Feynman rules map 1-1 in the limit: BtB propagators map to Feynman propagators, Btb propagators map to plane waves and interaction vertices are kept fixed in the limit. There were recent progress in setting up loop computation in AdS, see \cite{Banados:2022nhj} and references therein, and it would be interesting to combine the methods described there with the results we present here in order to obtain explicit loop-level results for flat space scattering amplitudes from AdS.

Note that the matching using the CFT 3-point function is non-perturbative, so if we know the 
coefficients of the low-energy effective action non-perturbatively this would provide a non-perturbative determination of the gyromagnetic and quadruple couplings. The coefficients in the low-energy effective action in $d+1$ dimension are linked to coefficients in the low-energy effective action in $10d$ and $11d$ supergravity via compactification, and some of these coefficient may be fixed non-perturbatively using U-duality. It would be interesting to track these relations in detail.

In flat space, we know that the gyromagnetic ratios can take two values ${ \alpha}=2$ or ${ \alpha}=1$ (see, e.g., \cite{Marotta:2019cip, 2102.13180} for recent works on this). Massive fields charged under the gauge fields, which arise from the closed string degrees of freedom (such as the graviton or the Kalb-Ramond field), have gyromagnetic ratio 1 whereas massive fields which are charged under the gauge fields arising from open strings have gyromagnetic ratio 2 \cite{Ferrara:1992yc, 2102.13180}. Now, the gyromagnetic ratio $\alpha$ appears in the 3-point function. Hence, noting that $\alpha$ is a constant at tree level, the exact matching of the 3-point amplitude implies that its value in AdS should also be 1 or 2. The fixing of the gyromagnetic ratio in AdS will have implications for the bootstrap program in the dual CFT as the constraints on the bulk coupling will restrict the OPE coefficients in the boundary CFT theory. 

We expect our analysis to extend to higher-point functions. As already noted, the perturbative Feynman rules map 1-1 between AdS and flat space, {\it i.e.} for each Witten diagram there is a corresponding flat space Feynman diagram. Moreover, as  we recover translational invariance  in the flat-space limit, the energy-preserving delta function should arise from the Bulk-to-boundary propagators. It would be interesting to work out the details. Non-perturbative things are less clear but also more interesting. The general CFT $n$-point function of scalar operators in momentum space is known  \cite{Bzowski:2019kwd, Bzowski:2020kfw} (but the corresponding answer for spinning operators is still missing). It would be interesting to analyze the flat-space limit of the general momentum-space CFT $n$-point functions, starting from scalar ones.

Another application of our analysis is in the context of higher spin theories. In 4-dimensional flat space, a fully consistent formulation of massive higher spin theories is still missing and is an active area of research (see e.g. the review \cite{2205.01567}). Holography allows us to construct the flat-space couplings from the CFT correlators as we have seen for the massive spin 1 case in this paper. Using this approach should be promising for constructing the consistent massive higher spin theories in the flat space.

\bigskip

{\bf Acknowledgement:} 
{ We are thankful to C. Corian\`o, P. Di Vecchia, D. Francia, C. Heissenberg,  Yue-Zhou Li, S. Lionetti and R. Loganayagam for useful discussions. KS and MV were supported in part by the STFC consolidated grant ST/T000775/1 “New Frontiers in Particle Physics, Cosmology and Gravity”. MV is also supported in part by the “Young Faculty Research Seed Grant Scheme'' of IIT Indore.}

\appendix
\section{Conventions and useful identities}
 \label{geo768}

In this appendix, we summarise our conventions and note some useful identities which have been used in this work. 
We denote the indices corresponding to the $d+1$ dimensional AdS directions by $M,\,N, P\dots$ which run from $0$ to $d$. On the other hand, the $d$ dimensional boundary indices are denoted by Greek letters $\mu,\,\nu,\rho,\cdots$ which run from $1$ to $ d$. The $d+1$ dimensional flat space indices have been denoted by $a,b,\cdots$ which run from $1$ to $d+1$. The anti-symmetrization of two fields is defined as
\begin{eqnarray}
A_{[M}\,B_{N]}=\frac{1}{2} \Big(A_M\,B_N-A_N\,B_M\Big). 
\end{eqnarray}
Throughout this paper, we have worked in the Euclidian AdS$_{d+1}$. Only after taking the flat limit, we perform a Wick rotation  $z\equiv x_0^E= i x^0$, with $x^0$ the time coordinate,  of the radial direction.  We use mostly positive signature convention 
for the Minkowski metric. The Wick-rotation  transforms the zero component of a generic  vector field $M$  in mostly positive 
metric  as\cite{9611043,9912205}:
\begin{eqnarray}
V_0=-
V^0=iV^E_{0}  \qquad ,\qquad {\cal V}_{0\mu}=\partial_0V_\mu-\partial_\mu V_0=i( \partial_0 V_\mu-\partial_\mu V_0^E)
\end{eqnarray} 
where $V_M$ can be either a massless or massive vector field.
According to this rule,  the square of the field strength of the vector field remains unchanged under the rotation.  
The Lorentzian action $e^{iS_L}$ is transformed in the  Euclidean one $e^{-S_E}$ getting the identity $S_E=-iS_L$.  The action of a massive complex vector field in mostly positive 
signature therefore transforms under the wick rotation as
\begin{eqnarray}
i S_{L}&=&i \int d^{d+1} x \left[ -\frac{1}{2} {\cal V}^\dagger_{MN}\,{\cal V}^{MN}- 
\,m^2 V^\dagger_M\,  V^M\right]\non\\
&=& \int d^{d+1} x_E \left[ -\frac{1}{4} {\cal V}_{MN}^\dagger\,{\cal V}^{MN} -m^2 V^\dagger_M\, V^M\right]_E\non\\
&=&-S_E
\end{eqnarray}  
 where we are treating $V_M$ and $V_M^\dagger$ as two independent fields.
 
Our convention for the Riemann tensor is
\begin{eqnarray}
R^P_{~LMN}=\partial_M\Gamma^P_{LN}-\partial_N\Gamma^P_{LM}+\Gamma^P_{QM}\Gamma^Q_{LN}-\Gamma^P_{QN}\Gamma^Q_{LN}~~;~~R_{MN}=g^{PQ}R_{PMQN}
\end{eqnarray}
For any tensor ${\cal T}_{PQ}$, we have 
\begin{eqnarray}
&&[\nabla_M,\nabla_N] {\cal T}_{PQ}=-R^L_{~PMN}{\cal T}_{LQ}-R^L_{~QMN}{\cal T}_{PL}\label{A.24}
\end{eqnarray}
 The AdS metric in the Poincar\'e coordinates is given by
\begin{eqnarray}
ds^2=\frac{L^2}{z^2} \big( dz^2+ \delta_{\mu\nu} \, dx^\mu \,dx^\nu\big)\quad;\qquad \sqrt{G} = \left(\f{L}{z}\right)^{d+1}\label{stanpoin54}
\end{eqnarray}
with $L$ being the AdS-radius. The Christoffel  symbols in this coordinates are
\begin{eqnarray}
\Gamma^z_{zz} = -\frac{1}{z}~~;~~\Gamma_{\mu z}^z=0~~;~~\Gamma^z_{\mu\nu}=\frac{1}{z} \delta_{\mu\nu}
~~;~~\Gamma^\mu_{zz}=0~~;~~\Gamma^\mu_{\nu z}=-\,\frac{\delta^\mu_\nu}{z}~~;~~\Gamma^\mu_{\nu\lambda}=0
\end{eqnarray}
The above equation can be compactly written as
\begin{eqnarray}
\Gamma^M_{NP}=-\frac{1}{z}\left(\delta^M_N\,\delta_{Pz} +\delta^M_P\,\delta_{Nz} -\delta^M_z\,\delta_{NP}\right)=-\frac{z}{L^2}\left(\delta^M_N\,g_{Pz} +\delta^M_P\,g_{Nz} -\delta^M_z\,g_{NP}\right)\label{6.21}
\end{eqnarray}
where $g_{MN}$ denotes the AdS-metric in the Poincar\'e coordinates.

For the purposes of holographic renomalization, it is convenient to use the Fefferman-Graham (FG) coordinates which is related to the Poincar\'{e} coordinates by $\rho=\frac{z^2}{L}$.\footnote{The purpose of keeping the AdS radius $L$ in $\rho=\frac{z^2}{L}$ is to give both $\rho$ and $z$  the dimension of length.  }  Thus, in FG coordinates, the metric takes the form
\begin{eqnarray}
ds^2=L^2\frac{d\rho^2}{4\rho^2}+L \frac{\delta_{\mu\nu}\, dx^\mu\,dx^\nu}{\rho} \quad;\qquad \sqrt{G} = \f{1}{2}\left(\f{L}{\rho}\right)^{\f{d+2}{2}}\label{B.26a}
\end{eqnarray}
The Christoffel symbols in this coordinates are given by
\begin{eqnarray}
\Gamma^{\rho}_{\rho\rho}=-\frac{1}{\rho}~~;~~\Gamma^\rho_{\mu\nu}=\f{2}{L}\delta_{\mu\nu}~~;~~\Gamma^\mu_{\rho\rho}=0~~;~~\Gamma^\nu_{\rho\mu}= -\frac{1}{2\rho}\delta_\nu^\mu~~;~~\Gamma_{\nu\mu}^\sigma=0
\end{eqnarray}
The Riemann tensor, Ricci tensor and the scalar curvature for the AdS can be expressed in coordinate independent manner as
\begin{eqnarray}
R_{MNPQ}= \frac{G_{MQ}\,G_{NP}-G_{MP}G_{NQ}}{L^2}~~;~~R_{MN}=-\frac{d}{L^2}\,G_{MN}\quad;\quad R=-\frac{d(d+1)}{L^2} \label{geomet56}
\end{eqnarray}
with $G_{MN}$ denoting the AdS metric in the corresponding coordinate system.

\section{ Limiting behaviours of modified Bessel functions}
\label{UnEx}
For the calculation of holographic renormalisation and taking the flat limit, we need the expressions of modified Bessel functions in various limits. In this appendix, we review the required results.

\subsection{Expansions for large and small arguments}

For the large arguments, the asymptotic expansions of the modified Bessel functions are given by
\be
I_\nu(z)\;  \rightarrow\; \f{e^z}{(2\pi z)^{\f{1}{2}}}\quad\mbox{and}\qquad K_\nu(z) \; \rightarrow \; \left(\f{\pi}{2 z}\right)^{\f{1}{2}}e^{-z}\qquad\mbox{as}\qquad z\rightarrow\infty\label{5.49b}
\ee
On the other hand, in the limit $z\rightarrow0$, we have following leading order approximations
\be
I_\nu(z)\;  \rightarrow\; \f{2^{-\nu}}{\Gamma(\nu+1)}z^\nu \quad\mbox{and}\qquad K_\nu(z) \; \rightarrow \; 2^{\nu-1}\Gamma(\nu) \, z^{-\nu}\qquad\mbox{as}\qquad z\rightarrow 0\label{5.49bf}
\ee
In the above equation \eqref{5.49bf}, the approximation for $I_\nu(z)$ is valid for $\nu\not=-1,-2,\cdots$ and the approximation for $K_\nu(z)$ is valid for $\nu >0$.  For the holographic renormalisation of the Proca field, we shall need the expansion of $K_\nu(z)$ in the limit $z\rightarrow0$ in more detail. For non-integer $\nu$ we have
\be
K_\nu(z)&=& \f{\pi}{2}\f{I_{-\nu}(z)-I_\nu(z)}{\sin (\pi\nu)}\;;\quad I_\nu(z) = \sum_{j=0}^\infty \f{1}{\Gamma(j+1)\Gamma(\nu+j+1)}\left(\f{z}{2}\right)^{\nu+2j}\, ,
\label{kiexpansion}
\ee
while for positive integer $n$ the expansion reads 
\be
K_n(x) &=& \f{1}{2}\left(\f{x}{2}\right)^{-n}\sum_{j=0}^{n-1} \f{\Gamma(n-j)}{\Gamma(j+1)} (-1)^j \left(\f{x}{2}\right)^{2j} +(-1)^{n+1} \ln \left(\f{x}{2}\right)I_n(x)+\non\\
&&+(-1)^n\f{1}{2}\left(\f{x}{2}\right)^{n}\sum_{j=0}^{\infty} \f{\psi(j+1)+\psi(n+j+1)}{\Gamma(j+1)\Gamma(n+j+1)} \left(\f{x}{2}\right)^{2j} 
\ee
where
\be
\psi(z) =\sum_{k=1}^\infty \left(\f{1}{k}-\f{1}{z+k-1}\right)-\gamma 
\ee
and $\gamma$ is the Euler Mascheroni constant.

\subsection{Uniform expansions}

The uniform expansion involves taking the argument as well as the order of the modified Bessel function to be large. Here, we review the derivation of such expansion following \cite{AsympOlver&74}. We start by noting that the modified Bessel functions satisfy the differential equation
\begin{eqnarray}
z^2\frac{d^2}{dz^2}F_\nu +z\frac{d}{d z} F_\nu -(z^2+\nu^2) F_\nu\; =\; 0\label{eqrefty6}
\end{eqnarray}
where $F_\nu$ can be $K_\nu(z)$ or $I_\nu(z)$.

Let us start by deriving the asymptotic expansion when $\nu$ is large and $z$ bounded.
To this end, it is convenient to first perform the Liouville-type transformation 
\begin{equation}
h_\nu(z) = z^{\frac{1}{2} }\, F\, ,
\end{equation}
 and rewrite the differential equation \eqref{eqrefty6} in the form \cite{AsympOlver&74}
\begin{eqnarray}
\frac{d^2}{d z^2} h_\nu(z)\;=\; \Bigl(\nu^2 f(z) +g(z)\Bigl)h_\nu(z), \qquad f(z) =\frac{1}{z^2}, \quad g(z) =1-\frac{1}{4\,z^2}\, . \label{B.92a}
 \end{eqnarray}
 We can remove the $z$-dependence from the coefficient of $\nu^2$ by further change of dependent and independent variables,
\begin{eqnarray}
\xi=\int f^{\frac{1}{2}}(z) \,dz \quad;\qquad h_\nu=f^{-\frac{1}{4}} (z) \, H_\nu(\xi)\label{B.93a}
\end{eqnarray}
In terms of them, equation \eqref{B.92a} can be expressed as
\begin{eqnarray}
\frac{d^2}{d\xi^2} H_\nu(\xi) = \left(\nu^2+\psi(\xi)\right)  H_\nu(\xi),\quad\qquad\psi(\xi) =
\frac{g(z)}{f(z)} -\frac{1}{f^{3/4}(z)}\frac{d^2}{dz^2} \left( \frac{1}{f^{1/4}(z)}\right) \label{B.94a}
\end{eqnarray} 
With $\nu$ large and $z$ bounded such that $\nu \gg \psi(\xi)$, the differential equation \eqref{B.94a} can be solved perturbatively in $1/\nu$,
\begin{eqnarray}  \label{B.95a}
H_\nu( \xi)= e^{-\nu\,\xi}\sum_{s=0}^\infty \frac{A_s(\xi)}{\nu^s } 
\end{eqnarray}
As \eqref{B.94a} is invariant under $\nu \to -\nu$, there is a second asymptotic expansion which is related to (\ref{B.95a}) by 
$\nu$ with $-\nu$. 
The coefficients $A_s$ in \eqref{B.95a} can be determined recursively by plugging the above series expansion in equation \eqref{B.94a}:
\begin{eqnarray}
	2A'_{s+1}\;=\;A''_{s} -\psi(\xi) A_s(\xi)\quad\Longrightarrow \quad	A_{s+1}\;\; =\;\; \frac{1}{2} f^{-1/2}(z) \frac{d A_s}{dz} -\frac{1}{2}\int dz\, \Lambda(z)\, A_s\,dz\label{zeroder}
	\end{eqnarray}
where
 \begin{eqnarray}
 	\Lambda(z) &=& f^{1/2}(z) \psi(\xi(z))\;\;=\;\;f^{1/2}(z) \left[\frac{g(z)}{f(z)}  -f(z)^{-1/2} \left( \frac{5}{16} \frac{(f'(z))^2}{f(z)^2} +\frac{1}{4} \frac{f''(z)}{f(z)}\right)\right]
 	\end{eqnarray}
Taking $s=-1$ in the differential equation in \eqref{zeroder}
we find that $A_0$ should be constant (since $A_{-1}=0$ -- there are no the coefficients with negative order in \eqref{B.95a}). One may recursively solve for the higher order coefficients. However, it turns out that the coefficients are, in general, divergent near $z\rightarrow\infty$ for the functions $f(z)$ and $g(z)$ given in equation \eqref{B.92a}, as explained in \cite{AsympOlver&74}. 

To discuss the case when both $\nu$ and $z$ going to infinity, we rescale $z$ to $z\nu$ \eqref{eqrefty6} and repeat the analysis. It turns out one gets the same equation as in  \eqref{B.92a} but with different $f(z)$ and $g(z)$, namely,
\begin{eqnarray}
\frac{d^2}{d z^2} h_\nu(\nu z)\;=\; \Bigl(\nu^2 f(z) +g(z)\Bigl)h_\nu(\nu z), \;\;\;\;\;\;\;\;\;f(z)= \frac{1+z^2}{z^2}, ~~~~g(z) =-\frac{1}{4\,z^2}
\end{eqnarray} 
Assuming $\nu$ to be real and positive  (more generally it suffices  for the real part of $\nu$ to be positive $|\arg (\nu)| <\frac{1}{2} \pi$), the above expression of $f(z)$ when used in equation \eqref{B.93a} gives
\begin{eqnarray}
\xi(z)=(1+z^2)^{1/2} +\ln \frac{z}{1+(1+z^2)^{1/2}}\quad;\qquad h_\nu =\left(\f{z^2}{1+z^2}\right)^{\f{1}{4}}H_\nu(\xi)\label{hyutred}
\end{eqnarray} 
In writing the expression of $\xi$, we have set the integration constant to zero.  This is allowed because equation \eqref{B.93a} is nothing but a change of variable. Finally, we can write a series solution of the modified Bessel function $K_\nu(\nu z)$ by using equation \eqref{B.95a}  and the relation between $h_\nu(\nu z), H_\nu(\nu z)$ and $K_\nu(\nu z)$
\begin{eqnarray}
K_\nu(\nu z)&=& (\nu\,z)^{-\frac{1}{2}} \,f^{-\frac{1}{4}}H_\nu(\nu z)\;\;=\;\;  \frac{e^{-\nu\xi(z)}}{(1+z^2)^{\frac{1}{4}}}\, \sum_{s=0}^\infty \frac{A_s}{\nu^s}
\label{B.101a}
\end{eqnarray} 
where $\xi(z)$ is given in \eqref{hyutred} and the overall factor $\sqrt{\nu}$ originates from the rescaling of the $z$-variable discussed before. 

Next, we want to find the leading order term of the above series solution. As before, the recursive relation \eqref{zeroder} again implies that $A_0$ is constant. To find its value, we make use of the fact that for large $z$, we have \cite{NIST, AsympOlver&74}
\begin{eqnarray}
K_\nu(\nu z) &\sim& \sqrt{\frac{\pi}{2\,\nu}}\,\frac{e^{-\nu \,z}}{z^{1/2}}\label{expectedgtyuh}
\end{eqnarray}
Now, the expression of $\xi(z)$ given in \eqref{hyutred} for large $z$ gives $\xi=z+{\cal O}(\f{1}{z})$. Hence, $ e^{-\nu \xi}\sim e^{- \nu z}$. Thus, the leading order term in \eqref{B.101a} for large $z$ becomes 
\be
K_\nu(\nu z)\;\; =\;\;   A_0\,\frac{e^{-\nu \,z}}{(\nu\,z)^{1/2}}
\ee
Comparing this with the expected result \eqref{expectedgtyuh}, we find $A_0 =\sqrt{\frac{\pi}{2}}$. Using this, we see that the leading order expression for the uniform expansion of the modified Bessel function is given by 
\be
K_\nu (\nu \,z)\Bigg|_{\nu\rightarrow\infty} &\simeq & \left(\f{\pi}{2\nu}\right)^{\f{1}{2}} \f{e^{-\nu \,\xi(z)}}{(1+z^2)^{\f{1}{4}}} \quad;\qquad \xi(z) = (1+z^2)^{\f{1}{2}} +\ln \left(\f{z}{1+(1+z^2)^{\f{1}{2}}}\right)\label{B.121}
\ee 
A similar analysis yields,
\be
I_\nu (\nu \,z)\Bigg|_{\nu\rightarrow\infty} &\simeq & \left(\f{1}{2\pi\,\nu}\right)^{\f{1}{2}} \f{e^{\nu \,\xi(z)}}{(1+z^2)^{\f{1}{4}}} 
\label{B.122}
\ee
with the same $\xi(z)$ as in equation \eqref{B.121}.  

\subsection{Expansion for $K_{\Delta-\f{d}{2}+\ell}(zk)$}
For taking the flat limit, we need to know the expansion of $K_{\Delta-\f{d}{2}+\ell}(zk)$ with $z$ parametrized by $z=Le^{\f{\tau}{L}}$ in the limit $\Delta,L\rightarrow\infty$. Using \eqref{3.13}, we find
\be
 \Delta- \f{d}{2}+\ell\;=\;\ell+mL\sqrt{1+\f{(d-2)^2}{4m^2L^2}}\;=\;mL+\ell+O\left(\f{1}{L}\right)\equiv mL + \beta\label{apm}
\ee
where $\beta=\ell +O\left(\f{1}{L}\right)$.

We have 
\be
K_{\Delta-\f{d}{2}+\ell}(zk) &=& K_{mL+\beta}\left(kL +k\tau+O(\f{1}{L})\right)\non\\[.3cm]
&=&K_{\nu+\beta}(p\,\nu + k \tau)+O\left(\f{1}{L}\right) \non\\[.3cm]
&=& K_{\nu+\beta}(p \nu) +k\tau K'_{\nu+\beta}(p\nu) +\f{(k\tau)^2}{2}K''_{\nu+\beta}(p\nu)+\f{(k\tau)^3}{3!}K'''_{\nu+\beta}(p\nu)+\cdots
\ee
where, we have defined $p= k/m$ and $\nu =mL$. The derivatives of modified Bessel functions can be expressed in terms of linear combinations of the modified Bessel functions with different orders. E.g., 
\be
\f{dK_{\sigma}(x)}{dx} &=& -\f{1}{2} \Bigl[K_{\sigma-1}(x)+K_{\sigma+1}(x)\Bigl]\non\\
\f{d^2K_{\sigma}(x)}{dx^2} &=& \f{1}{4} \Bigl[K_{\sigma-2}(x)+2K_{\sigma}(x)+K_{\sigma+2}(x)\Bigl]\non\\
\f{d^3K_{\sigma}(x)}{dx^3} &=& -\f{1}{8} \Bigl[K_{\sigma-3}(x)+3(K_{\sigma-1}(x)+K_{\sigma+1}(x) )+K_{\sigma+3}(x)\Bigl]
\ee
Now, using the identity \cite{besselratio}
\be
\f{K_{\nu+\alpha}(\nu z)}{K_\nu (\nu z)}&=& \left(\f{1+\sqrt{1+z^2}}{z}\right)^\alpha\left[1-\f{1-\alpha\sqrt{z^2+1}}{2(1+z^2)}\f{\alpha}{\nu}+O\left( \f{1}{\nu^2} \right)\right]\label{onebynu}
\ee
and the uniform expansion result for $K_\nu(\nu z)$ reviewed in the previous subsection, we find
\be
K_{\Delta-\f{d}{2}+\ell}(zk) &=& \left(\f{\pi}{2EL}\right)^{\f{1}{2}} \left(\f{k}{m+E}\right)^{-mL-\ell}\; e^{-EL}\;\left(1-E\tau+\f{E^2\tau^2}{2}-\f{E^3\tau^3}{3!}+\cdots\right)\left[1+O\left(\f{1}{L}\right)\right]\non
\ee
where $E=\sqrt{k^2+m^2}$.

In the above expression, we have kept only the leading order terms in the expansion in $1/L$. The $O(1/\nu)$ term in \eqref{onebynu} is of order $1/L$ does not contribute to the leading order term. All terms in the series in $E\tau$ present in the above expression are of the same order w.r.t. expansion in $1/L$ 
 and resum to give an exponential function. Hence, we get
\be
K_{\Delta-\f{d}{2}+\ell}(zk) &=& \left(\f{\pi}{2EL}\right)^{\f{1}{2}} \left(\f{k}{m+E}\right)^{-mL-\ell}\; e^{-EL-E\tau}\left[1+O\left(\f{1}{L}\right)\right]\label{unibese1}
\ee
Following a similar analysis and using \cite{besselratio}
\be
\f{I_{\nu+\alpha}(\nu z)}{I_\nu (\nu z)}&=& \left(\f{1+\sqrt{1+z^2}}{z}\right)^{-\alpha}\left[1-\f{1+\alpha\sqrt{z^2+1}}{2(1+z^2)}\f{\alpha}{\nu}+O\left( \f{1}{\nu^2} \right)\right]\label{onebynu1}
\ee
we also find
\be
I_{\Delta-\f{d}{2}+\ell}(zk) &=& \left(\f{1}{2\pi EL}\right)^{\f{1}{2}} \left(\f{k}{m+E}\right)^{mL+\ell}\; e^{EL+E\tau}\;\left[1+O\left(\f{1}{L}\right)\right]
\ee

\section{General cubic action in AdS for gauge and Proca fields} 
\label{appen:D}
In this appendix, we construct the general cubic action involving a gauge field and a complex Proca field in AdS$_{d+1}$. There are general group theoretic constructions of cubic interaction terms involving fields of arbitrary spins (see, e.g., \cite{Joung:2012hz, Sleight:2016dba
}). However, for our purposes, it would be sufficient to consider a perturbative effective field theory approach. 

If we are working at a fixed order in perturbation theory, we can eliminate those terms in the Lagrangian which are proportional to lowest order equation of motion. More precisely, we can use field redefinitions to transfer these terms to higher order terms in the perturbative expansion. We start by reviewing this procedure for a general action following \cite{Burgess:2007pt}. Suppose, we have an action $S[\phi]$ involving a generic field $\phi$ in which terms with different orders are parametrised by a parameter $\epsilon$
\be
S[\phi]= S_0[\phi] +\epsilon S_1[\phi]+\epsilon^2S_2[\phi]+\cdots\label{act54y}
\ee
Now, suppose at $O(\epsilon^n)$, the $S_n[\phi]$ includes a term $\mathcal{S}_n[\phi]$ which is proportional to the equation of motion for the lowest order action $S_0[\phi]$, {\it i.e.}, 
\be
\mathcal{S}_n[\phi]=\int d^dx\; f(x)\f{\delta S_0}{\delta\phi(x)}\quad ,
\ee
Here $f(x)$ denotes some arbitrary function of the field and its derivatives. We now make the field redefinition
\be
\phi(x) \rightarrow \tilde\phi(x) = \phi(x)-\epsilon^n\;f(x)\label{fiered}
\ee 
Under this redefinition, the action \eqref{act54y} becomes
\be
S[\phi] \rightarrow S[\tilde\phi] =S[\phi] -\epsilon^n \int d^dx f(x)\f{\delta S_0}{\delta\phi(x)} +O(\epsilon^{n+1})
\ee
The second term on the right hand side cancels $\mathcal{S}_n[\phi]$. This shows that the effect of the field redefinition \eqref{fiered} is to remove the term proportional to the lowest order equation of motion in the action without changing any other term up to $O(\epsilon^n)$. Thus, we can only focus on those terms which do not involve lower order equations of motion if we are working at a fixed order in perturbation theory. 
Note that the use of the lowest order equation of motion (instead of the full non-linear equations) in the field redefinition was useful in that the redefinition does not mix different orders in the perturbative expansion. Had we used the full non-linear equations, one would need to keep track of how nonlinearities mix different orders in the $\epsilon$ expansion.

We can now apply the above procedure to write the cubic action involving a gauge and the complex Proca field. Gauge invariance implies that the gauge field can appear only in terms of the field strength $F^{MN}$. Further, the complex Proca field is taken to be charged under this gauge field and the conservation of the charge implies that each term involving the Proca field $W_M$ must also have its complex conjugate $W^*_M$. Now, the kinetic terms of the action involving the gauge and complex Proca field are given by
\be
S_2=\int d^{d+1}x\sqrt{G}\biggl[\f{1}{4}F^{MN}F_{MN}+\f{1}{2}W^{*MN}W_{MN}+m^2W^{*M}W_M\biggl]
\ee
where, the indices $M,N$ run from $0$ to $d$ and $F_{MN}$ denotes the field strength of the gauge field $A_M$,
\be
F_{MN} = \nabla_M A_N -\nabla_N A_M = \p_M A_N -\p_N A_M\;\;.
\ee 
We  have also introduced $W_{MN} = D_M W_N -D_N W_M$ with
\be
D_M W_N =\nabla_M W_N +ig A_M W_N = \p_M W_N -\Gamma_{MN}^P W_P +ig A_M W_N\, .
\ee
This ensures that the kinetic term is invariant under the gauge transformation
\be
W_M \rightarrow e^{ig\lambda(x)}W_M\quad,\qquad W^*_M \rightarrow e^{-ig\lambda(x)}W^*_M\quad;\qquad A_M \rightarrow A_M-\p_M \lambda(x)\, .
\ee
The lowest order equations of motion of the gauge and the Proca field follow from the variation of the kinetic terms and are given by
\be
\nabla_M F^{MN}=0 \quad;\qquad D_M W^{MN} +m^2 W^N =0\quad;\qquad D_M W^{*MN} +m^2 W^{*N} =0\, .\label{low1}
\ee
An important condition on the massive Proca fields can be obtained by taking the divergence of their equations which gives 
\be
m^2D_M W^M =D_M D_N W^{NM} =D_{[M}D_{N]} W^{NM} = \f{ig}{2} F_{MN}W^{MN}\, . \label{low2}
\ee
This shows that the divergence $D_M W^M$ is actually quadratic in the fields. This will be useful below, as we shall see. Another set of useful equations are
\begin{equation} \label{U1_feq}
\nabla_M F^{MN} =0\ \implies\ \Box A^N = \nabla^N(\nabla\cdot A)+R^{NP}A_P \ \implies\ 
\Box F^{MN} \;
=\; \f{(2d+2)}{L^2}F^{MN}
\end{equation}
where the last equality holds in $AdS$.

Next, we want to write the cubic interaction terms. We shall write down all possible cubic terms and then eliminate the redundant terms using the procedure described above. We shall focus on terms with up to 3 derivatives. At the lowest order in derivatives ({\it i.e.} one derivative), there is only one possible term, 
\be
I_1 =i \;\f{a_1}{2} F_{MN}\bigl(W^{*M} W^N-W^{*N} W^M\bigl)\, . \label{315r}
\ee
An important point to note is that after integration by parts in the above term, its tensor structure matches with one of the terms in $W^{*MN}W_{MN}$. So, naively, it would seem as if we could forget about the $a_1$ term in \eqref{315r}. However, the structure of $W^{*MN}W_{MN}$ follows from the minimal coupling procedure when we promote the global phase invariance to local gauge invariance, while the term involving $a_1$ in \eqref{315r} is gauge invariant by itself and does not follow from minimal coupling. Hence, its coefficient is independent of the coefficient in the minimal coupling term in $W^{*MN}W_{MN}$. Thus, we must keep the $a_1$ term. The existence of a new gauge invariant term is responsible for the gyromagnetic coupling.

At the level of 3 derivatives, the terms need to be constructed using $F_{MN}, W_M, W^*_M$ and two derivatives $D_M$. Using an integration by parts we can ensure that $D_M$ acts only on the Proca fields. Using these rules, the most general cubic structure involving 3 derivatives can be written as 
\be
I_3
&=& F^{MN}\biggl[ \Bigl(c_0D_{M} W^*_{P} D^{P} W_{N} +c_0^*D_{M} W_{P} D^{P} W^*_{N} \Bigl)+\Bigl(c_1D_{P} W^*_{M} D^{P} W_{N} +c_1^*D_{P} W_{M} D^{P} W^*_{N}\Bigl)\non\\
&+& \Bigl(c_2D_{M} W^*_{P} D_{N} W^{P} +c_2^*D_{M} W_{P} D_{N} W^{*P} \Bigl)+\Bigl(c_3D_{P} W^{*P} D_{M} W_{N}+c_3^*D_{P} W^{P} D_{M} W^*_{N}\Bigl)\non\\
&+&\Bigl(c_4W^*_{M} D_{P} D^{P}   W_{N}+c_4^*W_{M} D_{P} D^{P}   W^*_{N}\Bigl)+\Bigl(c_5W^*_{P} D^{P} D_{M}   W_{N}+c_5^*W^{P} D_{P} D_{M}   W^*_{N} \Bigl)\non\\
&+&\Bigl(c_6W^*_{M} D_{N} D_{P}   W^{P}+c_6^*W_{M} D_{N} D_{P}   W^{*P}\Bigl)+\Bigl(c_7W^*_{P} D_{M} D_{N}   W^{P}+c_7^*W_{P} D_{M} D_{N}   W^{*P}\Bigl)\non\\
&+&\Bigl(c_8W^*_{P} D_{M} D^{P}   W_{N}+c_8^*W^{P} D_{M} D_{P}   W^*_{N} \Bigl)+\Bigl(c_9W^*_{M}  D_{P}  D_{N} W^{P}+c_9^*W_{M}  D_{P}  D_{N} W^{*P}\Bigl)\biggl]
\label{3der5}
\ee
The coefficients $c_i$ are in general complex. Now using integration by parts, the explicit form of the AdS curvature and the lower order equations of motion \eqref{low1}, \eqref{low2} and \eqref{U1_feq}, one can show that all terms except first one is either higher order in fields or give the same structures as either the first term in \eqref{3der5} or the term in \eqref{315r}. Hence, we can ignore all terms in \eqref{3der5} except the first one. Further, for the action to be real  the constants $c_0$ may be complex 
but an explicit computation shows that  the real part of $c_0$ does not contribute to the three-point amplitude on AdS backgrounds  (see appendix \ref{exact} for the similar result on flat background).
Hence, we shall take $c_0$ also to be purely imaginary and write $c_0 = i\beta$ with $\beta\in \mathbb{R}$. Thus, we can express the 3 derivative cubic terms in the form
\be
I_3
&=&igF^{MN}\biggl[\beta\bigl(\nabla_{M} W^*_{P} \nabla^{P} W_{N} -\nabla_{M} W_{P} \nabla^{P} W^*_{N}\bigl)
\biggl]
\label{hasdea1}
\ee

Thus, the most general cubic Lagrangian involving a gauge field and complex massive spin 1 field takes the form
\be
\mathcal{L} = i gF^{MN}\biggl[-\alpha W^*_MW_N +\beta\bigl(\nabla_{M} W^*_{P} \nabla^{P} W_{N} -\nabla_{M} W_{P} \nabla^{P} W^*_{N}\bigl)\biggl]
\ee
We shall work with the above form of cubic interaction terms in this paper.

\section{ Classical Solutions on AdS Background}
\label{Classical}
In this appendix, we summarise the classical solutions of the gauge and Proca fields in AdS background from the perspective of the AdS/CFT correspondence. 
\subsection{Classical Solution of Gauge Field} 
\label{Momentum gauge}

In this section, we give some details of the solution of the gauge field equation of motion obtained from the Euclidean massive spin-1 Lagrangian
\begin{eqnarray}
S&=&\!\!\!\!\!\int d^{d+1}x\sqrt{G} \Bigl[\frac{1}{4} F^{MN}F_{MN}+\frac{1}{2}W^{*}_{MN} W^{MN} +m^2 W^{*}_M W^M -ig\,\alpha F^{MN}W^*_MW_N \nonumber\\
&&+\,ig\beta F^{MN}\,\Big(  \nabla_{M} W^*_P\nabla^PW_{N} -\nabla_{M} W_P\nabla^PW_{N}^*\Big)
\Bigl]
\label{5.6a}
\end{eqnarray}
The length dimension of various quantities appearing in the above action are given by 
\be
[W_M] = \f{1-d}{2};\qquad  [A_M] = \f{1-d}{2}\;;\quad [g] = \f{d-3}{2};\quad  [\alpha] = 0;\qquad  [\beta] = 2
\ee
The gauge field equation of motion in the AdS background is given in equation \eqref{ytr5a}. In the Poincar\'e coordinates, the $z$ and $\mu$ components of this equation take the form
\begin{eqnarray}
\f{z^2}{L^2}\,\delta^{\mu\nu}\, k_\mu \,\partial_z\,A_\nu(z,\,k)=i\,J_z(z,\,k)\qquad;\qquad\frac{z^2}{L^2} \partial_z^2 A_\mu+(3-d) \frac{z}{L^2} \partial_zA_\mu-\frac{k^2}{L^2}\,\pi^{\;\;\nu}_\mu A_\nu=J_\mu\label{C.37}
\end{eqnarray}
where $k^2=\delta^{\mu\nu}\,k_\mu\,k_\nu$ and we have introduced the transverse projector
\be   
\pi_{\mu\nu}= \delta_{\mu\nu} -\frac{k_\mu\,k_\nu}{k^2}\quad;\quad \delta^{\mu\nu} k_\mu \pi_{\nu\sigma}=0\quad;\quad \pi_{\mu\nu} \,\delta^{\nu\tau}\pi_{\tau\sigma}=\pi_{\mu\sigma}\, .\label{pimunuq1}
\ee
In the following we shall solve the classical equations of motion of the gauge field perturbatively in $g$ as
\begin{eqnarray}
A_\mu(z,\,k)={\cal A}_\mu^{[0]}(z,\,k) +g\,{\cal A}_\mu^{[1]}(z,\,k)\, , \label{ftr5}
\end{eqnarray} 
where ${\cal A}_\mu^{[1]}(z,\,k)$  and ${\cal A}_\mu^{[0]}(z,\,k) $ satisfy \eqref{C.37} with and without the source term, respectively. The ${\cal A}_\mu^{[0]}(z,\,k)$  and ${\cal A}_\mu^{[1]}(z,\,k) $ can be solved easily in terms of the bulk-to-boundary (Btb) and bulk-to-bulk (BtB) propagators. This will be done below. However, before doing this, we note that for solving the equations of motion, it is convenient to split $A_\mu$ and $J_\mu$ in the transverse and longitudinal components as \cite{Liu:1998ty}
\begin{eqnarray} \label{decomp_A_J}
A_\mu =  A_\mu^\perp+i \,k_\mu\,A^{||} \qquad;\qquad J_\mu = \pi_\mu^{\;\;\nu}  J_\mu= J_\mu^\perp+i \,k_\mu\, J^{||}
\end{eqnarray}
where $A_\mu^\perp = \pi_\mu^{\;\;\nu} A_\nu, \ A^{||} = -i k^\mu A_\mu/k^2$ and similar for $J_\mu^\perp$ and  $J^{||}$ (indices are contracted with the flat metric $\delta_{\mu\nu}$). 

Using the two equations in \eqref{C.37}, the equations of motion for the longitudinal modes is found to be
\begin{eqnarray} \label{cons}
J^{||}= \frac{1}{k^2} \partial_zJ_z+\frac{(1-d)}{k^2} \frac{J_z}{z}\, .
\end{eqnarray}
This is same as the conservation condition $\nabla_MJ^M=0$ and hence it is identically satisfied. This also shows that the $z$ component of the equation of motion is satisfied automatically provided the current $J_M$ is conserved.

\subsubsection{Bulk-to-boundary propagator}

Substituting \eqref{ftr5} in \eqref{C.37}, we find that ${\cal A}_\mu^{(0)}$ satisfies \eqref{C.37} without the source terms $J_\mu$ and $J_z$ since the source term is linear in the coupling $g$. We can solve the resulting homogeneous equation by introducing the bulk-to-boundary (Btb) propagator $\mathbb{K}_{\mu}^{\;\;\nu}(z,k)$ defined as
\be
{\cal A}^{[0]}_\mu(z,k) = \mathbb{K}_{\mu}^{\;\;\nu}(z,k) A_{(0)\nu}(k)\, ,
\ee
where $A_{(0)\nu}(k)$ is the boundary value of the gauge field, i.e.,
\be
{\cal A}^{(0)}_\mu(z\rightarrow 0,k)=A_{(0)\nu}(k)\, . \label{C.61}
\ee
The $\mathbb{K}_{\mu}^{\;\;\nu}(z,k)$ satisfies the differential equation 
\begin{eqnarray}
\left(z^2 \partial_z^2 +(3-d)z \partial_z\right)\mathbb{K}_\mu^{~\nu}(z,\,k)-k^2\,\pi^{\;\;\sigma}_\mu\mathbb{K}_\sigma^{~\nu}(z,\,k)=0\, , \label{C.41a}
\end{eqnarray}
with the boundary condition
\begin{eqnarray}
\lim_{z\rightarrow 0} z^{\Delta-d+1}\,\mathbb{ K}_\mu^{~\nu}( z,\,k)=\delta_\mu^\nu\qquad;\qquad \Delta =d-1\, .\label{bound6}
\end{eqnarray}
The solution of \eqref{C.41a} is easily obtained by splitting the longitudinal and transverse parts as
\be
\mathbb{K}_\mu^{~\nu}(z,\,k) &=& \mathbb{K}^\perp(z,\,k) \pi_\mu^{~\nu} +\mathbb{K}^{||}(z,\,k)\f{k_\mu k^\nu}{k^2}
\ee
These longitudinal and transverse components satisfy decoupled differential equations
\be
z^2\p_z^2\mathbb{K}^{\perp}+(3-d)z\p_z\mathbb{K}^{\perp}-z^2k^2\mathbb{K}^{\perp} =0\quad;\quad z^2\p_z^2 \mathbb{K}^{||} +(3-d)z\p_z\mathbb{K}^{||}=0\, . \label{gft1}
\ee
These have the solution
\be
\mathbb{K}^{\perp} = c_0(k) z^{\f{d-2}{2}}K_{\f{d}{2}-1}(zk)\qquad,\quad  \mathbb{K}^{||}=c_1(k)z^{d-2}+c_2(k)\, .
\ee
Imposing the boundary condition \eqref{bound6}, we find
\be
c_0(k)=\f{2^{2-\f{d}{2}}}{\Gamma\left(\f{d}{2}-1\right)}k^{\f{d}{2}-1}\qquad,\quad c_1(k)=0\qquad,\quad c_2(k)=1\, . \label{C.131}  
\ee
Thus, the bulk-to-boundary propagator can be written as 
\be
\mathbb{K}_{\mu\nu}(z,k)&=&c_0(k)z^{\f{d-2}{2}}K_{\f{d}{2}-1}(zk)\pi_{\mu\nu}\;+\; \f{k_\mu k_\nu}{k^2}\, , \label{C.70}
\ee
where we have lowered the boundary indices using the flat metric $\delta_{\mu\nu}$. 

The leading order solution ${\cal A}^{(0)}_{\mu}$ is, thus, given by
\be
{\cal A}^{[0]}_{\mu}(z,k)&=&c_0(k)z^{\f{d-2}{2}}K_{\f{d}{2}-1}(zk)\pi_{\mu}^{\;\;\nu}(k)A_{(0)\nu}(k)\;+\; \f{k_\mu k^\nu}{k^2}A_{(0)\nu}(k)\non\\
&=&\mathcal{A}_\mu^{[0]\perp}(z,k) + ik_\mu \mathcal{A}_\mu^{[0]||}\label{C.85a}
\ee
It is straightforward to verify that the above solution automatically satisfies both the equations in \eqref{C.37} with $J_M=0$.

\subsubsection{Bulk-to-bulk  propagator}
\label{btbappx}
The solution of \eqref{C.37} at first order in the gauge coupling constant $g$ can be obtained using the bulk-to-bulk propagator ${\cal G}_{\mu\nu}(z,w;k)$ defined by
\begin{eqnarray}
\left[\left(\frac{z}{L^2}(3-d) \partial_z +\frac{z^2}{L^2}\partial^2_z\right)\delta^{\;\sigma}_\mu -\frac{k^2}{L^2}z^2\pi_\mu^{\;\;\sigma}\right] {\cal G}_{\sigma\nu}(z,\,w;\,k)&=&\frac{G_{\mu \nu}}{\sqrt{G}}\;\delta(z-w)\, , \label{C.41}
\end{eqnarray}
with the boundary condition at the conformal boundary,
\be \label{BtB_bdry}
  \lim_{z\rightarrow0} z^{\Delta-d+1} \mathcal{G}_{\mu\nu}(z,w;k)= 0\qquad,\qquad 
    \Delta=d-1\,
\ee
and the regularity in the interior. The solution of the gauge field equation to first order in the gauge coupling can now be expressed as 
\begin{eqnarray} \label{amu1}
{\cal A}^{[1]}_\mu(z,\,k)=\int dw \sqrt{G} \,{\cal G}_{\mu\nu}(z,\,w;\,k)\,J^\nu(w,\,k)\, .
\end{eqnarray}
Equation \eqref{C.41} can again be solved by splitting ${\cal G}_{\mu\nu}(z,w;k)$ in the transverse and longitudinal components as
 \begin{eqnarray}
 {\cal G}_{\mu\nu}(z,\,w;\,k)={\pi}_{\mu\nu} {\cal G}^\perp(z,\,w;\,k)+ \frac{k_\mu k_\nu}{k^2}{\cal G}^\parallel(z,\,w;\,k)\, .
 \end{eqnarray}
These components satisfy the equations
\begin{eqnarray}
\left[ \frac{d}{dz} \left(\hat{z}^{3-d}\frac{d}{dz}\right)-\hat{z}^{3-d}k^2
\right]{\cal G}^\perp=\delta(z-w)~~;~~ \left[ \frac{d}{dz} \left(\hat{z}^{3-d}\frac{d}{dz}\right)\right]{\cal G}^\parallel(z,w;k)=\delta(z-w)\, ,\label{N.5}
 \end{eqnarray}
where, to simplify the notation, we have introduced $\hat{z}=\frac{z}{L}$.

To solve the two equations in \eqref{N.5}, it is useful to recall the Green's function solution of first order inhomogeneous differential equations of the form 
 \begin{eqnarray}
{\cal L}\;y(z) =f(z)\qquad;\qquad {\cal L} = \frac{d}{dz} \left(p(z) \frac{d}{d z}\right) +q(z)\, , \label{L.1}
\end{eqnarray}
where ${\cal L}$ is a self-adjoint differential operator. The Green's function for this equation is defined by
\begin{eqnarray}
{\cal L}\,G(z,w)=\delta(z-w)\, ,
\end{eqnarray}
and its solution is obtained by following a standard procedure, see e.g., \cite{arfken}. The general solution, in an interval $(a,b)$, is given by
\begin{eqnarray}
G(z,\,w)=\left\{\begin{array}{ll}
A\,y_1(z)\,y_2(w),&\mbox{ for}\;\;z<w\\
A\,y_2(z)\,y_1(w),&\mbox{ for}\;\;z>w\end{array}\right.\label{3319u}
\end{eqnarray}
$y_1$ and $y_2$ satisfy ${\cal L}\; y_{1}=0={\cal L}\; y_{2}$, and $y_1(z)$ satisfies the suitable boundary condition at $z=a$ while $y_2(z)$ satisfies the suitable boundary condition at $z=b$. The coefficient $A$ is determined by requiring the Green's function to be continuous at $z=w$ but with a discontinuous derivative. This gives
\begin{eqnarray}
A\left[y'_2(w)\,y_1(w)- y'_1(w)\,y_2(w)\right]=\frac{1}{p(w)}\, .
\end{eqnarray}
Following this procedure to solve the two equations in \eqref{N.5}, we find that the solution of the homogeneous equation corresponding to the first equation in \eqref{N.5} is given by Bessel functions of the first and second kinds as 
\be
y_1(k,z) = \hat{z}^{\f{d}{2}-1} I_{\f{d}{2}-1}(kz)\qquad,\qquad y_2(k,z) = \hat{z}^{\f{d}{2}-1} K_{\f{d}{2}-1}(kz)
\ee
where $y_1$ satisfies the boundary condition at $z=0$ ({\it  i.e.} for $z<w$) and $y_2$ satisfies the boundary condition at $z=\infty$ ({\it i.e.} for $z>w$). The constant $A$ in \eqref{3319u} is evaluated to be $A=-1$. Thus, the transverse component $ \mathcal{G}^\perp(z,w;k)$ can be expressed as
\begin{eqnarray}
    \mathcal{G}^\perp(z,w;k)= 
-L\begin{cases}
    (\hat{z}\hat{w})^{\f{d}{2}-1}I_{\f{d}{2}-1}(k z)K_{\f{d}{2}-1}(k w),& \text{for } z< w\\[.3cm]
     (\hat{z}\hat{w})^{\f{d}{2}-1}I_{\f{d}{2}-1}(k w)K_{\f{d}{2}-1}(k z),              & \text{for } z > w
\end{cases}
\end{eqnarray}
Following similar steps, the longitudinal component is obtained to be
\be
\mathcal{G}^{||}_{\mu\nu}(z,w;k)&=& -\f{L}{d-2}\f{k_\mu k_\nu}{k^2}\begin{cases}
      \hat{z}^{d-2},& \text{if } z< w\\[.3cm]
     \hat{w}^{d-2},              & \text{if } z > w
         \end{cases}
\ee
Combining the transverse and longitudinal parts, the full bulk-to-bulk propagator for the gauge field is obtained to be
\be
 \mathcal{G}_{\mu\nu}(z,w;k) &=& -L
    \begin{cases}
      (\hat{z}\hat{w})^{\f{d}{2}-1}I_{\f{d}{2}-1}(k z)K_{\f{d}{2}-1}(k w)\pi_{\mu\nu}+\f{\hat{z}^{d-2}}{d-2}\f{k_\mu k_\nu}{k^2},& \text{if } z< w\\[.4cm]
     (\hat{z}\hat{w})^{\f{d}{2}-1}I_{\f{d}{2}-1}(k w)K_{\f{d}{2}-1}(k z)\pi_{\mu\nu}+\f{\hat{w}^{d-2}}{d-2}\f{k_\mu k_\nu}{k^2},              & \text{if } z > w
         \end{cases}\label{C.85}
\ee
By construction, the bulk-to-bulk propagator satify the second equation in \eqref{C.37}. Let us now verify that it satisfies the first equation as well. Using \eqref{amu1} we compute,
\begin{align}
 k^\mu {\cal A}^{[1]}_\mu(z,\,k)
&=\int dw \sqrt{G} k^\mu {\cal G}_{\mu\nu}(z,\,w;\,k)\,J^\nu(w,\,k) \nonumber \\
&=-\frac{L^{2}}{d-2} \int_{0}^{\infty} \frac{dw}{w^{d-1}} \left(\Theta(z-w) w^{d-2} + \Theta(w-z) z^{d-2}\right) k^\mu J_\mu(w,\,k)
\end{align}
where in the second equality we used \eqref{C.85}. Using \eqref{decomp_A_J} and \eqref{cons} we find
\begin{equation}
 k^\mu J_\mu(w,\,k) = i \left(\partial_w J_w + (1-d) \frac{J_w}{w}\right) 
 \quad \Rightarrow \quad \frac{k^\mu J_\mu(w,\,k)}{w^{d-1}} = i \partial_w \left(\frac{J_w}{w^{d-1}}\right)\, . 
\end{equation}
Thus,
\begin{align}
 k^\mu {\cal A}^{[1]}_\mu(z,\,k)
&= -i \frac{L^{2}}{d-2} \int_{0}^{\infty} dw \left(\Theta(z-w) w^{d-2} + \Theta(w-z) z^{d-2}\right) \partial_w \left(\frac{J_w}{w^{d-1}}\right) \nonumber \\
&=-i \frac{L^{2}}{d-2} \left(\left[ \left(\Theta(z-w) w^{d-2} + \Theta(w-z) z^{d-2}\right) \frac{J_w}{w^{d-1}} \right]_0^\infty\right. \nonumber \\
&\left.\qquad \qquad -\int_{0}^{\infty} dw \left(\delta(w-z) (z^{d-2} - w^{d-2}) -(d-2) w^{d-3} \Theta(z-w)\right) \frac{J_w}{w^{d-1}} \right) \nonumber \\
&=i L^2 \int_{0}^{\infty} dw \Theta(z-w) \frac{J_w}{w^2} \label{kA}
 \end{align}
where the vanishing of the boundary term at $w=0$ requires that $J_w$ goes to zero faster than $w$, which is guaranteed by the first of \eqref{C.37} and the boundary conditions in \eqref{BtB_bdry}. Differentiating \eqref{kA} w.r.t. $z$ and rearranging yields the first of \eqref{C.37}.

In computing the 3-point function, we need the expression of the bulk-to-bulk propagator near the boundary $z\rightarrow0$. In this limit, the expression \eqref{C.85} gives
\be
\mathcal{G}_{\mu\nu}(z\rightarrow0,w;k) &=&-\f{L}{2^{\f{d}{2}-1}\Gamma(\f{d}{2})} (k)^{\f{d}{2}-1}(\hat{z}^2w)^{\f{d}{2}-1}K_{\f{d}{2}-1}(k w)\pi_{\mu\nu}-L\,\f{\hat{z}^{d-2}}{d-2}\f{k_\mu k_\nu}{k^2}\non\\
  &=&-\f{L^{3-d}}{(d-2)} {z}^{d-2}\mathbb{K}_{\mu\nu}( w,k)\label{C.86}
\ee

\subsection{Classical Solution of Massive Spin-1 Field}
\label{C.4}

In this section, we review the solution of the massive spin-1 field following the approach given in \cite{9805145}. We are interested in getting the classical solution of the massive field at the leading order in the gauge coupling $g$. As we shall see below, this can be obtained in terms of the bulk-to-boundary propagator of the massive field. The equation of motion of the massive spin-1 field is given by
\begin{eqnarray}
2\nabla_M\nabla^{[M}W^{N]} -m^2 W^N=0+{\cal O}(g)\, . \label{wm67}
\end{eqnarray}
By acting with the covariant derivative $\nabla_N$, we obtain the following {\em subsidiary condition}
\begin{eqnarray}
 \nabla_MW^M=0+{\cal O}(g)\quad\implies\qquad \delta^{\mu\nu}\partial_\mu W_\nu +\partial_zW_z- \frac{(d-1)}{z} W_z=0+{\cal O}(g)\, . \label{C.42}
\end{eqnarray}
The classical profile of the massive spin-1 fields must satisfy this constraint at the leading order in the gauge coupling expansion. 

Fourier transforming the boundary directions and using the subsidiary condition \eqref{C.42}, the $z$ component of the equation of motion \eqref{wm67} gives in  Poincar\'e coordinates,
\begin{eqnarray}
&&z^2 \partial_z^2 W_z-(d-1) z \partial_z W_z -k^2 z^2  W_z+ \Bigl(d-1-\,m^2L^2\Bigl)W_z=0\, . \label{wzq1}
\end{eqnarray}
Demanding regularity at $z=\infty$, the above equation has the solution 
\begin{eqnarray}
W_z(z,\,k)=c(k)\, z^{\frac{d}{2}}\,K_\beta(z\,k)\qquad;\qquad \beta^2= \frac{(d-2)^2}{4}+m^2L^2 \quad;\quad \beta=\Delta-\frac{d}{2}\, ,\label{B.45}
\end{eqnarray}
 where $K_\beta(z\,k)$ is the modified Bessel function of the second kind and $c(k)$ is an arbitrary function. 

Similarly, the $\mu$ component of the equation of motion \eqref{wm67} on using \eqref{B.45} gives 
\begin{eqnarray}
z^2\partial_z^2W_\mu+(3-d)z\,\partial_zW_\mu-(z^2\,k^2+m^2L^2) W_\mu= 2iz k_\mu W_z=2i\,c(k) \,k_\mu z^{\frac{d}{2}+1}\,K_\beta(z\,k)\, .\label{wmuq1}
\end{eqnarray}
The solution of this equation has a homogeneous and an inhomegeneous part. The inhomogeneous part should be proportional to $k_\mu$. It is easy to see that the above equation has the following solution consistent with the constraint \eqref{C.42}
\be
W_\mu (z,k) &=&\left[\delta_\mu^\nu z^{\f{d-2}{2}}K_\beta(kz)+ \f{k^\nu k_\mu}{k(d-\Delta-1)} z^{\f{d}{2}} K_{\beta+1}(zk)\right]a_\nu(k)\, .
\label{3260n}
\ee
For later use, we note that the relation between $c(k)$ and $a_\mu$ following from the constraint \eqref{C.42} is  
\be
c(k) \Bigl(\f{d}{2}-\beta-1\Bigl)= ik^\mu a_\mu(k)\, .
\ee
We can obtain the bulk-to-boundary propagator of the massive spin-1 field using the above solution. For this, we need to relate $a_\mu(k)$ to the boundary value of the field $W_\mu(z,k)$. Writing $a_\mu= b_\mu +i \,k_\mu b$ and using the expression of the modified Bessel function in $z\rightarrow0$ limit given in equation \eqref{5.49bf}, we find
\be \label{massive_BC}
W_\mu(z\rightarrow0,k)
&\equiv& z^{d-\Delta-1}w_\mu(k)\, ,
\ee
where
\be
w_\mu(k) &=&\f{1}{2} \left(\f{k}{2}\right)^{\f{d}{2}-\Delta} \Gamma\Bigl(\Delta-\f{d}{2}\Bigl)\left[b_\mu+ k_\mu\left(\f{(\Delta-1)}{(d-\Delta-1)}b+\f{2k^\nu b_\nu \Bigl(\Delta-\f{d}{2}\Bigl)}{k^2(d-\Delta-1)}\right)  \right]\, .
\ee
We can get rid of term proportional to $k_\mu$ by choosing $b$ to be $ \f{(d-2\Delta)}{(\Delta-1)}\f{k^\nu b_\nu}{k^2}$. This allows us to relate the integration constant with the boundary value of the field. Collecting all results and using Bessel function identities, we can write
\be
W_\mu (z,k) 
&=&\f{2\;z^{\f{d-2}{2}}}{\Gamma(\Delta-\f{d}{2})}\Bigl(\f{k}{2}\Bigl)^{\Delta-\f{d}{2}}\left[\delta_\mu^\nu\; K_{\Delta-\f{d}{2}}(kz)+ \f{z\,k^\nu k_\mu}{k(\Delta-1)} K_{\Delta-\f{d}{2}-1}(zk)\right]
w_\nu(k)\label{3260n1y}
\ee
\begin{eqnarray}
W_z(z,\,k)=i  \frac{2^{\frac{d}{2}+1-\Delta }}{\Gamma(\Delta -\frac{d}{2})} \,\f{1}{\Delta-1}k^{\Delta-\frac{d}{2}}\, z^{\frac{d}{2}}\,K_{\Delta -\frac{d}{2}}(z\,k)\, k^\nu\,w_\nu(k)\label{C.124}
\end{eqnarray}
The bulk-to-boundary propagator $\mathcal{K}_M^{~\mu} (z,k)$ for the massive spin-1 field can now be defined by
\begin{eqnarray}
W_M(z,\,k) = \mathcal{K}_M^{~\mu} (z,k)\,w_\mu(k)\qquad;\qquad\lim_{z\rightarrow 0} z^{-d+\Delta +1} \, \mathcal{K}_M^{~\mu} (z,k)\,w_\mu(k)=\delta^\mu_M\, .\label{C.11}
\end{eqnarray}
Comparing \eqref{C.11} with \eqref{3260n1y} and \eqref{C.124}, we get 
\begin{eqnarray}
\mathcal{K}_\mu^{~\nu}(z,\,k)&=&  \frac{ 2^{\frac{d}{2}+1-\Delta}}{\Gamma\left(\Delta-\frac{d}{2}\right)} \,~k^{\Delta-\frac{d}{2}} \, z^{\frac{d}{2}-1} \left[\delta_\mu^\nu~K_{\Delta-\frac{d}{2}}(z k)+\frac{k_\mu\,k^\nu}{k}~\frac{z}{\Delta-1}~K_{\Delta-\frac{d}{2}-1}(zk)\right]\, ,\nonumber\\
\mathcal{K}_z^{~\nu} (z,\,k) &=& i\,\frac{2^{\frac{d}{2}+1-\Delta} }{\Gamma\left(\Delta-\frac{d}{2}\right)}~\frac{k^\nu~ k^{\Delta -\frac{d}{2}}}{\Delta-1}~z^{\frac{d}{2}}\, K_{\Delta -\frac{d}{2}}(zk)\, . \label{C.64}
\end{eqnarray}
We also need the bulk-to-boundary propagator of the complex conjugate field $W^*_M$. This  is considered independent of $W_M$ and its boundary Fourier transform is defined by
\begin{eqnarray}
W^*_M(z,\,x)=\int \frac{d^d k}{(2\pi)^d} \,e^{ik\cdot x} \, W^*_M(z,\,k)\,,\label{C.65}
\end{eqnarray}
The $W^*_M$ satisfies the same equation of motion as $W_M$. From this, we find that the bulk to propagator for $W^*_M$, denoted with $\bar{\mathcal{K}}_M^{~\mu}(z,\,k)$, coincides with equation \eqref{C.64}, {\it i.e.},
\begin{eqnarray}
\bar{\mathcal{K}}_M^{~\nu}(z,\,k)&=&  {\mathcal{K}}_M^{~\nu}(z,\,k) \;\,=\;\, {\mathcal{K}}_M^{*\nu}(z,\,-k) \label{C.66}
\end{eqnarray}
where ${\mathcal{K}}_M^{*\nu}$ denotes the complex conjugate of ${\mathcal{K}}_M^{\;\;\nu}$.

\section{Analysis in Lorenz Gauge \label{app:lorenz}} 

In the previous appendix and in the main text, we had worked with the axial gauge in which we set $A_z=0$. It is also instructive to consider the standard Lorenz gauge where we set $\nabla_M A^M=0$. This condition is imposed by adding to the action a gauge fixing term so that the gauge field action becomes  
\be
S&=& \int d^{d+1}x \sqrt{g}  \biggl[ \f{1}{4}F^{MN}F_{MN}+\f{1}{2\xi}(\nabla_MA^M)^2+A_NJ^N  \biggl]\, ,
\ee
and taking the limit $\xi \to 0$ (sometimes this is also referred as Landau gauge).
The equation of motion is given by
\be
\nabla_MF^{MN} +\f{1}{\xi}\nabla^N\nabla^MA_M=  \left(\Box+\f{d}{L^2}\right)A^N -\left(1-\f{1}{\xi}\right)\nabla^N\nabla_MA^M  =J^N\, , \label{403erg}
\ee
where the first equality is valid in AdS space. 
For $N=z$ and $N=\mu$, the above equation gives
\be
L^2J_z&=&-z^2 \delta^{\mu\nu}\p_z\p_\mu A_\nu +z^2\delta^{\mu\nu}\p_\mu\p_\nu A_z\non\\
&&+\f{1}{\xi}\biggl(z^2\p_z^2 A_z +(3-d)z\p_zA_z+z^2\delta^{\mu\nu}\p_\mu\p_z A_\nu+2z\delta^{\mu\nu}\p_\mu A_\nu +(1-d)A_z\biggl)\, ,\label{poytg}
\ee
\be
L^2J_\mu&=&\Bigl[z^2\p_z^2+(3-d)z\p_z+z^2\delta^{\nu\sigma}\p_\nu\p_\sigma \Bigl]A_\mu  -z^2\delta^{\nu\sigma}\p_\nu\p_\mu A_\sigma -z^2\p_z\p_\mu A_z -(3-d)z\p_\mu A_z \non\\
&&+\f{1}{\xi}\biggl( z^2\delta^{\sigma\nu}\p_\mu \p_\sigma A_\nu+z^2\p_\mu \p_z A_z +(1-d)z\p_\mu A_z \biggl)\, .
\label{poyt1g}
\ee
Next, we use the condition $\nabla_M A^M=0$. For the source free case, the above equations take the same form as in \eqref{wzq1} and \eqref{wmuq1} for the corresponding equations for the massive spin-1 fields but with $m=0$. Thus, we can immediately write down the solution 
\be
A_z(z,k) &=& e(k)z^{\f{d}{2}}K_{\f{d}{2}-1}(kz)\non\\
A_\mu (z,k) &=& e_\mu(k) z^{\f{d-2}{2}}K_{\f{d}{2}-1}(kz)-i \f{e(k)}{k}k_\mu z^{\f{d}{2}} K_{\f{d}{2}}(zk)\label{3260h}
\ee
Substituting the above solution in the Lorenz condition $\nabla_MA^M=0$ gives $e_\mu k^\mu=0$. Thus, we can parametrise $e_\mu$ as $e_\mu =\pi_{\mu}^{\;\;\nu}\alpha_\nu$ where $\pi_{\mu\nu}$ is the transverse projector defined in equation \eqref{pimunuq1}. Writing $\alpha_\mu = b_0 a_\mu+b_1 k_\mu$ where $a_\mu$ is the boundary value of the field and following the same manipulations we did for the Proca field, we can fix the constants in terms of the boundary value of the fields to be 
\begin{eqnarray}
&&A_\mu(z\,k)=\frac{2^{2-\frac{d}{2}}}{\Gamma\left[\frac{d}{2}-1\right]}\, (z\,k)^{\frac{d}{2}-1}\,\,\left[a^\nu\,{\pi}_{\nu\mu}K_{\frac{d}{2} -1}(z\,k)+\frac{(k\cdot a)\,z\,k_\mu}{(d-2)\,k}\,K_{\frac{d}{2}}(z\,k)\right]\, ,\nonumber\\
&&A_z(z,\,k)= i\frac{2^{2-\frac{d}{2}}}{\Gamma\left[\frac{d}{2}-1\right]}\,k^{\frac{d}{2}-1} \, z^{\frac{d}{2}} \,\frac{(k\cdot a)}{ (d-2)}\,K_{\frac{d}{2}-1}(z,k)\, . 
\end{eqnarray} {These expressions can also be obtained from the corresponding solutions of the Proca field given in equations \eqref{3260n1y} and \eqref{C.124} by substituting $\Delta=d-1$. The above expressions also give the bulk-to boundary propagator of the gauge field in the Lorenz gauge by writing $A_M = \mathcal{K}_M^{\;\;\mu}\;a_\mu$.}

\vspace*{.07in}This is not the end of the story. We still have a residual gauge freedom allowed by the Lorenz gauge condition. More specifically, even after fixing the Lorenz gauge, a further gauge transformation
\begin{eqnarray}
A_M\rightarrow A_M-i\partial_M g(z,x)& \Longleftrightarrow &A_\mu\rightarrow A_\mu +k_\mu g(z,k)~~\mbox{and}~~ A_z\rightarrow A_z- i \partial_z g(z,k)\, , \label{e.174q1}
\end{eqnarray}
will be a residual gauge transformation if the function $g(z,k)$ satisfies the condition $\Box g=0$, {\it i.e.}, 
\begin{eqnarray}
\partial_z^2g(z,k) -\frac{(d-1)}{z}\partial_z g(z,k) -k^2g(z,k)=0\label{9.86a}
\end{eqnarray}
The solution of this equation is given by
\begin{eqnarray}
g(z,k)= \lambda(k)\,z^{\frac{d}{2}}\, K_{\frac{d}{2}}(z\,k)
\end{eqnarray}
The $\lambda$ is an arbitrary function of the momentum. If we choose it to be
\begin{eqnarray}
\lambda(k) = -\frac{2^{2-\frac{d}{2}}\,k^{\frac{d}{2} -1}}{\Gamma\left[ \frac{d}{2} -1\right]}\,\frac{(k\cdot a)}{(d-2)\,k}
\end{eqnarray}  
then taking into account the residual gauge transformation \eqref{e.174q1}, the boundary component of the gauge field becomes completely transverse, i.e.,  
\begin{eqnarray}
&&A_\mu(z\,k)=\frac{2^{2-\frac{d}{2}}}{\Gamma\left[\frac{d}{2}-1\right]}\, (z\,k)^{\frac{d}{2}-1}\,\,a^\nu\,{\pi}_{\nu\mu}K_{\frac{d}{2} -1}(z\,k)~~;~~
A_z(z,\,k)= 0
\end{eqnarray} 
This also gives a proof of the result we have used, namely, we can choose the gauge field to be completely transverse in the axial gauge $A_z=0$.

\section{Holographic renormalization of massive spin-1 field \label{s3proca} }

In this appendix, we compute the 2-point function of the boundary operators that are dual to the Proca field by applying the holographic renormalization procedure to the Euclidean action on the AdS background given in Section \ref{Bulk theory}. 
This would be needed to fix the longitudinal part of our 3-point function as well as for verifying the conservation Ward  identity \eqref{decnb4e} using a bulk computation. We give the details for the $W_M$ field; the analysis for the corresponding complex conjugate field $W_M^*$ is identical. We start by solving the Proca equations of motion asymptotically.

\subsection{Asymptotic analysis}

We want to solve the free Proca field equation in AdS, which is given by
\be
  \left(\Box+\f{d}{L^2}\right)W_M -g^{PQ}\nabla_M\nabla_PW_Q-m^2W_M=0\;;\qquad \nabla^MW_M=0\, ,\label{F.197}
\ee
where the second equation (the subsidiary condition)  follows from the first upon contracting with $\nabla_M$, see section \ref{C.4}.
In Fefferman Graham coordinates, \eqref{B.26a}, we have
\be
L^2\Box W_M&=& \Bigl[4\rho^2 \p_\rho^2 +\rho L\delta^{\mu\nu}\p_\mu\p_\nu   +2(2-d)\rho \p_\rho \Bigl] W_M+\Bigl[4\rho (\p_\rho W_\nu -\p_\nu W_\rho) -d W_\nu\Bigl]\delta_M^\nu\non\\
L^2\nabla_M(\nabla_NW^N)&=& (4-2d)\rho \p_MW_\rho +4\rho^2 \p_M\p_\rho W_\rho+\rho L\delta^{\nu\sigma}\p_M\p_\nu W_\sigma\non\\
&& +\Bigl[8\rho\p_\rho W_\rho +L\delta^{\nu\sigma}\p_\nu W_\sigma+(4-2d)W_\rho\Bigl]\delta_{\rho M}\non\\[.3cm]
\nabla_MW^M &=& \f{\rho}{L} \delta^{\mu\nu}\p_\mu W_\nu +\f{4\rho^2}{L^2}\p_\rho W_\rho +\f{(4-2d)}{L^2} \rho W_\rho\label{3rt564yp}
\ee
Using these equations, the  boundary and radial components of the equation of motion can be expressed as  
\be
0&=&4\rho^2 \p_\rho^2W_\mu  +2(4-d)\rho \p_\rho W_\mu + \rho L\delta^{\sigma\nu}\p_\sigma\p_\nu W_\mu-m^2L^2W_\mu-4\rho L \partial_\nu W_\rho\, ,\nonumber\\
0&=&4\rho^2 \p_\rho^2W_\rho +2(4-d)\rho \p_\rho W_\rho 
+ \rho L\delta^{\sigma\nu}\p_\sigma\p_\nu W_\rho-m^2L^2W_\rho\, ,
\label{524erw}
\ee
where we have used the subsidiary condition, $\nabla_MW^M=0$, to simplify the expressions. The above equations can also be derived by transforming the equations of motion given in section \ref{C.4} in Poincar\'e coordinates to the Fefferman Graham coordinates. 

We want to obtain the general asymptotic solution of \eqref{524erw}. To this end, we first need to obtain the leading radial dependence as $\rho \to 0$, and to determine this it suffices to consider a solution that only depends on $\rho$.
In this case, both of the above equations take the same form 
\be
4\rho^2 \p_\rho^2W_M(\rho) +2(4-d)\rho \p_\rho W_M(\rho) -m^2L^2W_M(\rho)=0\, ,
\ee
with solution given by
\be
W_M(\rho) = c_M \rho^{\Delta_-} +e_M \rho^{\Delta_+}\, \label{ghuio}
\ee
where $c_M$ and $e_M$ are integration constants and 
\be
2\Delta_+ \;=\; \f{d-2}{2} +\sqrt{m^2L^2+\f{(d-2)^2}{4}} \; = \; \Delta-1 \;;\qquad 2\Delta_-= d-2-2\Delta_+=d-\Delta-1\, .  \non
\ee
The leading behaviour as $\rho \to $ is given by $\Delta_-$ and its coefficient  plays the role of the source for the dual boundary operator,  while the coefficient of $\Delta_+$ is linked with the 1-point function in the presence of sources \cite{0209067}, as will be seen below.

Next, we turn to obtain the general asymptotic solution by solving the equations order by order in the $\rho$ variable near $\rho=0$. This is achieved by factoring out the leading behavior, using 
\be \label{eq:sub}
W_\rho = \left(\f{\rho}{L}\right)^{\f{d-\Delta-1}{2}}\mathcal{W}_\rho(\rho,x) \quad;\qquad W_\mu = \left(\f{\rho}{L}\right)^{\f{d-\Delta-1}{2}}\mathcal{W}_\mu(\rho,x)
\ee
and then solving for $\mathcal{W}_\rho(\rho,x), \mathcal{W}_\mu(\rho,x)$. By construction, these variables are finite at $\rho=0$. 
Substituting these in the equations of motion one finds at leading order the relation  between the mass of the bulk field and the conformal dimension $\Delta$, namely $m^2L^2 = (\Delta-1)(\Delta-d+1)$. 
After cancelling an overall factor of $\rho$ the field equations become, 
\be
0&=&4\rho \p_\rho^2 \mathcal{W}_\mu +2(2+d-2\Delta)\p_\rho \mathcal{W}_\mu + L\delta^{\nu\sigma}\p_\nu\p_\sigma \mathcal{W}_\mu
- L \delta^{\nu\sigma}\p_\mu\p_\nu \mathcal{W}_\sigma -4\rho\p_\rho\p_\mu \mathcal{W}_\rho +2(\Delta-3)\p_\mu \mathcal{W}_\rho\, ;
\label{vector} \non \\
\ee
In addition, the subsidiary condition, $\nabla_M W^M=0$, shows that the asymptotic expansion of $W_\rho$ is fully determined in terms of that of $W_\mu$. Indeed, using \eqref{eq:sub} in \eqref{3rt564yp} and rearranging yields
\begin{equation} \label{eq:sub2}
  2(1-\Delta)  \mathcal{W}_\rho + 4 \rho \partial_\rho \mathcal{W}_\rho + L \delta^{\mu \nu} \partial_\mu \mathcal{W}_\nu =0\, .
\end{equation}
These equations are now solved by setting $\rho=0$ and solving them, then differentiating w.r.t. $\rho$, setting $\rho=0$ and solving them, and so on. This determines recursively the derivatives $\frac{\partial^n \mathcal{W}_\rho(\rho=0,x)}{\partial \rho^n}, \frac{\partial^n \mathcal{W}_\mu(\rho=0,x)}{\partial \rho^n}$ at $\rho=0$ in terms of lower order terms, provided that the coefficient that multiplies the term with the highest number of derivatives is non-zero. This is indeed the case till order $\lfloor \Delta-\f{d}{2} \rfloor$, where $\lfloor x \rfloor$ indicates the integer part of $x$. 
At order $\Delta-\f{d}{2}$, a new asymptotic solution appears, associated with the $\Delta_+$ solution in \eqref{ghuio}, and its coefficient is unconstrained by asymptotic analysis.  
When $\Delta-\f{d}{2}=n$ is an integer, the equations do not admit a solution unless there is a logarithmic term in the asymptotic solution. We thus obtain the asymptotic solution, 
\be
\mathcal{W}_\rho(\rho,x)&=& \sum_{j=0}^{\lfloor \Delta-\f{d}{2} \rfloor} \left(\f{\rho}{L}\right)^j \mathcal{W}^{(2j)}_\rho(x) \;\;+\;\;  \left(\f{\rho}{L}\right)^{\Delta-\f{d}{2}} \biggl(\mathcal{W}^{(2\Delta-d)}_\rho(x)+\delta_{\Delta, \frac{d}{2} + n}\mathcal{V}_\rho^{(2\Delta-d)}(x)\ln\f{\rho}{L}\biggl)\, ,\nonumber\\
\ee
\be
\mathcal{W}_\mu(\rho,x)&=& \sum_{j=0}^{\lfloor\Delta-\f{d}{2}\rfloor} \left(\f{\rho}{L}\right)^j \mathcal{W}^{(2j)}_\mu(x) \;\;+\;\;  \left(\f{\rho}{L}\right)^{\Delta-\f{d}{2}} \biggl(\mathcal{W}^{(2\Delta-d)}_\mu(x)+\delta_{\Delta, \frac{d}{2} + n}\mathcal{V}_\mu^{(2\Delta-d)}(x)\ln\f{\rho}{L}\biggl)\, . \nonumber\\\label{wmu789nj}
\ee 
The subsidiary conditions yields,
\begin{align}
&\mathcal{W}_\rho^{(2j)}= -\frac{ L}{2(1-\Delta +2j)}\delta^{\mu\nu} \partial_\mu {\cal W}_\nu^{(2 j)}, \qquad j \le \Delta-\f{d}{2} \label{eq:sub3} \\
&(2 + 2 n - d) \mathcal{W}_\rho^{(2\Delta -d)} + 4 \mathcal{V}_\rho^{(2\Delta -d)}=- L \delta^{\mu\nu}\partial_\nu\mathcal{W}_\mu^{(2\Delta-d)}, \qquad
(2 + 2 n - d) \mathcal{V}_\rho^{(2\Delta -d)}= -L \delta^{\mu\nu}\partial_\nu\mathcal{V}_\mu^{(2\Delta-d)}\, , \non
\end{align}
where the second line holds only when $\Delta=d/2+n$, with $n$ an integer.  
In our case, we are considering generic $\Delta$, not satisfying this condition, and the logarithmic terms will not be important, but for completeness we quote them below. The coefficients appearing in the above expansion are given by 
\be
\mathcal{W}_\mu^{(2j)} &=& B_{j} (L^2\Box_0)^{j-1} L^2 \left(\Box_0 \delta_\mu^\nu  +\frac{2j}{(1-\Delta)} \delta^{\kappa \nu} \p_\mu \p_\kappa \right) \mathcal{W}_\nu^{(0)}
\;;\quad j< \Delta-\f{d}{2}
\non\\[.2cm]
\mathcal{V}_\mu^{(2\Delta -d)} &=& -\f{1}{2^{2n}\Gamma(n)\Gamma(n+1)}   
(\Box_0)^{j-1} L^2 \left(\Box_0 \delta_\mu^\nu  +\frac{4n}{(2 - d- 2n)} \delta^{\kappa \nu} \p_\mu \p_\kappa \right) \mathcal{W}_\nu^{(0)}\, , \label{eq:mucoef}
\ee
where $\Box_0=\delta^{\mu\nu}\p_\mu\p_\nu$,  the last equation holds only when $\Delta=d/2+n$, and \be
B_j = \prod_{q=1}^{j}\f{1}{2q(2\Delta-d-2q)}\, .
\label{Bmwe}
\ee
For completeness, we also quote the asymptotic coefficients of $\mathcal{W}_\rho$ as they directly follow from \eqref{524erw},
\be
\mathcal{W}_\rho^{(2j)}(x)&=&   B_{j}(L^2\Box_0)^j\mathcal{W}_\rho^{(0)}(x)\;;\quad j< \Delta-\f{d}{2}\non\\[.2cm]
\mathcal{V}^{(2\Delta -d)}_\rho(x) &=& -\f{1}{2^{2n}\Gamma(n)\Gamma(n+1)}(L^2\Box_0)^n\mathcal{W}_\rho^{(0)} 
\ee
These values are consistent with \eqref{eq:mucoef} and \eqref{eq:sub3}.

The asymptotic analysis does not determine $\mathcal{W}_\mu^{(2 \Delta -d)}$ and $\mathcal{W}_\rho^{(2 \Delta -d)}$, but when $\mathcal{W}_\mu^{(2 \Delta -d)}$ is known $\mathcal{W}_\rho^{(2 \Delta -d)}$ follows from its divergence.
This is expected since these play the role of 1-point functions in the presence of sources,  and higher point correlators can be determined by differentiating them. However, the asymptotic analysis should not be enough to completely fix the full correlators, and this is reflected by the fact that these coefficients are undetermined from the asymptotic analysis. To find their expression, we need to use the exact solution of the free Proca field given in equations \eqref{3260n1y} and \eqref{C.124} and expand them near the boundary. The relation between the solutions in the FG and Poincar\'e coordinates is given by
\be
W_\rho(\rho,x) = \f{\p z}{\p \rho} W_z(z(\rho),x)=\f{1}{2}\sqrt{\f{L}{\rho}}\;W_z(z(\rho),x)\quad;\qquad  W_\mu (\rho,x)= W_\mu(z(\rho),x)
\ee
This gives in momentum space
\be
W_\rho(\rho,{\bf k}) &=& i \f{2^{\f{d}{2}-\Delta}}{\Gamma\bigl(\Delta-\f{d}{2}\bigl)}\f{1}{\Delta-1} k^{\Delta-\f{d}{2}}\sqrt{\f{L}{\rho}}(\rho L)^{\f{d}{4}} K_\beta (k\sqrt{\rho L}){\bf k}^\nu w_\nu\non\\
W_\mu(\rho,{\bf k})&=& \f{2}{\Gamma\bigl(\Delta-\f{d}{2}\bigl)} \left(\f{k}{2}\right)^{\Delta-\f{d}{2}}(\rho L)^{\f{d-2}{4}}\biggl[ \delta_\mu^\nu K_\beta (k\sqrt{\rho L})+\f{{\bf k}^\nu {\bf k}_\mu}{k(\Delta-1)}(\rho L)^{\f{1}{2}}K_{\beta-1} (k\sqrt{\rho L})\biggl]w_\nu\non
\ee
Using the asymptotic expansion of Bessel function given in \eqref{kiexpansion}, the above exact solution can be expanded near the boundary as
\be
W_\mu(\rho,{\bf k})&=& (\rho L)^{\f{d-\Delta-1}{2}}\Biggl[w_\mu+\cdots+ (-1)^j
(\rho Lk^2)^j 
\,{B_j}\,
\Bigl(w_\mu-\f{2j w_\nu {\bf k}^\nu {\bf k}_\mu}{k^2(\Delta-1)}\Bigl)+\cdots\biggl]\non\\
&&+(\rho L)^{\f{d-\Delta-1}{2}}\Biggl[ \left(\f{k}{2}\right)^{2\Delta-d}   \f{\pi\;(\rho L)^{\Delta-\f{d}{2}}\mbox{cosec}\Bigl(\f{(d-2\Delta)\pi}{2}\Bigl)}{\Gamma\Bigl(\Delta-\f{d}{2}\Bigl)\Gamma\Bigl(\Delta-\f{d}{2}+1\Bigl)} \Bigl(w_\mu+\f{w_\nu {\bf k}^\nu {\bf k}_\mu(d-2\Delta)}{k^2(\Delta-1)}\Bigl) +\cdots\Biggl]\non
\ee
\be
W_\rho(\rho,{\bf k})&=&\f{i\pi L\;\mbox{cosec}\Bigl(\f{(2\Delta-d)\pi}{2}\Bigl) }{(\Delta-1)\Gamma\Bigl(\Delta-\f{d}{2}\Bigl)} (\rho L)^{\f{d-\Delta-1}{2}}\Biggl[\f{1}{2\Gamma\Bigl(\f{d}{2}-\Delta+1\Bigl)}+ \f{k^2(L\rho)}{2^3\Gamma\Bigl(\f{d}{2}-\Delta+2\Bigl)}+\cdots\non\\
&& -\f{(\rho L)^{\Delta-\f{d}{2}}}{2\Gamma\Bigl(\Delta-\f{d}{2}+1\Bigl)}\left(\f{k}{2}\right)^{2\Delta-d}-\f{(\rho L)^{\Delta-\f{d}{2}+1}}{2^5\Gamma\Bigl(\Delta-\f{d}{2}+2\Bigl)}\left(\f{k}{2}\right)^{2\Delta-d+2}+\cdots\Biggl]{\bf k}^\nu w_\nu
\ee
with $B_j$ given in \eqref{Bmwe}.
Comparing the above expansion with asymptotic analysis and using the identity $\Gamma(x)\Gamma(-x)=-\f{\pi}{x}\mbox{cosec}(\pi x)$, we find
\be
\mathcal{W}^{(0)}_\mu({\bf k})&=& L^{d-\Delta-1}\;w_\mu({\bf k})\non\\
\mathcal{W}^{(2\Delta-d)}_\mu({\bf k}) 
&=& \left(\f{k}{2}\right)^{2\Delta-d}   \f{\Gamma\Bigl(\f{d}{2}-\Delta\Bigl)\; L^{2\Delta-d}}{\Gamma\Bigl(\Delta-\f{d}{2}\Bigl)} \Bigl(\delta^\nu_\mu+\f{ {\bf k}^\nu {\bf k}_\mu(d-2\Delta)}{k^2(\Delta-1)}\Bigl)\mathcal{W}^{(0)}_\mu({\bf k})\label{wnutyui}
\ee
\be
\mathcal{W}^{(0)}_\rho({\bf k})&=& \f{i \, L^{d-\Delta}}{2(\Delta-1)} \;{\bf k}^{\nu}w_\nu({\bf k})\non\\
\mathcal{W}^{(2\Delta-d)}_\rho({\bf k}) 
&=&-\f{L^{2\Delta-d}\Gamma\Bigl(\f{d}{2}-\Delta+1\Bigl)}{\Gamma\Bigl(\Delta-\f{d}{2}+1\Bigl)}\left(\f{k}{2}\right)^{2\Delta-d} \mathcal{W}^{(0)}_\rho({\bf k})
\ee
Note that the length dimension of $w_\mu$ and $\mathcal{W}^{(0)}_\mu$ are different. The $\mathcal{W}^{(0)}_\mu$ has the same dimension as the Proca field $W_M$, namely $\f{1-d}{2}$ whereas the length dimension of $w_\mu$ is $\f{3(1-d)}{2}+\Delta$.

\subsection{Regularised action and counter terms }

The bulk on-shell action diverges due to the near boundary contributions. To regularise the action, we first place the AdS boundary at $\rho=L\epsilon$ and then evaluate the resulting on-shell action. Doing an integration by parts and using the free field equation of motion of the Proca field, we find
\be
S_{\rm reg}&=& \int_{\rho\ge L\epsilon} d^{d+1}x \sqrt{g}  \Bigl[ \f{1}{2}W^*_{MN}W^{MN}  +m^2W^{*}_MW^M \Bigl]\non\\
&=& \int_{\rho= L\epsilon} d^{d+1}x \sqrt{\gamma}\;n_M (W^*_NW^{MN})\non\\
&=&-2\int_{\rho=L\epsilon} d^{d}x\;\left(\f{\rho}{L}\right)^{\f{d}{2}-\Delta}\left[\delta^{\mu\nu}\mathcal{W}^*_\mu\Bigl(\f{(d-\Delta-1)}{2L}\mathcal{W}_\nu+\f{\rho}{L}\p_\rho \mathcal{W}_\nu-\f{\rho}{L}\p_\nu \mathcal{W}_\rho\Bigl)\right]\, .\label{531wew}
\ee
In going to the last equality, we have used the fact that the boundary hypersurface is defined by $\rho=$ constant. Hence, the induced metric and the unit normal space like vector on this hypersurface are given by 
\be
n_M =-\f{\p_M\rho}{\sqrt{g^{PQ}n_Pn_Q}} =-\f{\p_M\rho}{\sqrt{g^{\rho\rho}}}=-\f{L}{2\rho} \delta_M^\rho\;;\quad
\gamma_{\mu\nu}(x) =\f{L}{\rho}\delta_{\mu\nu}\;;\quad \sqrt{\gamma} =\left(\f{L}{\rho}\right)^{\f{d}{2}}\, .
\ee
Now, using the asymptotic solution, we get 
\be
S_{\rm reg}
&=&-2\int d^dx \biggl[\epsilon^{\f{d}{2}-\Delta}a_{(0)} + \epsilon^{\f{d}{2}-\Delta+1}a_{(2)}+\cdots + 
\ln\epsilon\; b_{(2\Delta-d)}+O(\epsilon)\biggl]\, , \label{regact56yht}
\ee
where
\be
a_{(2j)}&=& \delta^{\mu\nu}\mathcal{W}^{*(0)}_\mu T^{(2j)}_\nu+ \delta^{\mu\nu}\mathcal{W}^{*(2)}_\mu T^{(2j-2)}_\nu+\cdots+\delta^{\mu\nu}\mathcal{W}^{*(2j-2)}_\mu T^{(2)}_\nu+\delta^{\mu\nu}\mathcal{W}^{*(2j)}_\mu T^{(0)}_\nu\non\\[.3cm]
b_{(2\Delta-d)}&=& \f{\Delta-1}{2L}\delta^{\mu\nu}\mathcal{W}^{*(0)}_\mu \mathcal{V}^{(2\Delta-d)}_\nu+ \f{d-\Delta-1}{2L}\delta^{\mu\nu}\mathcal{V}^{*(2\Delta-d)}_\mu \mathcal{W}^{(0)}_\nu
\ee
with $T^{(2j)}_\nu$ 
given by 
\be
T^{(2j)}_\nu&=& \f{d-\Delta-1+2j}{2L}\mathcal{W}_\nu^{(2j)} +\f{L}{2(2j-\Delta-1)} \delta^{\sigma\tau}\p_\nu\p_\sigma\mathcal{W}_\tau^{(2j-2)}\;;\quad 0\le j <\Delta-\f{d}{2}\, . 
\ee
We can express the regularised action in \eqref{regact56yht} in a covariant form. For this, we need to invert the asymptotic expansion of $\mathcal{W}_M$ to express $\mathcal{W}^{(0)}_M$ in terms of $\mathcal{W}_M$. Up to $O(\rho^{j})$, this is given by
\be
\mathcal{W}_\mu^{(0)}(x)&=& \left(\f{\rho}{L}\right)^{\f{\Delta-d+1}{2}}\sum_{q=0}^{j} \Biggl[b_q (L^2\Box_\gamma)^qW_\mu(\rho,x) + d_q L^2(L^2\Box_\gamma)^{q-1}\gamma^{\sigma\nu}\p_\mu\p_\sigma W_\nu(\rho,x)\Biggl]\, ,\non\\[.3cm]
\mathcal{W}_\rho^{(0)}(x)&=& \left(\f{\rho}{L}\right)^{\f{\Delta-d+1}{2}}\sum_{q=0}^{j} b_q (L^2\Box_\gamma)^qW_\rho(\rho,x)\, , \label{f202rt}
\ee
where for completeness we have also given the expression of $\mathcal{W}^{(0)}_\rho$ in terms of $W_\rho$. The notation $\Box_\gamma$ denotes $\Box_\gamma=\gamma^{\mu\nu}\p_\mu\p_\nu$ and the coefficients $b_q$ and $d_q$ are determined recursively by
\be
b_q&=& -\sum_{m+n=q\atop m\ge 1;n\ge 0} B_mb_n\;\;;\qquad b_0=1\, ,\non\\
d_q&=& -\sum_{m+n=q\atop m\ge 1;n\ge 0} \left[\left(1+\f{2m}{1-\Delta}\right)B_md_n+\f{2m}{1-\Delta}B_mb_n\right]\;\;;\qquad d_0=0\, ,\label{F.229}
\ee
with $B_m$ defined in equation \eqref{Bmwe}. Here, $m,n$ are integer numbers less than $\frac{(2\Delta -d)}{2}$.

Using \eqref{f202rt}, we can express the divergent terms in the regularised action in covariant form. 
The role of counterterms is to get rid of the divergent terms in the above expression when we take the limit $\epsilon\rightarrow0$. Hence, the counterterms are simply given by the negative of the divergent terms in the regularised action,
\be
S_{\rm ct}&=& 2\int d^dx\sqrt{\gamma}\biggl[\f{d-\Delta-1}{2L}\gamma^{\mu\nu}W^*_\mu W_\nu \nonumber \\
&& \qquad +\f{L}{2(2\Delta-d-2)}  W^*_\mu \left(
\gamma^{\mu\nu} \Box_\gamma 
- \f{(2\Delta-d)}{(\Delta-1)} \gamma^{\mu \sigma} \gamma^{\tau \nu} \p_\sigma \p_\tau
\right)W_\nu +\cdots\biggl]\, ,\label{F.218}
\ee
and the renormalised action is then the sum of regularised and counterterm actions, namely
\be
S_{\rm ren}=\lim_{\epsilon \to 0} \left(S_{\rm reg} +S_{\rm ct}\right)\, .
\ee
The above counterterm agrees with the first counter term obtained in equation \eqref{G.252} for the gauge field when we substitute $\Delta = d-1$
(and in particular it is manifestly gauge invariant).

\subsection{Two-point function}

To derive the correctly normalised 2-point function of the operators dual to the Proca field, we first note that the one point function of the boundary operator $\mathcal{O}^\mu$ which is dual to $W^*_\mu$ is given by
\be
\langle \mathcal{O}^{\mu}(x) \rangle &=&
\f{\delta S_{\rm ren}}{\delta \mathcal{W}^{(0)*}_{\mu}(x)}\;=\; \lim_{\epsilon\rightarrow 0} \f{1}{\epsilon^{\f{\Delta+1}{2}}}\f{1}{\sqrt{\gamma}} \f{\delta S_{\rm ren}}{\delta W^*_\mu(x,\epsilon)}\label{def6ty}
\ee
Using the renormalised action obtained in the previous subsection, we find
\be
\langle \mathcal{O}^{\mu}(x) \rangle  &=&  -\delta^{\mu\nu}\f{\bigl(2\Delta-d\bigl)}{L} \mathcal{W}^{(2\Delta-d)}_{\nu}(x)\label{F.221}
\ee
The 2-point function can be computed by differentiating the above expression with respect to the source $\mathcal{W}^{(0)}_\mu$. Using the expression of $\mathcal{W}_\nu^{(2\Delta-d)}$ given in \eqref{wnutyui}, we find at $O(g^0)$ in coupling\footnote{Note that to compute the correlators in momentum space we need to multiply by the factor of $(2\pi)^d$. 
To see this, we note that the expression of the generating functional is given by
\be
Z[J] &=&\biggl\langle  \exp\Bigl(-\int d^dx J(x)\phi(x)\Bigl)\biggl\rangle\;\; =\;\;\biggl\langle  \exp\Bigl(-\int \f{d^dp}{(2\pi)^d} J(-p)\phi(p)\Bigl)\biggl\rangle
\ee
Thus, in momentum space we have
\be
\bigl\langle \phi(p_1)\cdots \phi(p_n) \bigl\rangle_c \; =\; (-1)^n \f{(2\pi)^d\delta}{\delta J(-p_1)}\cdots  \f{(2\pi)^d\delta}{\delta J(-p_n)}W[J]\Bigl|_{J=0}
\ee
where the subscript $c$ denotes the connected part of the correlator and we have used the definition $W[J]= \ln Z[J]$. In writing \eqref{def6ty}, we have used the relation $Z\simeq e^{-S_{\rm ren}}$ which gives $W\simeq -S_{\rm ren}$ for the boundary CFT correlators.  
}
\be
\langle \mathcal{O}^{*\nu}({\bf p})\mathcal{O}^{\mu}({\bf k}) \rangle  \;=\; - (2\pi)^d\f{\delta \langle\mathcal{ O}^{\mu}({\bf k}) \rangle }{\delta \mathcal{W}^{(0)}_\nu(-{\bf p})}\;=\;  (2\pi)^d\delta^d({\bf p}+{\bf k})a_0\Bigl(\delta^{\mu\nu}+\f{ {\bf k}^\nu {\bf k}^\mu(d-2\Delta)}{k^2(\Delta-1)}\Bigl)k^{2\Delta-d}
\label{2pty786}
\ee
where 
\be
a_0\; \equiv\; \bigl(2\Delta-d\bigl) \left(\f{1}{2}\right)^{2\Delta-d}   \f{\Gamma\Bigl(\f{d}{2}-\Delta\Bigl)\; }{\Gamma\Bigl(\Delta-\f{d}{2}\Bigl)} \;L^{2\Delta-d-1}
\ee
and we have used the functional identity 
\be
\f{\delta \mathcal{W}^{(0)}_\mu({\bf p})}{\delta \mathcal{W}^{(0)}_\nu({\bf q})}&=& \delta^\nu_\mu\;\delta^d({\bf p}-{\bf q})
\ee

\section{Holographic renormalization of gauge field}
\label{s3}
In this appendix  we discuss how to obtain from the gravity dual the CFT three-point correlator involving two (non-conserved) spin-one operators and one conserved current using the holographic renormalization procedure. The UV divergences of the boundary conformal theory (which arise when two operators approach each other), manifest themselves as long range IR divergences in the bulk gravity theory when we approach the boundary. The standard procedure to deal with these divergences in the bulk theory is to first obtain the near boundary solution of the bulk equations of motion and then regularise the action by introducing a radial cutoff $\epsilon$ \cite{Henningson:1998gx}. The action diverges as we remove the boundary cut-off, and these divergences may be cancelled by adding boundary covariant counterterms. The full renormalized on-shell action is the sum of the regularized action plus the counterterms in the limit $\epsilon\rightarrow 0$
\begin{eqnarray}
S_{\rm ren}=\lim_{\epsilon\rightarrow 0}\big[ S_{\rm reg}+S_{\rm ct}\big]
\end{eqnarray}
The connected correlators can now be computed by taking the functional derivative of the renormalised action with respect to the bulk sources. Below, we shall describe this procedure in detail for the case of interest. For further information on holographic renormalization, see \cite{deHaro:2000vlm, Bianchi:2001de, 0112119, Papadimitriou:2004ap,Papadimitriou:2004rz, Papadimitriou:2010as, Papadimitriou:2016yit}.

 \subsection{Asymptotic Solution}
 The first step in obtaining the renormalised correlators in AdS is to obtain the asymptotic solution of the equations of motion. We are interested in computing the 3-point function that 
only involves a single insertion of the symmetry current and two insertions of the non-conserved vector operator.
For the purpose of computing this 3-point function, the 1-point function of ${\cal J}^\mu$ in the presence of sources, as defined in equation \eqref{pi314}, may be computed from the bulk action truncated to quadratic order in the number of bulk gauge fields. The non-conserved operators are irrelevant operators, and correlators of the irrelevant operators have complicated UV structure. To avoid this problem we will only consider the irrelevant operators inserted at non-coincident points. This can be achieved by working with infinitesimal sources which have support at separated points \cite{9802150, deHaro:2000vlm}. This then implies that 
most of the contributions of  the cubic interaction terms to the gauge field equation vanish because they are quadratic in the source part of  the massive vector field, which vanish when the sources have support on non-coincident points. 

Thus, for the purpose of this analysis, it is sufficient to use the equation of motion,
 \begin{eqnarray}
\frac{1}{\sqrt{G}}\partial_M\left( \sqrt{G} F^{MN}\right)\;=\; \left(\Box +\frac{d}{L^2}\right)  A^N- \nabla^N\nabla_MA^M\;=\;J^N\, ,
\end{eqnarray}
where $J^N$ is constructed from the massive vector fields and is given in \eqref{B.26aa}.
In axial gauge $A_0=0$ and in Fefferman Graham coordinates (defined in equation \eqref{B.26a}) the above equation gives
\begin{eqnarray}
-\rho L\delta^{\mu\nu}\partial_\rho\partial_\mu A_\nu= L^2J_\rho
\;;\quad
4\rho^2 \partial_\rho^2 A_\mu+2(4-d) \rho \partial_\rho A_\mu+ \rho L\Box_0 A_\mu-\rho\,L\delta^{\nu\sigma}\partial_\mu \partial_\nu A_\sigma= L^2J_\mu \, .
\label{7.29}
\end{eqnarray}
where $\Box_0\equiv\delta^{\mu\nu}\partial_\mu\partial_\nu$.  

We need to obtain the asymptotic solution of the above equation without splitting the gauge field in the transverse and longitudinal components. This is due to the fact that the projection operators (which project onto these components) are non-local whereas locality is essential for renormalisation. To obtain the asymptotic solution, the general strategy is to solve the equations order by order in the radial direction $\rho$. Setting $\rho=0$, we find that the equations are satisfied. Next, we take the derivative of these equations with respect to $\rho$ and then set $\rho=0$. For the next order, we take the second derivative of the equations with respect to $\rho$ and then set $\rho=0$ and so on. We need to treat the case of even and odd dimensions separately. 
\subsubsection*{Even Dimensions}

Following the procedure described above and solving the resulting equations, we find that the asymptotic solution for gauge field for even $d$ has the following structure
\begin{eqnarray}
A_\mu(\rho,x)\;=\; \sum_{j=0}^{\frac{d}{2}-2} A^{(2j)}_\mu(x) \frac{\rho^j}{L^{j}}\;\,+\;\,   \frac{\rho^{\frac{d}{2}-1}}{L^{\frac{d}{2}-1}}\left(A_\mu^{(d-2)}(x)+B^{(d-2)}_\mu(x)\,\log\frac{\rho}{L}\right) +\dots  \label{5.11}
\end{eqnarray} 
 where the dots denote higher powers in $\rho$  which are irrelevant in the forthcoming discussion. 
  
 The need to introduce the $\log$ term at $O(\rho^{\f{d}{2}-1})$ is due to the fact that the equations of motion at this order develop a pole and the resulting equations cannot be satisfied without the $\log$ term. The equations of motion give the following solutions
\begin{eqnarray}
A_\mu^{(2j)}&=&  
\frac{L^2 }{4 j\Big(\frac{d}{2} -1 -j\Big)}
\left(\Box_{0} \delta^\mu_\nu - \p_\mu \p^\nu\right) A_\nu^{(2j-2)}
\quad,\qquad 1 \le j < \frac{d}{2}-2\label{A2j}\\[.3cm]
B_\mu^{(d-2)}&=&
\frac{L^2}{2(2-d)}
\left(\Box_{0} \delta^\mu_\nu - \p_\mu \p^\nu\right) A_\nu^{(d-4)}\, ,
\label{jhyt675}
\end{eqnarray}
where indices are raised with $\delta^{\mu \nu}$.
It follows from these equations that $\partial^\mu\,A_\mu^{(2j)}=\partial^\mu\,B_\mu^{(d-2)} =0$. In addition
the field equations determine the divergence of $A_\mu^{(d-2)}$,
\begin{eqnarray}
\p^\mu A_\mu^{(d-2)} &=& \frac{2ig}{(d-2)} \left(\Delta -\frac{d}{2}\right)
\delta^{\mu\nu}\left(\mathcal{W}_\mu^{*(0)} \mathcal{W}_\nu^{(2\Delta-d)}-
{\rm c.c.}
\right)\, , \label{fgtyho}
\end{eqnarray}
where c.c. stands for complex conjugate.
Equation \eqref{fgtyho} comes from solving the first equation in \eqref{7.29} near the boundary, and the r.h.s. comes from $J_\rho$.
Note that it involves a source times a vev piece: all other terms that are quadratic in the sources are automatically zero since by assumption the sources have disjoint support.  The r.h.s. of \eqref{fgtyho} is crucial for deriving the correct conservation Ward identity from the bulk. 
The solution shows that all the coefficients except the transverse part of  $A_\mu^{(d-2)}$ are locally determined in terms of $A_\mu^{(0)}$. 

In deriving \eqref{fgtyho}, we have assumed an arbitrary value of $\Delta$, which is appropriate when the mass of the bulk Proca fields is taken to be arbitrary. For special values of $\Delta$ the solution is modified:
if $\Delta=d/2+n$, for some integer $n$, then the r.h.s. of \eqref{fgtyho} is modified as follows
\begin{equation}
\p^\mu A_\mu^{(d-2)}
= \frac{2ig}{(d-2)} 
\delta^{\mu\nu}\Bigl( \mathcal{W}_\mu^{*(0)} \left(\left(\Delta -\frac{d}{2}\right) \mathcal{W}_\nu^{(2\Delta-d)} +\mathcal{V}_\nu^{(2\Delta-d)}\right)- {\rm c.c.} \Bigl)\non
\end{equation}
However, we shall not make use of this since we work with arbitrary $\Delta$ in this paper.

The solutions in \eqref{jhyt675} can also be written using the field strength for $A_\mu^{(0)}$, which makes gauge invariance manifest,
 \begin{eqnarray}
 A_\mu^{(2j)}&=&   \frac{L^{2j}\;\Box_0^{j-1}\partial^\nu F_{\nu\mu}^{(0)}   }{2^{2j} \Gamma[j+1]\, \prod_{n=1}^j\Big(\frac{d-2}{2}  -n\Big)};\qquad 1 \leq j\leq \frac{d}{2}-2\label{a2j_gi} \\[.2cm]
 B_\mu^{(d-2)}&=&-\frac{2^{2(1-\frac{d}{2})}\,L^{d-2}\;\Box_0^{\frac{d}{2}-2} \partial^\nu F_{\nu\mu}^{(0)}}{\Gamma\left[\frac{d}{2}\right]\prod_{n=1}^{\frac{d}{2} -2}\left(\frac{d-2}{2} -n\right)}\, .\label{bmu9}
 \end{eqnarray}
 
 \subsubsection*{Odd Dimensions}

The procedure for the case of odd $d$ is similar to the case of even $d$ considered above. The main difference is that the expansion no longer has a logarithmic term,
\be
 A_\mu(\rho,x)\;=\; \sum_{j=0}^{\frac{d-3}{2}}   \left(\frac{\rho}{L}\right)^{j}\; A^{(2j)}_\mu(x)\;\,+\;\,   \left(\frac{\rho}{L}\right)^{\frac{d-2}{2}}A_\mu^{(d-2)}(x) +\dots  \label{5.11gft}
 \ee
The coefficients $A^{(2j)}_\mu,\ 1\le j\le (d-3)/2$, are the same as in the \eqref{A2j} (or equivalently \eqref{a2j_gi}), and only the longitudinal part of $A^{(d-2)}_\mu$ is determined by the asymptotic analysis and is given by \eqref{fgtyho}.

  \subsection{Regularization}

We regularize the action by introducing a small cut-off $L\,\epsilon$ on the radial coordinate $\rho$ close to the boundary. The action will be  evaluated on this regularized action on the asymptotic solution. The cubic terms do not contribute to the divergences when the sources of the massive vector fields have disjoint support, so it suffices to consider the
regularised gauge kinetic term only, 
\begin{eqnarray}
S_{\rm reg}\;=\;\frac{1}{4} \int_{\rho\geq L \epsilon}d^{d+1} x\,\sqrt{G}\;   F_{MN}F^{MN}\;=\;\frac{1}{2}\int_{\rho=L\epsilon} d^d x\sqrt{\gamma}\;n_M A_N F^{MN}\, . 
\end{eqnarray}
In going to the second equality, we have used integration by parts and used the gauge field equation of motion at leading order in the coupling constant to set the bulk term to zero. The $\gamma_{\mu\nu}= \frac{1}{\epsilon} \delta_{\mu\nu}$ and $n_M$ represent the induced metric and the spacelike normal vector on the boundary hypersurface $\rho=L\epsilon$. Evaluating the above regularised action on the asymptotic solution, we find for even $d$
\begin{eqnarray}
S_{\rm reg}\;=\;-\f{1}{L}\int_{\rho=L\epsilon} d^dx\,\delta^{\mu\nu} \Bigg[ \sum_{k=0}^{\frac{d}{2}-2-j} \sum_{j=0}^{\frac{d}{2}-2} k\,\epsilon^{j+k+1-\frac{d}{2}} A_\mu^{(2j)}A_\nu^{(2k)}+\left(\frac{d}{2}-1\right)\log\epsilon\; A_\mu^{(0)} B_\nu^{(d-2)}+\dots\Bigg]\nonumber\\\label{G.250}
\end{eqnarray}
The $\cdots$ terms are non singular and hence are irrelevant for our discussion. 
For odd $d$, we have the same expression but the factors $(d-2)/2$ in the upper limit of the summations over  $j$ and $k$ are replaced by $(d-3)/2$ and there is no logarithmic term.
From the above expressions, we see that the number and the structure of the divergent terms depend on the dimensionality of the space-time. In $d=3$, {\it e.g.}, there are no divergences. In $d=4$, there is only a logarithmic divergence and so on. 

 \subsection{Counterterms}
We need to add counterterms to cancel the divergences,
\be
S_{\rm ct} &=& -\mbox{divergent terms of } S_{\rm reg}
\ee
Equation \eqref{G.250} shows  the relationship between the number of counterterms and the dimension $d$ of the AdS boundary. These counterterms  are obtained by expressing
 the divergent terms appearing in $S_{\rm reg}$ on the right hand side in terms of  the induced metric $\gamma_{\mu\nu}$ defined  on the regularized surface and express the coefficients $A_\mu^{(2j)}$ {\it etc.} in terms of the bulk field $A_\mu(\rho,x)$ by inverting the series in \eqref{5.11} and \eqref{5.11gft}. For even $d$ up to $d=10$, this procedure yields
\begin{eqnarray}
S_{\rm ct}&&= -\frac{L}{2}\,\int_{\rho=L\epsilon} d^{d}x\sqrt{\gamma}\,\gamma^{\mu\nu}\,\,\gamma^{\alpha\sigma}\, F_{\sigma \mu}\Bigg[C_1 + (2C_2-C_1^2)  \,(L^2\Box_\gamma) \nonumber\\
&&+\left(C_1(2\tilde{b}_2+\tilde{b}_1^2)+4 C_2\, \tilde{b}_1 + 3 C_3+ 2 C_1^2\,\tilde{b}_1+3 C_1\,C_2 \right) (L^2\Box_\gamma)^2+\left( \frac{d}{2} -1\right)\,C_{\frac{d}{2}-1}\,L^{d-4}\log\epsilon\, \Box_\gamma^{\frac{d}{2}-2}\Bigg]F_{\alpha\nu}\nonumber\\
&&\label{G.252}
\end{eqnarray}
where the first term appears at $d=6$ dimensions ($d=5$ for odd dimension). The second term is necessary from $d=8$ ( respectively $d=7$) dimensions, and the third one is present from $d=10$ dimensions ($d=9$ in odd-dimensions). 
Here, the coefficients are defined as:
\begin{eqnarray}
C_{\frac{d}{2}-1}= -\frac{ 2^{2-d}}{\Gamma\big[\frac{d}{2}\big]\prod_{n=1}^{\frac{d}{2} -2}\left( \frac{d-2}{2}-n\right)}
\quad;\quad C_j=\frac{ 2^{-2j}}{\Gamma\big[j+1\big]\prod_{n=1}^j\left( \frac{d-2}{2} -n\right)}\quad,\quad 1\le j \le \f{d}{2}-2:
\end{eqnarray}
Additionally, we  introduce  the inversion coefficients for the gauge field as given in \eqref{F.229} for Proca's field:
\begin{eqnarray}
\tilde{b}_q= -\sum_{\begin{array}{c}m+n=q\\ m\geq 1;\,n\geq 0\end{array}}C_m\,\tilde{b}_n~~;~~\tilde{b}_0=1
\end{eqnarray}

The explicit expression for all quantities introduced in Eq.\eqref{G.252} are:
\begin{eqnarray}
C_1= \frac{1}{2(D-5)}~~;~~2C_2-C_1^2=\frac{1}{2(D-5)^2(D-7)}~~;~~D=d+1\nonumber\\
C_1(2\tilde{b}_2+\tilde{b}_1^2)+4 C_2\, \tilde{b}_1 + 3 C_3+ 2 C_1^2\,\tilde{b}_1+3 C_1\,C_2=\frac{1}{(D-5)^3(D-7)(D-9)}
\end{eqnarray}
These expressions agree with those provided in \cite{2308.00476}\footnote{The logarithmic counterterm in equation \eqref{G.252} is in agreement with the corresponding one given in v3 of Ref. \cite{2308.00476}, where several misprints were addressed through email correspondence. We are also in agreement with \cite{Taylor:2000xw}, which derived the logarithmic term for $d=4$.}. 

For odd $d$,  the structure of the counterterms remains the same  but without the logarithmic contribution.

\subsection{Renormalised Correlators }
\label{renormalised}
After computing the regularised action, we now have all the ingredients to write down the expression of 1-point function of the gauge field. For this, we define the renormalised on-shell action as
\be
S_{\rm ren} = S_{\rm reg} + S_{\rm ct}
\ee
The exact renormalized 1-point function is obtained  by considering functional derivatives of the renormalized action with respect to the bulk sources and then removing the IR cutoff. More precisely, we have 
\begin{eqnarray}
\langle \mathcal{J}^\mu(x)\rangle=\frac{\delta S_{\rm ren}}{\delta A_{(0)\mu}(x)}=
\lim_{\epsilon \rightarrow 0}\frac{1}{\epsilon^{\frac{d}{2}}\sqrt{\gamma}}\frac{\delta (S_{\rm reg}+S_{\rm ct})}{\delta A_\mu(\epsilon ,x)} 
\end{eqnarray}
By construction, this limit is finite. Due to the counterterms, all the divergent terms cancel and we are left with the finite non-vanishing result in the limit $\epsilon\rightarrow 0$. Using the expressions of $S_{\rm reg}$ and $S_{\rm ct}$ given in previous sections, we find
\begin{eqnarray}
\langle \mathcal{J}^\mu(x)\rangle= -\frac{2}{L}\,
\delta^{\mu\nu}\left[ \left(\frac{d}{2} -1\right) A_\nu^{(d -2)}+B_\nu^{(d-2)} \right]\label{8.7}
\end{eqnarray}
for $d$ even and 
\begin{eqnarray}
\hspace*{-2cm}\langle \mathcal{J}^\mu(x)\rangle= -\frac{2}{L}\,
\,\delta^{\mu\nu}\ \left(\frac{d}{2} -1\right) A_\nu^{(d -2)}\label{8.8}
\end{eqnarray}
for $d$ odd. 

The coefficient $B_\mu^{(d-2)}$ present in \eqref{8.7} was determined in terms of $A_\mu^{(0)}$ (see equation \eqref{bmu9}). It turns out that this term contributes only a contact term (which is related linked with a conformal anomaly \cite{0209067}). Ignoring this term, we see that the exact 1-point function of the gauge field has same expression in both even as well as odd $d$. Further, the 1-point function is given in terms of the coefficient $A_\mu^{(d-2)}$ which was undetermined from the asymptotic analysis. We can determine this coefficient by solving the field equations perturbatively. Up to the first order in the gauge coupling constant $g$ (which is needed for the three-point correlation function), we have 
\begin{eqnarray}
A_\mu(z,x)=  \int d^d y\sqrt{G}\, {\mathbb K}_\mu^{~\,\nu}(z, x,y)A_\nu(x)+ \,\int d^{d} y\,dw \sqrt{G}\, {\cal G}_{\mu\nu}(z,w;x,y)J^\nu(y,w)\label{amu98}
\end{eqnarray}
In the above equation, $\mathbb{ K}_{\mu\nu}$ and  ${\cal G}_{\mu\nu}$ are the bulk- to-boundary and the bulk-to-bulk propagators of the gauge field given in equations  \eqref{C.70} and \eqref{C.85}, respectively,  in the Poincar\'e coordinates where $z=\sqrt{L\,\rho}$. The source $J^\mu$ at $\mathcal{O}(g)$ is defined in equation \eqref{B.26aa}.

To determine $A_\mu^{(d-2)}$, we note that it is the coefficient of $\rho^{\f{d}{2}-1}$ in the asymptotic expansion. The contribution of the first term in \eqref{amu98} to $\rho^{\f{d}{2}-1}$ can be obtained by expanding the Bessel function in the definition of $\mathbb{ K}_{\mu\nu}$. However, this term does not contribute to the 3-point function since it is independent of the massive spin-1 fields. To obtain the contribution of the second term of \eqref{amu98} to $\rho^{\f{d}{2}-1}$, we note that the relation between Poincar\'e and Fefferman Graham coordinates implies $\rho^{\frac{d}{2}-1}\equiv \,\frac{z^{d-2}}{L^{\frac{d}{2}-1}}$. Using this in \eqref{C.86}, we see that the near boundary expansion of ${\cal G}_{\mu\nu}$ precisely gives the correct power of $\rho$ to contribute to $A_\mu^{(d-2)}$ up to $O(g)$. Thus, we have 
\be
\langle \mathcal{J}^\mu(x)\rangle= 
\,\delta^{\mu\tau}\int d^{d} y\,dw \sqrt{G}\,{\mathbb K}_{\tau}^{\;\;\nu}(w;x,y)J_\nu(y,w)  \label{8.k8}
\ee
This expression is valid for both even as well as odd $d$. The desired 3-point function can now be obtained by differentiating the above expression with respect to sources for the massive fields, {\it i.e.}, 
\begin{eqnarray}
\langle {\cal O}^{*\nu}(x_1)\,{\cal J}^\mu(x_2)\, {\cal O}^\sigma(x_3)\rangle&=& \,\frac{\delta^2 \langle \mathcal{J}^\mu(x_2)\rangle}{\delta \mathcal{W}^{(0)}_\nu(x_1)\,\delta \mathcal{W}^{*(0)}_\sigma(x_3)}\non\\
&=&\;\delta^{\mu\tau}
\int d^{d}y\,dw\sqrt{G}  \, \,{\mathbb K}_\tau^{~\lambda}(w;x_2,y)\, \frac{\delta^2 J_\lambda(y)}{\delta { \mathcal{W}^{(0)}_\nu(x_1)\,\delta \mathcal{W}^{*(0)}_\sigma(x_3)}}
\label{frtyuir}\end{eqnarray}
In section \ref{s3a}, we work in Fourier space and the above expression \eqref{frtyuir} in Fourier space yields \eqref{4.42nju}.

\section{Expected 3-point amplitude in flat space}
\label{exact}
In this appendix, we summarize the computation of the 3-point function in Minkowski space with mostly minus metric, involving a photon  $\gamma$ and the massive charged spin one field $W$ in $d+1$ dimensional flat spacetime. Before discussing the computation of the 3-point scattering amplitude we comment on the kinematics of the process $W \to \gamma + W$. Energy and momentum conservation yields,
\begin{align}
    \sqrt{m^2 + k_1^2} &= k_2 + \sqrt{m^2+k_3^2} \\
    \vec{k}_1 &= -\vec{k}_2 - \vec{k}_3
\end{align}
where $\vec{k}_i$ are the spatial momenta and $k_i=\sqrt{\vec{k}_i\cdot \vec{k}_i}$. A short computation shows that for generic momenta these above equations imply,
\begin{equation}
    \cos \theta = \sqrt{1+ \frac{m^2}{k_3^2}},
\end{equation}
where $\cos \theta=\vec{k}_2\cdot \vec{k}_3/(k_2 k_3)$, which cannot be satisfied with real momenta, unless $k_3 \to \infty$, or $k_2 \to 0$.
The $k_2=0$ case may be thought as a special case of the Breit or brick-wall frame (sometimes also called infinite momentum frame). In this frame (that we also discuss in the next appendix) the spatial momentum of the massive vector is reversed after the scattering and thus $\vec{k}_2=0$. For a real photon this then implies $k_2=0$. The special kinematics is a peculiarity of 3-point functions, and if one wants to work with generic momenta one may either work with complex momenta or consider the mass of the incoming particle to be different than that of the out-going. In this case one would get the same tree-level scattering amplitude \eqref{F.233} by introducing two massive vectors, $W_a$ and $W'_a$ in \eqref{act34w} below, with masses $m^2$ and $m'{}^2$, respectively,  and change the interaction terms by replacing $W_a^*$ by $W_a'{}^*$.
In the remainder of this appendix, we will assume that one may use either options 
to ensure that the 3-point function is kinematically allowed for generic momenta, but we will not explicitly implement the one or the other option, to keep the discussion similar to that of higher-point scattering amplitudes.

The action describing these fields in flat space is given by 
\be
S=\int d^{d+1}x\biggl[-\f{1}{4}F^{ab}F_{ab}-\f{1}{2}W^{*ab}W_{ab}+m^2W^{*a}W_a+i \hat {g}\hat\alpha\; F_{ab}W^{*a}W^{b}+F_{ab}T^{ab}\biggl]
\label{act34w}
\ee
where $F_{ab} = \p_a A_b -\p_b A_a$ and
\be
W_{ab} = D_a W_b -D_b W_a\quad;\quad D_a W_b =\p_a W_b +ig A_a W_b
\ee
The cubic interaction terms can again be found by following the same procedure as described in the appendix \ref{appen:D}. As discussed there, the general form of the last term can be written as
\be
F_{ab}T^{ab}&=&F^{ab}\Bigl(c_0\p_{a} W^*_{c} \p^{c} W_{b} +c_0^*\p_{a} W_{c} \p^{c} W^*_{b} \Bigl)\label{dfrty11rt}
\ee
However, as we shall discuss below, the real part of $c_0$ does not contribute to the amplitude. Hence, we can write the above expression as 
\be
F_{ab}T^{ab}=i \hat g\hat\beta F^{ab}\bigl(\p_{a} W^*_{c} \p^{c} W_{b} -\p_{a} W_{c} \p^{c} W^*_{b}\bigl)\label{dfrty1}
\ee
To proceed further, we denote the momenta and polarisation vector of $W^*_a$ by $(k_1, \varepsilon_{1a})$, those of the gauge field by $(k_2, \varepsilon_{2a})$ and those of $W_a$ by $(k_3, \varepsilon_{3a})$. The equation of motion of the massive fields imply the transversality condition $\p_aW^a =0+ O(\hat g)$. Using this and the transversality of the gauge field, we find
\be
\varepsilon_1\cdot k_1 =O(\hat g)\quad;\qquad \varepsilon_2\cdot k_2=0 \quad;\qquad \varepsilon_3\cdot k_3 =O(\hat g) \label{ew34}
\ee
where the inner products are computed using the flat space metric. 

Taking all the momenta to be ingoing in the cubic vertex, the momentum conservation condition $k_1^a+k_2^a+k_3^a=0$ gives
\be
k_1\cdot k_3 = m^2\quad;\qquad
k_1\cdot k_2=  0\quad;\qquad
k_2\cdot k_3= 0\label{ew35}
\ee
Next, we consider the Feynman rules. We shall only need the expression of the momentum space cubic vertex describing the interaction between the gauge field and the massive charged spin one field. This is given by 
\be
V^{abc}(k_1,k_2,k_3) &=& - \hat g\biggl[\eta^{ba}\eta^{cd}k_{1d}- \eta^{bd}\eta^{ca}k_{1d}- \eta^{bc}\eta^{ad}k_{3d}+ \eta^{bd}\eta^{ca}k_{3d} + \hat\alpha  \Bigl(\eta^{ad}\eta^{bc}k_{2d}-  \eta^{ab}\eta^{dc}k_{2d} \Bigl)\non\\
&& -\hat\beta\; k_{2d}\; k_{1e}\;k_{3f}\Bigl(\eta^{af}(\eta^{ed}\eta^{bc}-\eta^{eb}\eta^{dc})-\eta^{ce}(\eta^{fd}\eta^{ba}-\eta^{fb}\eta^{ad})\Bigl) \biggl]
\ee
Using \eqref{ew34} and \eqref{ew35}, the desired 3 point function is obtained to be 
\be
{A}_3
&=& \varepsilon_{1\,a}(k_1)\varepsilon_{2\,b}(k_2) \varepsilon_{3\,c} (k_3) V^{abc}(k_1,k_2,k_3)\non\\[.3cm]
&=& \hat g\biggl[   2(\varepsilon_1 \cdot \varepsilon_3)(\varepsilon_2\cdot k_1)+(1+\hat\alpha)(\varepsilon_2\cdot \varepsilon_1)(\varepsilon_3\cdot k_2)-(1+\hat{\alpha} ) (\varepsilon_2\cdot \varepsilon_3)(\varepsilon_1\cdot k_2)\non\\[.1cm]
&&+2\hat\beta(\varepsilon_2\cdot k_1)(\varepsilon_3\cdot k_2)(\varepsilon_1\cdot k_2)
\biggl].\label{F.233}
\ee
Gauge invariance implies that the amplitude vanishes when $\varepsilon_2$ is replaced by $k_2$.

We shall now show that the real part of $c_0$ in \eqref{dfrty11rt} does not contribute to the amplitude. For this, we just focus on the terms containing $c_0$ and its complex conjugate in the action. In momentum space, this is given by
\begin{eqnarray}
I= \int \prod_{i=1}^3\frac{d^{d+1}p_i}{(2\pi)^{d+1}}\,W^{*}_a(p_1)\,A_b(p_2)\, W_c(p_3) A_3^{abc}(p_1,p_2,p_3)
 \end{eqnarray}
with
\begin{eqnarray}
A_3^{abc}(p_1,p_2,p_3)=-i \left[c_0\eta^{bc}\,p_3^a\,(p_1\cdot p_2) +c_0^*\, p_1^c\, (p_2\cdot p_3)\, \eta^{ab} -c_0p_2^c\, p_1^b\,p_3^a-c_0^* p_2^a\,p_3^b\,p_1^c\right] 
\end{eqnarray}
The above expression implies 
\begin{eqnarray}
A_3^{abc}(p_1,p_2,p_3)+ A_3^{cba}(p_3,p_2,p_1)= -i (c_0+c_0^*)\left[\eta^{bc} p_3^a(p_2
\cdot p_1) +p_1^c(p_3\cdot p_2) \eta^{ab}-p_2^cp_1^bp_3^a-p_2^ap_3^bp_1^c\right]\nonumber\\
\label{dftr56r}
\end{eqnarray}
Now, the on shell momenta and polarizations satisfy
\begin{eqnarray}
p_2\cdot p_3=0~~;~~p_1\cdot p_2=0~~;~~\varepsilon_a\cdot p_a=0~~,~~a=1,2,3
\end{eqnarray}
The condition on momenta imply that the first two terms in RHS of \eqref{dftr56r} vanish. Next, noting that the amplitude is given by dressing $A_3^{abc}$ with the external polarization vectors, we contract \eqref{dftr56r} with polarization vectors and use the transversality condition on polarization vectors to obtain
\begin{eqnarray}
&&\hspace*{-1.3cm}\varepsilon_{1a}\,\varepsilon_{2b}\,\varepsilon_{3c}\left[A_3^{abc}(p_1,p_2,p_3)+ A_3^{cba}(p_3,p_2,p_1)\right]\non\\
&=& i (c_0+c_0^*)\left[(\varepsilon_1\cdot p_3)(\varepsilon_2\cdot p_1)(\varepsilon_3\,p_2)+(\varepsilon_1\cdot p_2)(\varepsilon_2\,p_3)(\varepsilon_3\cdot p_1)\right]\nonumber\\
&=&2i \,{\mbox{ Re}}(c_0)\left[-(\varepsilon_1\cdot p_2)(\varepsilon_2\cdot p_1)(\varepsilon_3\cdot p_2)+(\varepsilon_1\cdot p_2)(\varepsilon_2\cdot p_1)(\varepsilon_3\cdot p_2)\right]\non\\
&=&0
\end{eqnarray}
Thus, we have 
\begin{eqnarray}
\varepsilon_{1a}\,\varepsilon_{2b}\,\varepsilon_{3c}A_3^{abc}(p_1,p_2,p_3)=-\varepsilon_{1a}\,\varepsilon_{2b}\,\varepsilon_{3c}   A_3^{cba}(p_3,p_2,p_1) 
\label{11.314}
\end{eqnarray}
From this, it is clear that the real part of the coefficient $c_0$ does not appear in the flat space amplitude. Further, it also shows the antisymmetry of the on shell amplitude under the exchange $(\varepsilon_1,\,p_1)\leftrightarrow (\varepsilon_3,\,p_3)$. This is also the property of the CFT 3-point function reviewed in section \ref{sec2review3er}. Hence, the flat space result is consistent with this expectation.
  
  \section{Multipole moments} \label{app: multipoles}

The coupling constants (Wilson coefficients) appearing in an effective field theory involving an Abelian gauge field encode information about how the gauge field interacts with the other massive fields. These coupling constants can be related to the electromagnetic moments. Here, we summarise some results regarding this relation in 4 dimensional Minkowski space following \cite{Lorce::1979}.

We start by recalling some basic facts about electromagnetic form factors and multipole expansion of the electromagnetic currents. This is a topic with a long history, see  \cite{Schwartz_1955, Glaser_Jaksic_1957,  Durand:1962zza, Aronson:1969ltq, Kim:1973ee, Fearing, Lorce::1979} for a selection of early papers. The main object is the expectation value of the electromagnetic current in a single-particle state 
\be \label{app_Multipole: current}
J^a(x) \;\equiv\; \f{\hat g}{2m} \langle p',s|J^a(x)|p,s\rangle = e^{i q \cdot x} \f{\hat g}{2m} \langle p',s|J^a(0)|p,s\rangle
\ee
where $\hat g$ and $m$ denote the charge and mass of the massive particle and $q=p+p'$ is the momentum transfer\footnote{\label{incoming} We are taking all the momenta to be ingoing. In \cite{Lorce::1979}, the definitions of $q^\mu$ and $P^\mu$ (appearing below in (\ref{form factor})) are interchanged as compared to the definitions given above since \cite{Lorce::1979} takes the momenta of the initial state to be in-going and of the final state to be out-going.} . The state $|p,s\rangle$ is the spin-$s$ single particle state with 4-momentum $p^a$. The dependence of $J^a$ on $x$ is simple because translational invariance implies that $J^a(x)=e^{i x \cdot \hat{P}} J^a(0) e^{-i x \cdot \hat{P}}$, where $\hat{P}$ is the 4-momentum operator, and we take the expectation value between momentum eigenstates. 

We now consider (\ref{app_Multipole: current}) in the Breit (or brick-wall) frame. In this frame, there is no energy transfer from the photon to the system, {\it i.e.} $q =(0, \bf{q})$ and $J^a(0, \bf{r})=(\rho(\bf{r}), \bf{J}({\bf r})) $ is static. The electric and magnetic multipoles can be obtained from the moments of the electric density, $\rho(\bf{r})$, and magnetic density, $\rho_M({\bf r}) = \nabla \cdot \left( \bf{J}({\bf r})  \times {\bf r} \right)$, using standard results from electrostatics and magnetostatics. 

With no loss of generality we may impose azimuthal symmetry, 
Fourier transform and expand in the spherical harmonics 
to obtain \cite{Lorce::1979}
\begin{eqnarray}
\rho(\textbf{q})=J^0(\textbf{q})&=& \hat g \sum_{l=0\atop l\; \mbox{\tiny even} }^{2s} (-\tau)^{\f{l}{2}} \sqrt{\f{4\pi}{2l+1}}\;\f{l!}{(2l-1)!!} \; G_{El}(Q^2) Y_{l0}(\Omega_q) \label{el}  \\
&=&\hat g\biggl[ G_{E0}(Q^2)-\f{2}{3}\tau G_{E2}(Q^2)\sqrt{\frac{4\pi}{5}}\,Y_{20}(\Omega_q)+\cdots   \biggl] \non\\
\rho_M(\textbf{q})=\vec{\nabla}\cdot (\textbf{J}(\textbf{q})\times\textbf{q})&=& i\,\hat g\sqrt{\tau} \sum_{l=1\atop l\; \mbox{\tiny odd} }^{2s} (l+1)(-\tau)^{\f{l-1}{2}} \sqrt{\f{4\pi}{2l+1}}\;\f{l!}{(2l-1)!!} \; G_{Ml}(Q^2) Y_{l0}(\Omega_q) \label{magn}\\
&=&2i\,\hat g\sqrt{\tau}\biggl[ G_{M1}(Q^2)\sqrt{\frac{4\pi}{3}} Y_{10}(\Omega_q)-\f{4}{5}\tau G_{M3}(Q^2)\sqrt{\frac{4\pi}{7}}Y_{30}(\Omega_q)+\cdots   \biggl]\nonumber
\end{eqnarray}
where  $G_{El}$ and $G_{Ml}$ are the electric and magnetic multipoles, $Y_{lm}(\Omega_q)$ are spherical harmonics, $\Omega_q$ denotes the solid angle associated with the vector $\textbf{q}$, $Q^2=-q^2$ denotes the momentum transfer squared
($Q^2=-{\bf q}^2$ in the Breit frame) and $\tau =\f{Q^2}{4m^2}$. As mentioned we consider a system with azimuthal symmetry and we took the symmetry axis to be the $z$-axis, so only the $m=0$ components of the spherical harmonics 
$Y_{lm}$ contribute in \eqref{el} and \eqref{magn}. The $l^{\mbox{\footnotesize th}}$ electric moment $Q_l$ and the $l^{\mbox{\footnotesize th}}$ magnetic moment $\mu_l$ are given by the $Q^2=0$ value of the multipoles, 
\be
Q_l =\f{\hat g}{m^l} \,\f{(l!)^2}{2^l} G_{El}(0);\qquad\qquad \mu_l =\f{\hat g}{2\,m^l} \,\f{(l!)^2}{2^{l-1}} G_{Ml}(0)\label{h232r}
\ee

The form of the electromagnetic current $J^a(q)$ in terms of form factors follows from Lorentz covariance, and goes back to \cite{Glaser_Jaksic_1957,  Durand:1962zza}. The analysis is similar to the determination of the form factor decomposition of the 3-point functions reviewed  in section \ref{sec2review3er}. For the case of massive spin-1 particles, this has also been discussed in the context of tri-linear gauge coupling in the standard model \cite{Gaemers:1978hg,Gounaris:1996rz}. When $d=4$ and $s$ is integer or half-integer, the combination of Lorentz covariance, gauge invariance (conservation of the current) and  parity and time-reversal symmetries imply that the current 
involves $(2 s+1)$ form factors \cite{Durand:1962zza}. The connection to CFT 3-point functions we discuss in this paper suggests that this number continues to be the same for any $d$. It would be interesting to show this explicitly.
The explicit form of the electromagnetic current for the case of integer $s$ (in   the form given in \cite{Lorce::1979}\footnote{See footnote \ref{incoming} regarding our conventions relative to that of \cite{Lorce::1979}.}) is as follows:
\be \label{form factor}
J^a_{(s)} \;=\; (-1)^s \varepsilon^*_{b_1\cdots b_s}(p')\biggl[P^a\sum_{(k,s)}F_{2k+1}(Q^2) \;+\; (g^{ac_s}q^{b_s}-g^{ab_s}q^{c_s} )\sum_{(k,s-1)}F_{2k+2}(Q^2)\biggl]\varepsilon^*_{c_1\cdots c_s}(p)
\ee
where $\epsilon_{a_1\cdots a_s}$ denote the polarisation tensor of the spin $s$ particle, $P^a = p^a-p'^a$ and $q^a=p^a+p'^a$ and
\be
\sum_{(k,s)}\;\equiv\; \sum_{k=0}^s\biggl[\prod_{i=1}^k\left(-\f{q^{b_i}q^{c_i}}{2m^2}\right)\prod_{i=k+1}^sg^{b_ic_i}\biggl]
\ee
For $s=1$, the above expressions give
\be
J^a &=&- W^*_b(p')\biggl[g^{bc}P^a F_1(Q^2) + \bigl(g^{a c}q^b-g^{a b}q^c\bigl)F_2(Q^2) -\f{q^bq^c}{2m^2}P^a F_3(Q^2)\biggl]W_c(p)\, . \label{ja1}
\ee

Comparing (\ref{el}), (\ref{magn}) and \eqref{form factor} shows that the electric and magnetic multipoles $G_{El}(Q^2) $ and $G_{Ml}(Q^2)$ are linear combinations of the form factors $F_i(Q^2)$. {\it E.g.}, for $s=1$, we have \cite{Lorce::1979}
\be \label{Gs}
G_{E0}(Q^2) -\f{2}{3}\tau\;G_{E2}(Q^2) &=& \sqrt{1+\tau} \; F_1(Q^2)\non\\
G_{E2}(Q^2) &=& \sqrt{1+\tau}\Bigl[ F_1(Q^2) -F_2(Q^2) +(1+\tau)F_3(Q^2) \Bigl]\non\\
G_{M1}(Q^2)&=& \sqrt{1+\tau} \; F_2(Q^2)
\ee

In perturbation theory, the electromagnetic form factor captures the lowest order terms in the scattering amplitude of the photon with the massive vector boson.
Now however the photon must me on-shell, $Q^2=0$. To avoid using special kinematics we may work with complex momenta (one may check that one can reach the Breit frame with complex momenta and non-trivial on-shell momentum for the photon). We can now see how the Wilson coefficients $\hat\alpha$ and $\hat \beta$ appearing in the flat space action are related to the electromagnetic moments. Stripping \eqref{F.233} of the gauge field polarisation and comparing with \eqref{ja1} we find 
\be \label{Fs}
F_1(0)= 1;
\quad\qquad  F_2(0)= 
(1+\hat\alpha);\quad\qquad F_3(0)= -2m^2\hat 
\beta\, .
\ee
For virtual photon the form factors may have $q^2$-dependence, see for example
\cite{Kim:1973ee}. Also, in general, the electromagnetic form factors for hadronic higher spin states are non-trivial functions of $Q^2$.
Using \eqref{Fs} in \eqref{Gs} we find,
\be
G_{E0}(0)\;=\;1; \qquad \qquad G_{E2}(0)\;=\; -
(\hat\alpha +2m^2\hat\beta); \qquad\qquad G_{M1}(0)= 
(1+\hat\alpha)\, ,
\ee
and then  \eqref{h232r} leads to the electric and magnetic moments,
\be
Q_0=\hat g;  \qquad\qquad  Q_2 \;=\; -\f{\hat g}{m^2} (\hat\alpha+2m^2\hat\beta); \qquad\qquad \mu_1 \;=\; \f{\hat g}{2\,m} (1+\hat\alpha)\, .
\ee
This is in exact agreement with the results in \cite{Kim:1973ee}.

\end{document}